\documentclass[prd,twocolumn,showkeys,preprintnumbers,floatfix,nofootinbib,superscriptaddress]{revtex4-1}
\usepackage[T1]{fontenc} 
\usepackage[utf8]{inputenc}
\usepackage{amsfonts} % AMS
\usepackage{amssymb} % AMS
\usepackage{amsmath} % AMS
\usepackage{graphicx} % Include figure files
\usepackage{subfigure} % Include figure files
\usepackage{array} % array
\usepackage{dcolumn} % Align table columns on decimal point
\usepackage{bm} % bold math
\usepackage{latexsym} % latex symbols
\usepackage{longtable} % long tables
\usepackage{hyperref} % hypertext links 
\usepackage{bbold}
\usepackage[dvipsnames]{xcolor}
\usepackage{multirow}
\usepackage{rotating}
\usepackage{comment}
\usepackage{bm}
\usepackage{placeins}
\usepackage{lipsum}
\usepackage{blkarray}

\graphicspath{{./figs/}}
\allowdisplaybreaks
\nopagebreak

\providecommand{\wbar}[1]{\overline#1}
\providecommand{\abs}[1]{\lvert#1\rvert}

\providecommand{\bra}[1]{\langle#1\rvert}
\providecommand{\ket}[1]{\lvert#1\rangle}
\providecommand{\matrixe}[3]{\langle#1\lvert#2\rvert#3\rangle}
\providecommand{\expv}[1]{\langle#1\rangle}

\providecommand{\GeV}{\mathrm{GeV}}
\providecommand{\MeV}{\mathrm{MeV}}

\providecommand{\fm}{\mathrm{fm}}

\providecommand{\CL}{\nonumber\\}

\newcommand{\Dslash}{\ensuremath{{D\kern -0.65em /}}}
\renewcommand{\Re}{\mathop{\mathrm{Re}}}
\renewcommand{\Im}{\mathop{\mathrm{Im}}}

\newcommand{\Ssim}{{\mathcal{S}_\text{sim}}}
\newcommand{\Sfour}{{\mathcal{S}_{A4}}}
\newcommand{\Stwo}{{\mathcal{S}_\text{2pt}}}

\newcommand{\GP}{\widetilde{G}_P}
\newcommand{\FP}{\widetilde{F}_P}

\newcommand{\gsim}{\raisebox{-0.7ex}{$\stackrel{\textstyle >}{\sim}$ }}

\definecolor{green}{rgb}{0.1, 0.8, 0.1}

\providecommand{\CL}{\nonumber\\}

\begin{document}

%%%

\title{Nucleon Isovector Axial Form Factors}

\author{Yong-Chull Jang}
\email{integration.field@gmail.com}
\affiliation{Electron-Ion Collider, Brookhaven National Laboratory, Upton, New York 11973}
\affiliation{Physics Department, Brookhaven National Laboratory, Upton, New York 11973}
\affiliation{Department of Physics, Columbia University, 538 West 120th Street, New York, NY 10027, USA}

\author{Rajan Gupta}
\email{rajan@lanl.gov}
\affiliation{Los Alamos National Laboratory, Theoretical Division T-2, Los Alamos, NM 87545, USA}

\author{Tanmoy Bhattacharya}
\email{tanmoy@lanl.gov}
\affiliation{Los Alamos National Laboratory, Theoretical Division T-2, Los Alamos, NM 87545, USA}

\author{Boram Yoon}
\email{byoon@nvidia.com}
\affiliation{NVIDIA Corporation, Santa Clara, CA 95051, USA}

\author{Huey-Wen Lin}
\email{hwlin@pa.msu.edu}
\affiliation{Department of Physics and Astronomy, Michigan State University, MI, 48824, USA}

\collaboration{Precision Neutron Decay Matrix Elements (PNDME) Collaboration}

\preprint{LA-UR-23-25225}
%
%\pacs{11.15.Ha, % Lattice gauge theory
%      12.38.Gc  % Lattice QCD calculations
%}
%
%\keywords{.}
%
\date{\today}
\begin{abstract}
 We present results for the isovector axial vector form factors
 obtained using thirteen 2+1+1-flavor highly improved staggered quark
 (HISQ) ensembles generated by the MILC collaboration. The calculation
 of nucleon two- and three-point correlation functions has been done
 using Wilson-clover fermions. In the analysis of these data, we
 quantify the sensitivity of the results to strategies used for
 removing excited state contamination and invoke the partially
 conserved axial current relation between the form factors to choose
 between them. Our data driven analysis includes removing
 contributions from multihadron $N \pi$ states that make significant
 contributions.  Our final results are: $g_A = 1.292
 (53)_\text{stat}\,(24)_\text{sys}$ for the axial charge; $g_S = 1.085
 (50)_\text{stat}\, (103)_\text{sys}$ and $g_T = 0.991
 (21)_\text{stat}\, (10)_\text{sys}$ for the scalar and tensor
 charges; $\expv{r_A^2} = 0.439 (56)_\text{stat}
 (34)_\text{sys}$~fm${}^2$ for the mean squared axial charge radius,
 $g_P^\ast = 9.03(47)_\text{stat}(42)_\text{sys} $ for the induced
 pseudoscalar charge; and $g_{\pi NN} =
 14.14(81)_\text{stat}(85)_\text{sys}$ for the pion-nucleon coupling.
 We also provide a parameterization of the axial form factor
 $G_A(Q^2)$ over the range $0 \le Q^2 \le 1$~GeV${}^2$ for use in
 phenomenology and a comparison with other lattice determinations. We
 find that the various lattice data agree within 10\% but are significantly
 different from the extraction of $G_A(Q^2)$ from the $\nu$-deuterium
 scattering data.
\end{abstract}

\maketitle

%/************************************************
\section{Introduction}   %S01
\label{sec:intro}

In ongoing neutrino scattering experiments (T2K, NOvA, MINERvA,
MicroBooNE, SBN), the lack of precise reconstruction of the final
state of the struck nucleus gives rise to uncertainty in the
cross-section.  Theoretical calculations of the cross-section for 
targets, such as ${}^{12}C$, ${}^{16}O$, and  
${}^{40}Ar$, being used in experiments  take as input axial-vector form factor of the
nucleon and build in nuclear effects using nuclear many body
theory~\cite{Ruso:2022qes,Kronfeld:2019nfb,Meyer:2022mix}.  Both of these steps, 
calculating nucleon axial form factors using 
lattice QCD and including nuclear effects using many-body theory, have
uncertainties.  Incorporating nuclear effects involves modeling of
the complex physical phenomena (quasi-elastic, resonance, deep
inelastic scattering) that contribute when considering incoming
neutrino energies up to 5~GeV relevant for ongoing and future (DUNE)
experiments. These complex phenomena make it hard to reconstruct the
incident neutrino energy or the cross-section from the imprecise
knowledge of the final state of the struck nucleus. On the other hand, 
the cleanest
experimental measurements of the nucleon axial-vector form factor from
scattering neutrinos off liquid hydrogen targets are not being carried
out due to safety concerns.  

The MINER$\nu$A experiment~\cite{MINERvA:2023avz} has recently shown
that the axial-vector form factor of the nucleon can be extracted from
the charged current elastic scattering process ${\overline{\nu}}_\mu H
\to \mu^+ n$ in which the free proton in hydrogen (H) gets converted
into a neutron. This opens the door to direct measurements of the
nucleon axial-vector form factor without the need for extraction from
scattering off nuclei, whose analysis involves nuclear corrections
which have unresolved systematics, and making detailed comparisons with
predictions from lattice QCD. For example, our result for the axial
charge radius, $\expv{r_A^2} = 0.439(56)_\text{stat}
(34)_\text{sys}\ {\rm fm}^2$, given in Eq.~\eqref{eq:gArAfinal}, is
consistent, within one combined sigma, with the MINER$\nu$A result,
$\sqrt{\expv{r_A^2}} = 0.73\pm 0.17$~fm.

Similarly, recent advances in simulations of lattice QCD have enabled
robust results for the nucleon charges that have been reviewed by the
Flavor Lattice Averaging Group (FLAG) in their 2019 and 2021
reports~\cite{FlavourLatticeAveragingGroupFLAG:2021npn,FlavourLatticeAveragingGroup:2019iem}).
Results for axial-vector form factors~\cite{Park:2021ypf} are now
available with $\lesssim 10\%$ uncertainty as we show in this work.
At the same time, there continues to be progress in nuclear many-body
theory for the calculation of the neutrino-nucleus
cross-section~\cite{Ruso:2022qes}.

In this work, we present lattice
QCD results for the isovector axial, $G_A(Q^2)$, induced pseudoscalar,
$\widetilde{G}_P(Q^2)$, and pesudoscalar $G_P(Q^2)$ form factors, the
axial, scalar and tensor isovector charges $g_A^{u-d}$, $g_S^{u-d}$
and $g_T^{u-d}$, the axial charge radius squared $\expv{r_A^2}$, the
induced pseudoscalar coupling $g_P^\ast$, and the pion-nucleon
coupling $g_{\pi NN}$.

The calculation has been done using thirteen ensembles generated with
2+1+1-flavors of highly improved staggered quarks (HISQ) by the MILC
collaboration~\cite{Bazavov2012:PhysRevD.87.054505}. The construction
of nucleon two- and three-point correlation functions has been done
using Wilson-clover fermions as described in~\cite{Rajan:2017lxk}. The
analysis of the data generated using this clover-on-HISQ formulation
includes a study of excited state contributions (ESC) in the
extraction of ground state matrix elements (GSME) and a simultaneous
chiral-continuum-finite-volume (CCFV) fit to obtain results at the
physical point, which throughout the paper will be defined as taking
the continuum ($a=0$) and infinite volume ($M_\pi L \to \infty$)
limits at physical light quark masses in the isospin symmetric limit,
$m_u = m_d$, which are set using the neutral pion mass ($M_{\pi^0} =
135$~MeV). The masses of the strange and charm quarks in the lattice
generation have been tuned to be close to their physical values in
each of the thirteen ensembles~\cite{Bazavov2012:PhysRevD.87.054505}.

The three form factors $G_A(Q^2)$, $\widetilde{G}_P(Q^2)$ and
$G_P(Q^2)$ must, up to discretization errors, satisfy the constraint
in Eq.~\eqref{eq:PCAC2} imposed by the partially conserved axial
current (PCAC) relation $\partial_\mu A_\mu = 2 m P$ between the axial
and pseudoscalar currents.  The decomposition of the matrix elements (ME)
into form factors, given in Eqs.~\eqref{eq:aff-a} and~\eqref{eq:aff-ps},
assumes that they are GSME. Post-facto, deviations from the PCAC
relation larger than those expected due to lattice discretization
artifacts are indicative of residual ESC in the extraction of ME
from the spectral decomposition of the three-point
correlation functions.  They point to the need for reevaluation of the
key inputs in this analysis---the number and energies of the excited
states that contribute significantly to the three-point functions.
The strategies used to remove ESC are described in
Secs.~\ref{ssec:ff-corfit-stg} and~\ref{sec:charges}, and the use
of the PCAC relation to evaluate how well ESC have been controlled is
discussed in Sec.~\ref{ssec:ppd-pcac}.

In Ref.~\cite{Jang:2019vkm}, we showed that the standard method of
taking the excited-state spectrum from fits to the nucleon two-point
correlation function to analyze the three-point functions lead to form
factors that fail the PCAC test by almost a factor of two on the
physical pion mass ensemble $a09m130W$, and identified the cause as
enhanced contributions to ME from multihadron, $N \pi$, excited states
that have mass gaps smaller than of radial
excitations~\cite{Bar:2018akl,Bar:2018xyi}. These contributions had
been missed in all prior lattice calculations. Including $N \pi$
excited states in the analysis reduces the disagreement to within
10\%, an amount that can be attributed to discretization effects. In
this paper, we include $N\pi$ states in the analysis of all thirteen
ensembles described in Table~\ref{tab:ens}. Data from various analyses
discussed in Secs.~\ref{ssec:aff-extrap},~\ref{ssec:aff-GA},
and~\ref{sec:extrap-gpstar} are then extrapolated to the physical
point using simultaneous CCFV fits and results compared to understand
systematics..

\begin{table*}[tbp]    %T01
\begin{center}
\renewcommand{\arraystretch}{1.2} % Change horizontal spacing
\begin{ruledtabular}
\begin{tabular}{l|ccc|cc|cccc|cc}
Ensemble ID & $a$  & $M_\pi^{\rm sea}$ & $M_\pi^{\rm val}$ & $L^3\times T$    & $M_\pi^{\rm val} L$ & $\tau/a$ & $N_\text{conf}$  & $N_{\rm meas}^{\rm HP}$  & $N_{\rm meas}^{\rm LP}$  & $Q^2|_{\rm max}^{n^2 = 6}$  & $Q^2|_{\rm max}^{n^2 = 10}$   \\
            & (fm) & (MeV)             & (MeV)             &                  &                     &          &                  &                          &                          & (GeV)${}^2$  \\
\hline
$a15m310 $      & 0.1510(20) & 306.9(5) & 320.6(4.3) & $16^3\times 48$ & 3.93 &  $\{5,6,7,8,9\}$    & 1917 & 7668  & 122,688   & 1.297   & 1.92   \\
\hline                                                                                                                                           
$a12m310 $      & 0.1207(11) & 305.3(4) & 310.2(2.8) & $24^3\times 64$ & 4.55 &  $\{8,10,12\}$      & 1013 & 8104  &  64,832   & 0.920  & 1.435  \\
$a12m220S$      & 0.1202(12) & 218.1(4) & 225.0(2.3) & $24^3\times 64$ & 3.29 & $\{8, 10, 12\}$     & 946  & 3784  &  60,544   & 0.909  & 1.358  \\
$a12m220 $      & 0.1184(10) & 216.9(2) & 227.9(1.9) & $32^3\times 64$ & 4.38 & $\{8, 10, 12\}$     & 744  & 2976  &  47,616   & 0.568  & 0.884  \\
$a12m220L$      & 0.1189(09) & 217.0(2) & 227.6(1.7) & $40^3\times 64$ & 5.49 & $\{8,10,12,14\}$    & 1000 & 4000  & 128,000   & 0.374  & 0.595  \\
\hline                                                                                                                                           
$a09m310 $      & 0.0888(08) & 312.7(6) & 313.0(2.8) & $32^3\times 96$ & 4.51 & $\{10,12,14,16\}$   & 2263 & 9052  & 114,832   & 0.961  & 1.421  \\
$a09m220 $      & 0.0872(07) & 220.3(2) & 225.9(1.8) & $48^3\times 96$ & 4.79 & $\{10,12,14,16\}$   & 964  & 7712  & 123,392   & 0.470  & 0.736  \\
$a09m130W$      & 0.0871(06) & 128.2(1) & 138.1(1.0) & $64^3\times 96$ & 3.90 & $\{8,10,12,14,16\}$ & 1290 & 5160  & 165,120   & 0.277  & 0.443  \\
\hline                                                                                                                                           
$a06m310 $      & 0.0582(04) & 319.3(5) & 319.6(2.2) & $48^3\times 144$& 4.52 & $\{16,20,22,24\}$   & 1000 & 8000  &  64,000   & 0.840  &        \\
$a06m310W$      &            &          &            &                 &      & $\{18,20,22,24\}$   & 500  & 2000  &  64,000   & 0.846  &        \\
$a06m220 $      & 0.0578(04) & 229.2(4) & 235.2(1.7) & $64^3\times 144$& 4.41 & $\{16,20,22,24\}$   & 650  & 2600  &  41,600   & 0.504  &        \\
$a06m220W$      &            &          &            &                 &      & $\{18,20,22,24\}$   & 649  & 2596  &  41,546   & 0.509  &        \\
$a06m135 $      & 0.0570(01) & 135.5(2) & 135.6(1.4) & $96^3\times 192$& 3.7  & $\{16,18,20,22\}$   & 675  & 2700  &  43,200   & 0.294  & 0.475  \\
\end{tabular}
\end{ruledtabular}
\caption{The parameters of the 2+1+1-flavor HISQ ensembles generated
  by the MILC collaboration and analyzed in this study are quoted from
  Ref.~\cite{Bazavov2012:PhysRevD.87.054505}.  On two ensembles,
  $a06m310$ and $a06m220$, a second set of calculations labeled
  $a06m310W$ and $a06m220W$, have been done with a larger smearing
  size $\sigma$ as described in Ref.~\protect\cite{Gupta:2018qil}.  In
  this clover-on-HISQ study, all fits are made versus $M_\pi^{\rm
    val}$, which is tuned to be close to the Goldstone pion mass
  $M_\pi^{\rm sea}$. The finite-size effects are analyzed in terms of
  $M_\pi^{\rm val} L$.  Columns 7---10 give the values of the
  source-sink separation $\tau$ used in the calculation of the
  three-point functions, the number of configurations analyzed, and
  the number of measurements made using the high precision (HP) and
  the low precision (LP) truncation of the inversion of the
  Wilson-clover operator~\protect\cite{Yoon:2016dij}. The last column
  gives the value of $Q^2|_{\rm max}$ for two cuts, ${n^2\le 6}$
  (${n^2\le 5}$ for the four $a06m310$ and $a06m220$ ensembles) and
  ${n^2\le 10}$ used in the analysis. The full set of $Q^2$ values
  simulated on each ensemble are given in
  Tables~\protect\ref{tab:ff-a15m310}---\protect\ref{tab:ff-a06m135}. }
\label{tab:ens}
\end{center}
\end{table*}

In order to extract $g_A$ and $\expv{r_A^2}$, we parameterize the
$Q^2$ behavior of $G_A(Q^2)$ using the
dipole and the model independent
$z$-expansion. We find that the dipole ansatz does not provide a good
fit and our final results are obtained using the model independent
$z$-expansion.  We show that the pion-pole dominance (PPD) hypothesis, Eq.~\eqref{eq:PPD}, 
tracks the improvement observed in satisfying the PCAC relation when
$N \pi$ states are included in the analysis. We, therefore, use it to 
parameterize $\widetilde{G}_P(Q^2)$ and extract $g_P^\ast$ and $g_{\pi NN}$ in
Sec.~\ref{sec:aff-gpx}. Similarly, the analysis of the ESC in
isovector charges extracted from the forward matrix elements is
carried out using information from both the 2- and 3-point correlation
functions and the noninteracting energy of the lowest $N\pi$ state.

Our final result for the axial form factor, parameterized using the
$z^2$ truncation, is given in Eq.~\eqref{eq:gAz2final}; the axial
charge obtained from extrapolating it to $Q^2 = 0$ in 
Eq.~\eqref{eq:gA-z}, and the charge radius in Eq.~\eqref{eq:rA-z}. The
results for the induced pseudoscalar charge $g_P^\ast$ and $g_{\pi
  NN}$ are given in Eqs.~\eqref{eq:gPstar-z}
and~\eqref{eq:gpiNN-z}. Lastly, the results for the three isovector
charges $g_{A,S,T}^{u-d}$ from the forward matrix elements are given
in Eq.~\eqref{eq:summary-charge-3RD}.

This paper is organized as follows.  In Sec.~\ref{sec:ff}, we
briefly review the notation and the methodology for the extraction of
the three form factors: the axial, $G_A$, the induced pseudoscalar
$\widetilde{G}_P$,  and the pseudoscalar, $G_P$, from matrix elements
of the axial and pseudoscalar currents within ground state nucleons.
In Sec.~\ref{ssec:ff-corfit-stg}, we explain the three strategies used
to remove the ESC to the three-point functions. The analysis of the
form factors with respect to how well they satisfy the relations
imposed between them by PCAC relation and the PPD 
hypothesis is presented in Sec.~\ref{ssec:ppd-pcac}. Based on this
analysis, we present our understanding of the excited states that
contribute in Sec.~\ref{ssec:ESS}.  The parameterization of the axial
form factors as a function of $Q^2$ and the extraction of the axial
charge $g_A$ and the charge radius squared $\expv{r_A^2}$ is carried
out in Sec.~\ref{sec:aff-gA-rA}.  Parameterization of the induced
pseudoscalar form factor, $\widetilde{G}_P$, and the extraction of the
induced pseudoscalar coupling $g_P^\ast$ and the pion-nucleon coupling
$g_{\pi NN}$ is carried out in Sec.~\ref{sec:aff-gpx}.  The calculation of
the isovector charges $g_{A,S,T}^{u-d}$ from forward matrix elements
is described in Sec.~\ref{sec:charges}. A summary of our results and a
comparison with previous lattice calculations is presented in the
concluding Sec.~\ref{sec:con}. Six appendices give further details
of the analysis and the data.\looseness-1

\section{Methodology for Extracting the Form Factors}  %S02
\label{sec:ff}
The matrix elements of the axial $A_\mu = \bar{u}\gamma_\mu\gamma_5 d$
and pseudoscalar $P = \bar{u}\gamma_5 d$ currents between the ground
state of the nucleon can be decomposed, in the isospin symmetric
limit, into the axial $G_A$, induced pseudoscalar $\widetilde{G}_P$,
and pseudoscalar $G_P$ form factors as
\begin{align}
  & \matrixe{N(\vec{p}_f)}{A_\mu (\vec{Q})}{N(\vec{p}_i)} \CL &=
  {\overline u}(\vec{p}_f)\left[ G_A(Q^2) \gamma_\mu \gamma_5
  + q_\mu \gamma_5 \frac{\widetilde{G}_P(Q^2)}{2 M}\right] u(\vec{p}_i) \,,
  \label{eq:aff-a}
  \\
  & \matrixe{N(\vec{p}_f)}{P (\vec{q})}{N(\vec{p}_i)}  =
  {\overline u}(\vec{p}_f)\left[ G_P(Q^2) \gamma_5 \right]u(\vec{p}_i) \,,
  \label{eq:aff-ps}
\end{align}
where $u(\vec{p}_i)$ is the nucleon spinor with momentum $\vec{p}_i$,
$q = p_{f} - p_{i}$ is the momentum transferred by the current, $Q^2 =
-q^2 = \vec{p}_f^2 - (E({\bm p}_f)-E({\bm p}_i))^2$ is the spacelike
four momentum squared transferred.  The spinor normalization used is
\begin{align}
  \sum_s u(\bm{p},s) \wbar{u}(\bm{p},s) &= \frac{E(\bm{p})\gamma_4 - i\gamma\cdot\bm{p} +M}{2E(\bm{p})} \,.
\end{align}
The process of obtaining the GSME needed in Eqs.~\eqref{eq:aff-a}
and~\eqref{eq:aff-ps} from fits to 2- and 3-point correlation
functions is described next.

\subsection{Two- and three-point correlation functions}

The lattice calculation starts with the measurement and analysis of
the two- and three-point correlation functions
$C^\text{2pt}(\bm{p};\tau)$ and $C_J({\bm q},t;\tau)$ constructed using the
nucleon interpolating operator ${\cal N}$,
\begin{align}
  {\cal N}(x) =& \epsilon^{abc} \left[ q_1^{aT}(x) C \gamma_5 \frac{1\pm\gamma_4}{2} q_2^b(x) \right] q_1^{c}(x) \,,
\label{eq:Nchi}
\end{align}
where the $\pm$ sign give  positive parity
states propagating forward/backward in time. 
The spectral decompositions of the two time-ordered correlation functions are 
\begin{align}
  C^\text{2pt}(\bm{p};\tau) \equiv& \matrixe{\Omega}{{\cal T}({\cal N}(\tau) \wbar{{\cal N}}(0))}{\Omega} = \sum_{i=0} \abs{A_i^\prime}^2 e^{-E_i \tau} \,, 
  \label{eq:C2pt}
\end{align}
and
\begin{align}
  C_J(\bm{q};t,\tau) \equiv& \matrixe{\Omega}{{\cal T}({\cal N}(\tau) J_\Gamma (t)\wbar{{\cal N}}(0))}{\Omega} \,, \CL
  =& 
  \sum_{i,j=0} {A_i^\prime}^\ast {A_j} \matrixe{i^\prime}{J_\Gamma}{j} e^{-E_i t -M_j(\tau-t)} \,, 
  \label{eq:C3pt-decomp}
\end{align}
where $J_\Gamma = A_\mu$ or $ P$ is the quark bilinear current 
inserted at time $t$ with momentum $\bm q$, and $\ket{\Omega}$ is the
vacuum state.  In our set up, the
nucleon state $\ket{j}$ is, by construction,  projected to zero momentum, i.e., $p_j =
(M, \bm{0})$, whereas $\bra{i^\prime}$ is projected onto definite
momentum $p_i = (E,\bm{p})$ with $\bm{p} = - \bm{q}$ by momentum
conservation.  Consequently, the states on the two sides of the
inserted operator $J$ are different for all ${\textbf q} \neq 0$. The prime in $\langle i^\prime|$
indicates that this state can have nonzero momentum.

For large time separations, $\tau$ and $\tau-t$, only the ground state
contributes and the GSME, 
$\matrixe{0^\prime}{J}{0}$, whose Lorentz covariant decomposition is
given in Eqs.~\eqref{eq:aff-a} and~\eqref{eq:aff-ps}, can be extracted
reliably. Assuming this is the case, and choosing the nucleon spin projection
to be in the ``3'' direction, the decompositions become
\begin{align}
  C_{A_i}(\bm{q}) \to& K^{-1} \left[ -q_i q_3 \frac{\GP}{2M} + \delta_{i3} (M+E)G_A \right] 
  \label{eq:Ai-ff-decomp} \,,\\
  C_{A_4}(\bm{q}) \to& K^{-1} q_3 \left[(M-E)\frac{\GP}{2M} + G_A\right] \,,\\
  C_{P}(\bm{q}) \to& K^{-1} q_3 G_P \,,
\end{align}
where $i \in 1,2,3$ and the kinematic factor $K^{-1} \equiv
\sqrt{2E(E+M)}$. These correlation functions are complex valued, and
the signal, for the CP symmetric theory, is in $\Im C_{A_i}$, $\Re
C_{A_4}$, and $\Re C_{P}$.

For a given $\bm{q}^2 \neq 0$, the correlators with momentum
combinations ${\bf{q}} = (2 \pi/L) {\bf n} \equiv  (2 \pi/L) (n_1, n_2,
n_3)$ related by cubic symmetry can be averaged to increase the statistics before 
making fits. We construct the following averages $\wbar{A}_\mu$ and $\wbar{P}$:
\begin{align}
  \wbar{A_i}(\bm{q^2}) \equiv& \frac{1}{\alpha_1\bm{q}^2} \sum_{\bm{q}} \mathrm{sgn}(q_i q_3) C_{A_i}(\bm{q}) \CL
  &\to K^{-1} \frac{\GP}{2M} \,, \quad (i=1,2) \,, \label{eq:avg-a}\\
  \wbar{A_{3,L}}(\bm{q^2}) \equiv& \frac{1}{\alpha_3 \bm{q}^2} \sum_{q_3\neq 0} C_{A_3}(\bm{q}) \CL
  \to&  K^{-1} \left[ - \frac{\GP}{2M} + \frac{(N-\beta)}{\alpha_3 \bm{q}^2}(M+E) G_A \right] \,, \label{eq:avg-b}\\
  \wbar{A_{3,T}}(\bm{q^2}) \equiv& \frac{1}{\beta} \sum_{q_3=0} C_{A_3}({\bm{q}}) \to K^{-1} (M+E) G_A \,, \label{eq:avg-c}\\
  \wbar{A}_4(\bm{q^2}) \equiv & \frac{1}{\alpha_3 \bm{q}^2} \sum_{\bm{q}} q_3 C_{A_4}(\bm{q}) \CL
  \to& K^{-1} \left[ (M-E)\frac{\GP}{2M} + G_A \right]  \,, \label{eq:avg-d}\\
  \wbar{P}(\bm{q^2}) \equiv & \frac{1}{\alpha_3 \bm{q}^2} \sum_{\bm{q}} q_3 C_{P}(\bm{q}) 
  \to K^{-1} G_P  \label{eq:avg-e}\,,
\end{align}
where $\mathrm{sgn}(x)=x/\abs{x}$ is a sign function with
$\mathrm{sgn}(0)=0$, $\alpha_1 \equiv \sum \abs{n_1 n_3}/\bm{n}^2$ ,
$\alpha_3 \equiv \sum_{q_3} {n_3^2} / \bm{n}^2 = N/3$, $\bm{q} =
(2\pi/L)\bm{n}$, $\beta \equiv \sum_{q_3=0} 1$, and $N \equiv
\sum_{\bm{q}} 1$ is the number of equivalent (under the cubic group) momenta averaged.

The pseudoscalar form factor, $G_P$, is given uniquely by
Eq.~\eqref{eq:avg-e}.  For a subset of momenta, $G_A$ and
$\widetilde{G}_P$ are determined uniquely from Eqs.~\eqref{eq:avg-a}
and~\eqref{eq:avg-c}. In general, we solve the over-determined
system of equations, Eqs.~\eqref{eq:avg-a}--\eqref{eq:avg-d}.  Of
these, correlators $\wbar{A}_{3,L}$ and $\wbar{A}_4$ are nonvanishing
for all $\bm{q}$, and are thus sufficient to solve for $G_A$ and
$\widetilde{G}_P$. In practice, the $A_4$ correlator has a poor signal
and is dominated by excited states contributions, which we exploit 
to determine the relevant low-lying excited states. These 
turn out to be towers of multihadron $N \pi$ and $N\pi \pi$
states. We find that including these states in fits to the spectral
decompositions given in Eqs.~\eqref{eq:C2pt} and~\eqref{eq:C3pt-decomp} 
is essential for extracting the GSME. With the GSME in hand, the form
factors $G_A$ and $\widetilde{G}_P$ are determined using
Eqs.~\eqref{eq:avg-a}--\eqref{eq:avg-c}.

\subsection{Strategies to extract ground state matrix elements}
\label{ssec:ff-corfit-stg}

Calculations of nucleon correlation functions face two key
challenges. First, the statistical signal-to-noise ratio decays
exponentially with the source-sink separation $\tau$ as $e^{-(M_N -
  1.5M_\pi) \tau}$~\cite{Parisi:1983ae,Lepage:1989hd}. This limits
current measurements of two-point (three-point) functions to $\lesssim
2$ ($\lesssim 1.5$)~fm. Second, at these $\tau$, the residual
contribution of many theoretically allowed radial and multihadron
excited states can be significant. These states arise because the
standard nucleon interpolating operator ${\cal N}$, defined in
Eq.~\eqref{eq:Nchi}, used to construct the correlation functions in
Eqs.~\eqref{eq:C2pt} and~\eqref{eq:C3pt-decomp}, couples to nucleons
and all its excitations with positive parity including multihadron
states, the lowest of which are $N({\bf p}) \pi(-{\bf p})$ and $N({\bf
  0}) \pi({\bf 0}) \pi({\bf 0})$.  The goal is to remove the
contributions of all these excited states to three-point functions to
obtain the GSME, $\matrixe{0^\prime}{J}{0}$, which we do by fitting
the averaged correlators $\wbar{A}_\mu$ and $\wbar{P}$ using
Eq.~\eqref{eq:C3pt-decomp}.

An important note applicable to all fits used to remove excited
states. For all our ensembles, the energy of the two lowest in these
tower of positive parity states, $N({\bf p}=1) \pi({\bf p}=-1)$ and
$N({\bf 0}) \pi({\bf 0}) \pi({\bf 0})$, are approximately the
same. Since fits to Eqs.~\eqref{eq:C2pt} and~\eqref{eq:C3pt-decomp}
depend only on $E_i$ and not on the nature of the states, the
contribution of both states is taken into account using the $E_1$
calculated for either state. Thus, the reader
should understand that the contribution of both states are being
included when, for brevity, we say $N \pi$ state.

To extract the GSME, we need to address two
questions: (i) which excited states make large contributions and (ii)
how large is this contribution to various observables. The most direct
(and statistically the best motivated assuming that a common set of states dominate the ESC in all 
correlators) way to get the ground-state
matrix element that addresses these questions is to simultaneously
fit, with the full covariance matrix, all five nucleon three-point
functions including two or more excited states, and then solve the linear
set of Eqs.~\eqref{eq:avg-a}--\eqref{eq:avg-e}.

In general, simultaneous fits to the current data (3-point or a combination of 2- and 3-point) do not resolve the excited
states, i.e., there are large regions of parameter space where fits
give similar $\chi^2/dof$. The one exception is fits to the
correlation functions $\langle N A_4 N\rangle$ that, as discussed
below, play a central role in our analysis as they expose the large
contribution of $N \pi$ states.  Unfortunately, even 2-state simultaneous fits to
all five nucleon three-point functions are not stable for all momentum
channels and ensembles. We, therefore, resort to taking the energies
and amplitudes, especially those of the ground state, from separate,  
but within a single overall jackknife process, fits to the
2-point function.

The analysis of the nucleon two-point functions and the extraction of
the spectrum is presented in Appendix~\ref{sec:meff-gen} and the
extrapolation of the data for the nucleon mass, $M_N$, to the
continuum limit in Appendix~\ref{sec:spectrum}. Results for the
excited state parameters, i.e., the energies $E_i(\bm{q})$, the masses
$M_i$, and the amplitudes $A_i$, have large uncertainty. For example,
in a four state fit, there is a large region of parameter space where
fits have similar $\chi^2/dof$. 

In short, statistical precision of current data does not allow
simultaneous fits to all five nucleon three-point functions using a
3-state (or higher) fit with the full covariance matrix. Even a robust
determination of the energies, $E_i$, and amplitudes, $A_i$, of excited states that make
significant contributions from fits to 2-point functions is
lacking. Our best approach is a hybrid of using information from fits
to 2- and 3-point functions.  To this end, we construct three
strategies with different estimates of excited-state parameters to fit
the three-point data using Eq.~\eqref{eq:C3pt-decomp}. These are
described next.

In the standard approach, labeled $S_{\rm 2pt}$, we take $E_i$, $M_j$, 
$A_0^\prime$ and $A_0$ from $4$-state fits to $C^\text{2pt}$, and 
input them into an $m$-state truncation ($m\leq n$) of
Eq.~\eqref{eq:C3pt-decomp} to extract the matrix element
$\matrixe{0^\prime}{J}{0}$. In this paper, we truncate the spectral
decompositions given in Eqs.~\eqref{eq:C3pt-decomp}
and~\eqref{eq:C2pt} at $m=3$ and $n=4$, respectively. 

The second strategy, labeled $\Sfour$, was proposed in
Ref.~\cite{Jang:2019vkm}. Again $E_0$, $M_0$, $A_0^\prime$ and $A_0$
are taken from $4$-state fits to $C^\text{2pt}$, however, $E_1$ and
$M_1$ are determined from two-state fits to the three-point correlator
$\wbar{A}_4$.
% using Eq.~\eqref{eq:avg-d}.  
The output $E_1$ and $M_1$ are then fed into the fits to the other
four correlation functions defined in
Eqs.~\eqref{eq:avg-a}--\eqref{eq:avg-c},~\eqref{eq:avg-e}. This
strategy assumes that the same [first] excited state parameters apply
to all five correlation functions, and these are given by fits to
$\wbar{A}_4$.

The third strategy $\Ssim$ is similar to $\Sfour$ except that $E_1$
and $M_1$ are outputs of simultaneous two-state fits to all five
three-point correlators defined in
Eqs.~\eqref{eq:avg-a}--\eqref{eq:avg-e}. It is, from a statistical
point of view, better motivated than $\Sfour$ because the
underlying assumption in both cases is that the same excited states 
contribute to all five correlators. It avoids the two-step procedure
used in $\Sfour$, i.e., first obtain $E_1$ and $M_1$ from fits to
$\wbar{A}_4$ and then use them in fits to the other four correlators.
In $\Ssim$, we used the averaged correlator  $\wbar{A}_{xy} =
(\wbar{A}_1+\wbar{A}_2)/2$, since these two correlators are equivalent
under cubic rotational symmetry, thus reducing the number of
correlators fit simultaneously to four. 

We used the full covariance matrix for all fits to the 2-point and
3-point functions with the $\Stwo$ and $\Sfour$ strategies.  In the
$\Ssim$ strategy, the covariance matrix was restricted to be block
diagonal in each correlation function.  In the $O(1000)$ fits made to remove ESC 
(ensembles $\otimes$ $Q^2$ values $\otimes$ correlation functions 
$\otimes$ strategies) the selection of parameters was done individually due to the large differences in
ESC behavior versus ensembles, $Q^2$ values, and correlation functions. 

We emphasize from the very outset that in all fits with each of the
three strategies, the excited state amplitudes, $A_i^{(\prime)}$ and
$A_j$, are not needed since these arise only in the combinations
$\abs{A_i^\prime}\abs{A_j} \matrixe{i^\prime}{J}{j}$, which are 
fit parameters but are not used thereafter in the analysis.  Second, the 
ground state parameters, $M_0$, $E_0$, ${A_0^{\prime}}$ and
${A_0}$ are common for all three strategies and are taken from
four-state fits to the two-point correlators.

%\clearpage
%\newpage
\begin{figure*}[!h]  %F01
  \centering
 \hspace{-2em}
  \includegraphics[trim=0 15 0 0, clip, width=0.50\columnwidth]{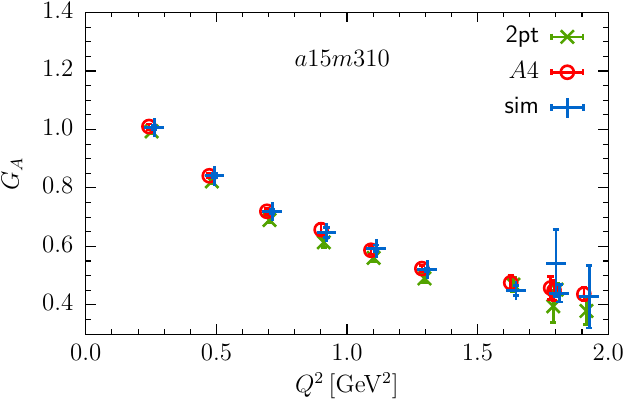}
  \includegraphics[trim=0 15 0 0, clip, width=0.50\columnwidth]{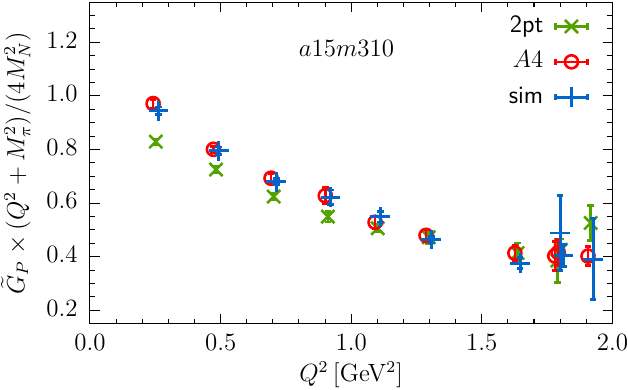}
  \includegraphics[trim=0 15 0 0, clip, width=0.50\columnwidth]{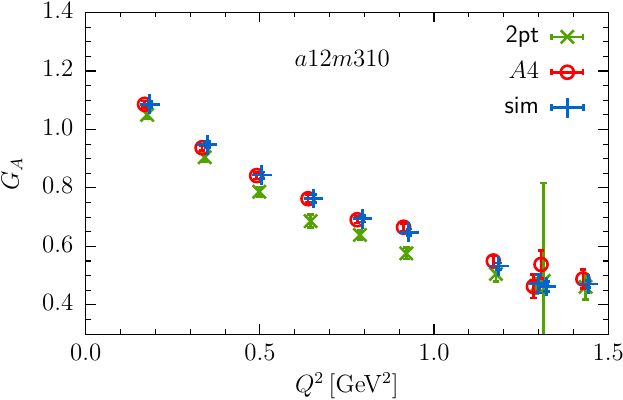}
  \includegraphics[trim=0 15 0 0, clip, width=0.50\columnwidth]{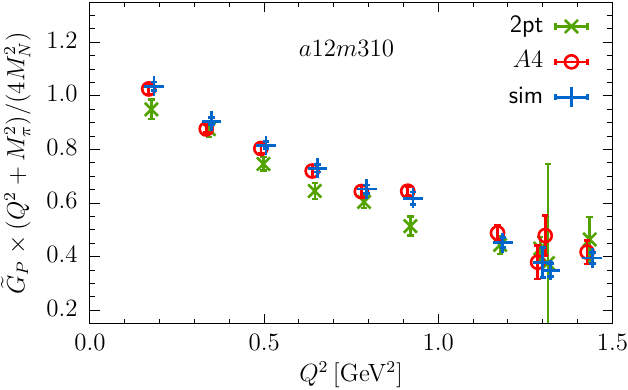}
  \\ %\vspace{0.5em}
  \hspace{-2em}
  \includegraphics[trim=0 15 0 0, clip, width=0.50\columnwidth]{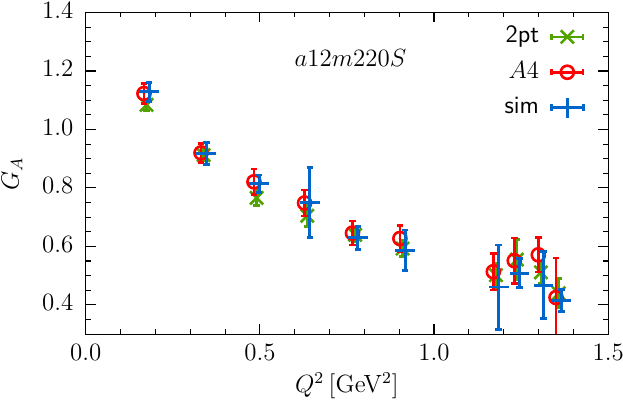}
  \includegraphics[trim=0 15 0 0, clip, width=0.50\columnwidth]{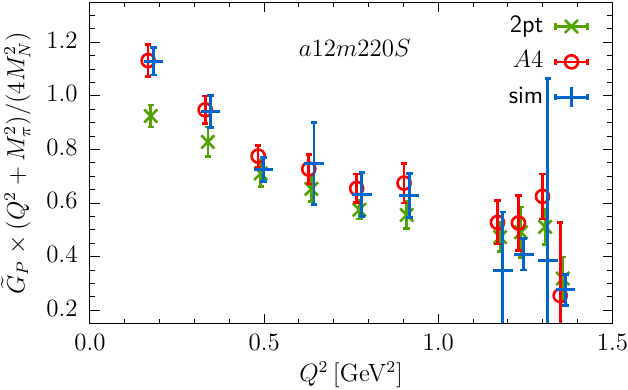}
  \includegraphics[trim=0 15 0 0, clip, width=0.50\columnwidth]{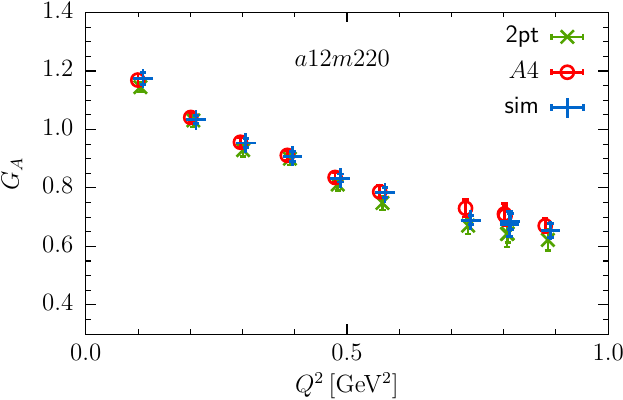}
  \includegraphics[trim=0 15 0 0, clip, width=0.50\columnwidth]{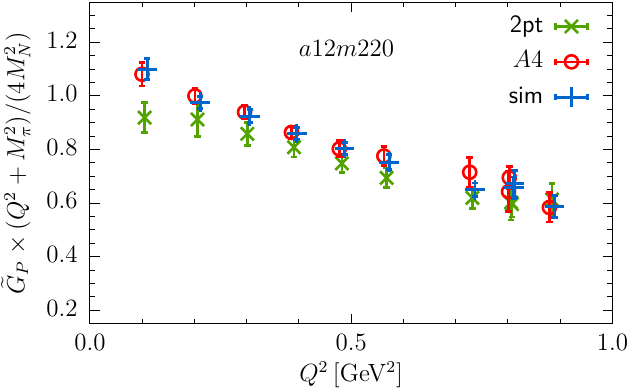}
  \\ %\vspace{0.5em}
  \hspace{-2em}
  \includegraphics[trim=0 15 0 0, clip, width=0.50\columnwidth]{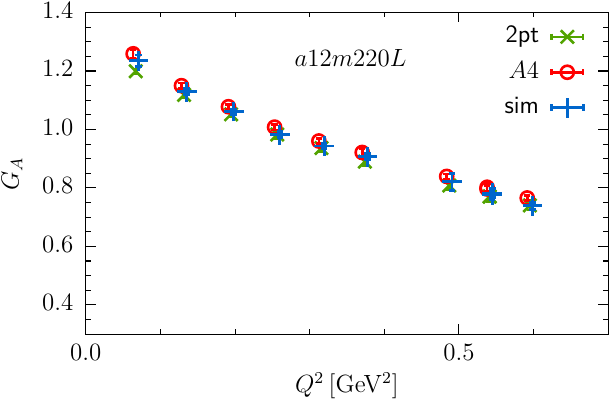}
  \includegraphics[trim=0 15 0 0, clip, width=0.50\columnwidth]{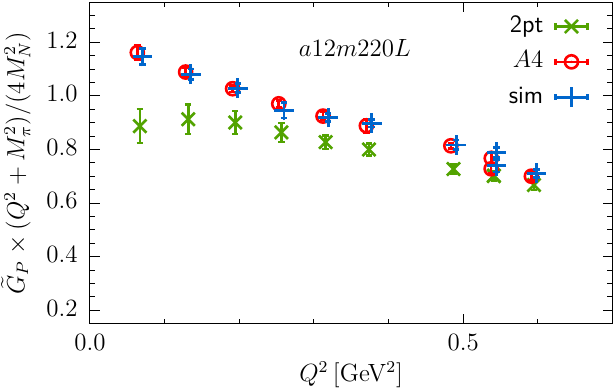}
  \includegraphics[trim=0 15 0 0, clip, width=0.50\columnwidth]{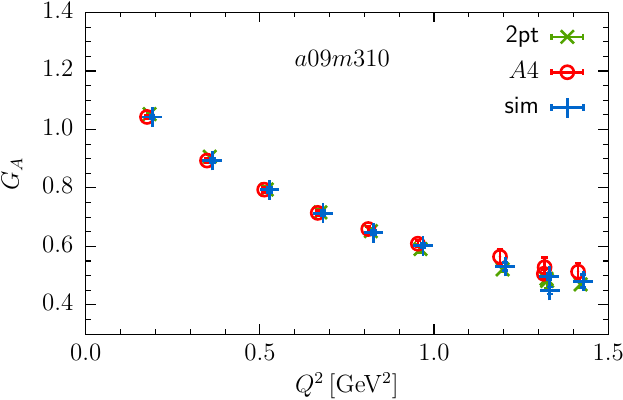}
  \includegraphics[trim=0 15 0 0, clip, width=0.50\columnwidth]{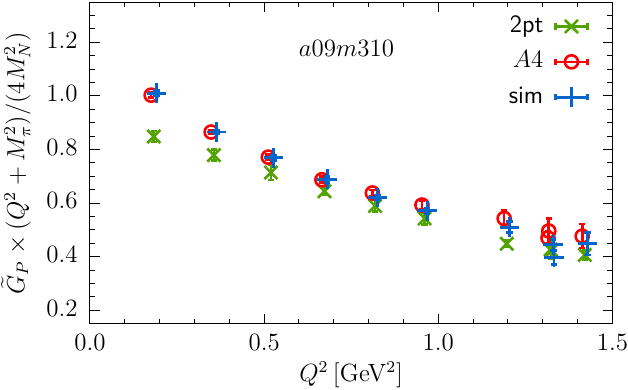}
  \\ %\vspace{0.5em}
  \hspace{-2em}
  \includegraphics[trim=0 15 0 0, clip, width=0.50\columnwidth]{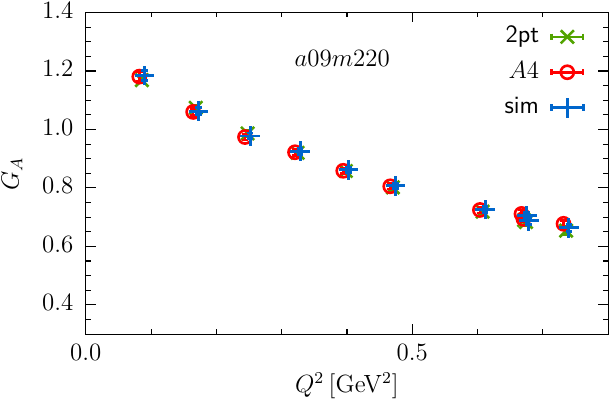}
  \includegraphics[trim=0 15 0 0, clip, width=0.50\columnwidth]{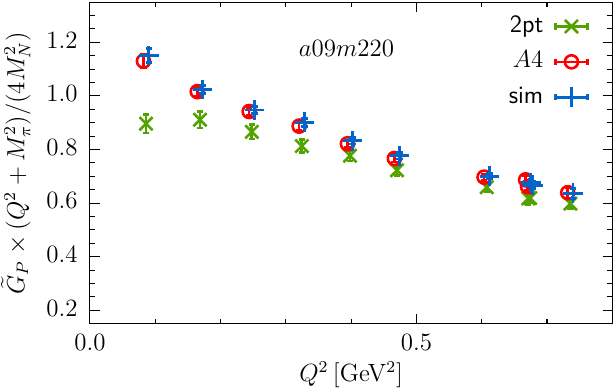}
  \includegraphics[trim=0 15 0 0, clip, width=0.50\columnwidth]{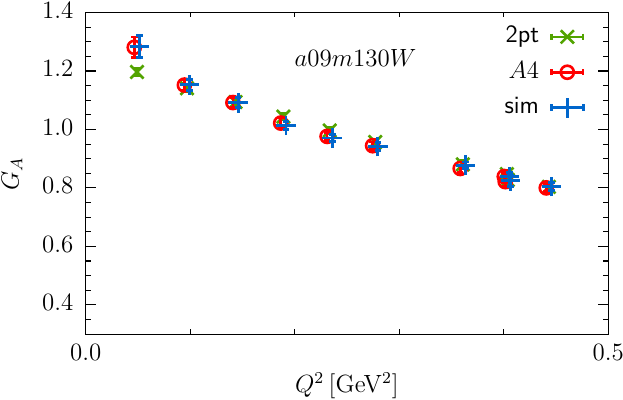}
  \includegraphics[trim=0 15 0 0, clip, width=0.50\columnwidth]{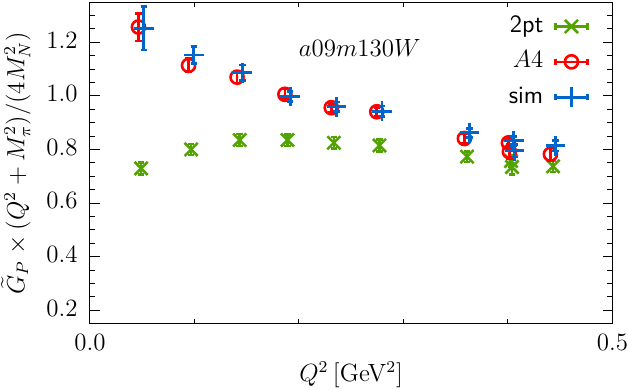}
  \\ %\vspace{0.5em}
  \hspace{-2em}
  \includegraphics[trim=0 15 0 0, clip, width=0.50\columnwidth]{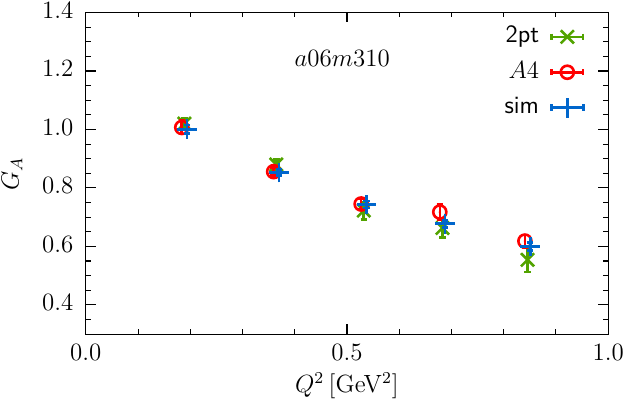}
  \includegraphics[trim=0 15 0 0, clip, width=0.50\columnwidth]{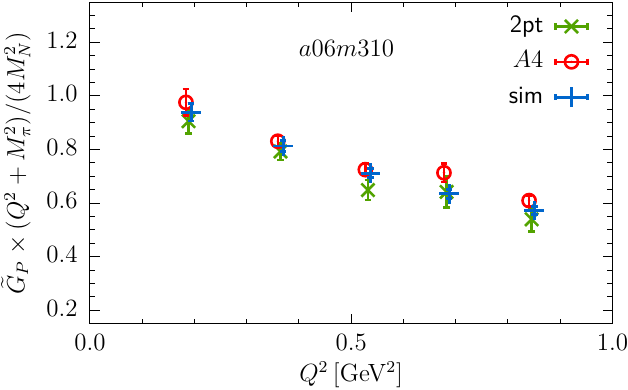}
  \includegraphics[trim=0 15 0 0, clip, width=0.50\columnwidth]{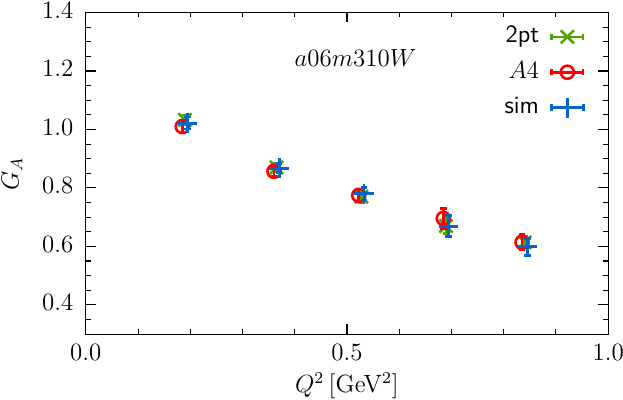}
  \includegraphics[trim=0 15 0 0, clip, width=0.50\columnwidth]{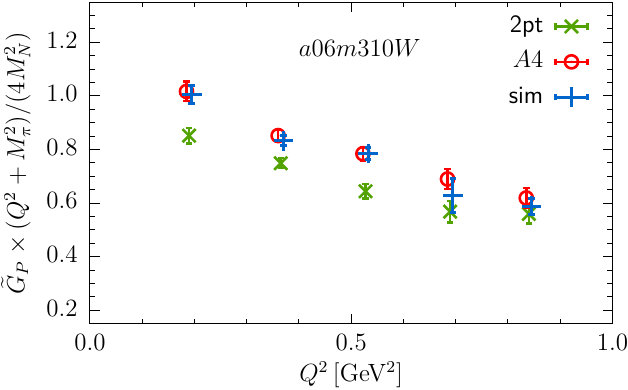}
  \\ %\vspace{0.5em}
  \hspace{-2em}
  \includegraphics[trim=0 15 0 0, clip, width=0.50\columnwidth]{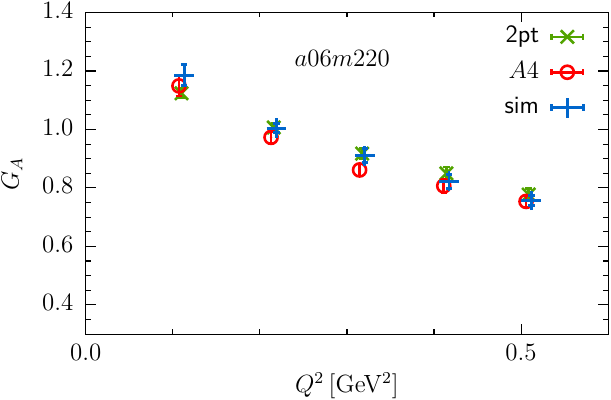}
  \includegraphics[trim=0 15 0 0, clip, width=0.50\columnwidth]{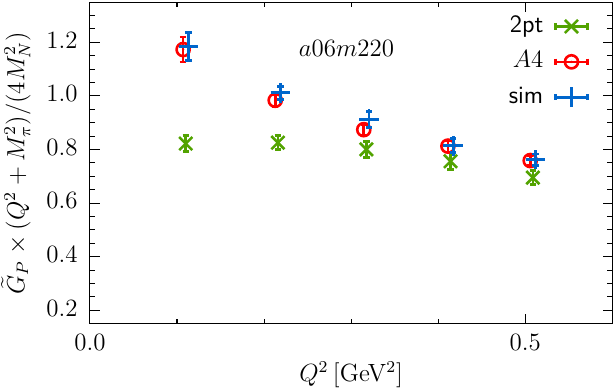}
  \includegraphics[trim=0 15 0 0, clip, width=0.50\columnwidth]{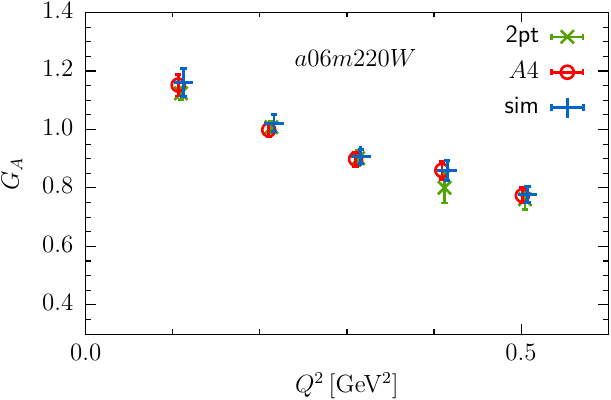}
  \includegraphics[trim=0 15 0 0, clip, width=0.50\columnwidth]{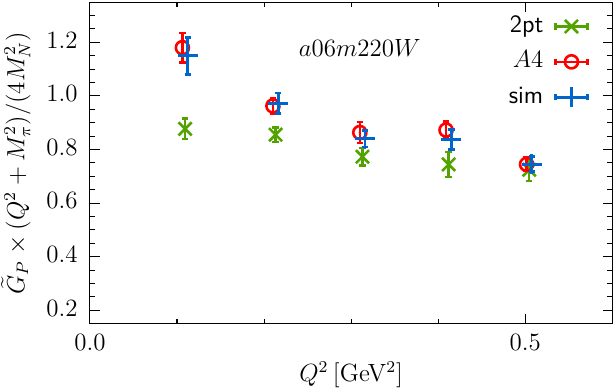}
  \\ %\vspace{0.5em}
%  \hspace{-2em}
  \includegraphics[trim=0 4 0 0, clip, width=0.50\columnwidth]{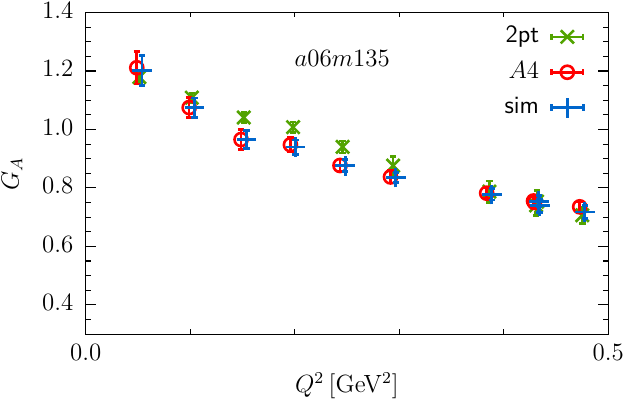}
  \includegraphics[trim=0 4 0 0, clip, width=0.50\columnwidth]{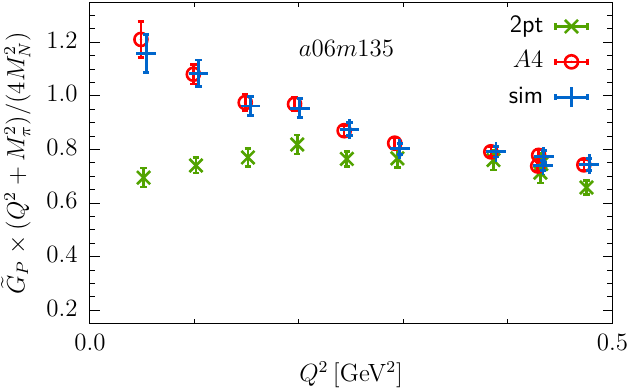}
  \phantom{
  \includegraphics[trim=0 4 0 0, clip, width=0.50\columnwidth]{figs/man/GA_Qsq_a06m135}
  \includegraphics[trim=0 4 0 0, clip, width=0.50\columnwidth]{figs/man/GP_Qsq_a06m135}
  }

  \caption{Data from the 13 ensembles with $a\approx
    0.15,\ 0.12,\ 0.09,$ and $ 0.06\,\fm$ for the unrenormalized axial
    form factor $G_A(Q^2)$ (first and third columns) and
    $(Q^2+M_\pi^2) \widetilde{G}_P(Q^2) /(4M_N^2)$ (second and fourth
    columns) plotted versus $Q^2$.  Each panel compares the data
    obtained using the three strategies $\Stwo$, $\Sfour$, and $\Ssim$
    for controlling ESC. }
  \label{fig:GA-Scomp-all}
\end{figure*}

The unrenormalized values of the three form factors at various values
of $Q^2$ simulated, for each of the three strategies and for the 13
ensembles are given in
Tables~\ref{tab:ff-a15m310}--\ref{tab:ff-a06m135} in
Appendix~\ref{sec:ff-tabs}. The size of the effect of $N\pi$ state can
already be inferred from the difference between the $\Stwo$ and
$\Ssim$ data even though $\Ssim$ includes only the lowest
$N(-1)\pi(1)$ state in the fit.  Overall, this comparison shows that
the contribution of the $N\pi$ state to ${\widetilde G}_P$ and $G_P$
is enhanced, reaching $\sim 45\%$ at the physical pion mass. The
roughly $5\%$ effect observed in $G_A$ is important phenomenologically
and needs to be made more precise.  Later, in
Sec.~\ref{ssec:ppd-pcac}, we choose the $\Ssim$ strategy to present
the final results based on the three form factors satisfying 
the PCAC relation.

A comparison of $G_A(Q^2)$ and the combination $\widetilde{G}_P(Q^2)
\times (Q^2+M_\pi^2)/(4M_N^2)$, which should be proportional to $G_A$
according to the PPD hypothesis, obtained using the three strategies
$\Stwo$, $\Sfour$, and $\Ssim$, is shown in
Fig.~\ref{fig:GA-Scomp-all}.
Results for both form factors are consistent between $\Sfour$ and
$\Ssim$ for each of thirteen ensembles with errors from $\Ssim$ being
slightly larger.  On the other hand $\GP$ (and $G_P$) from strategy
$\Stwo$ show noticeable differences that increase as $Q^2 \to 0$ and
$M_\pi \to 135$~MeV (see also the data in
Tables~\ref{tab:ff-a15m310}---\ref{tab:ff-a06m135} in
Appendix~\ref{sec:ff-tabs}). This effect is correlated with the
increase in the difference between $\Delta E_1^{\rm 2pt}$ (used in
$\Stwo$ fits) compared to $\Delta E_1^{\rm A4}$ and $\Delta M_1^{\rm
  A4}$ (output of $\Ssim$ fits) in the same two limits as shown later
in Fig.~\ref{fig:esc-axial}. Also, from Eq.~\eqref{eq:C3pt-decomp} it
is obvious that a smaller $\Delta E_1$ implies a larger ESC.

\begin{figure*}[!tbh]  %F02
  \centering
  \includegraphics[width=0.48\textwidth]{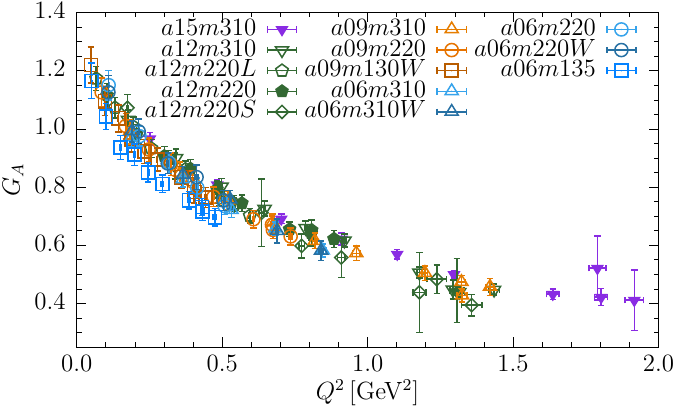} 
  \includegraphics[width=0.48\textwidth]{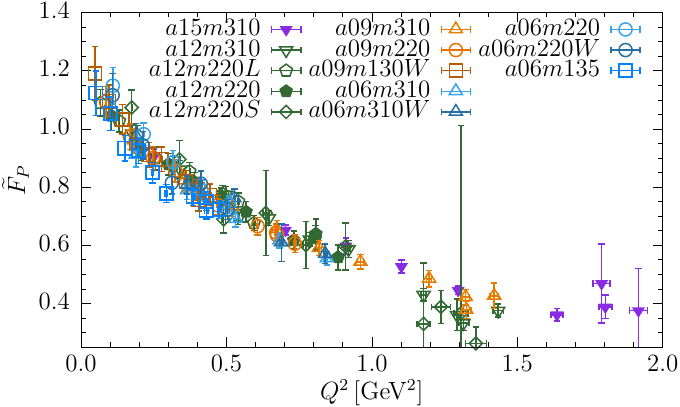} 
  \caption{The data for renormalized $G_A(Q^2)$ (left) and
    $\widetilde{F}_P(Q^2) \equiv \widetilde{G}_P(Q^2) \times
    (Q^2+M_\pi^2)/(4M_N^2)$ (right) from the 13 HISQ ensembles
    analyzed. No significant variation with respect to the lattice
    spacing $a$ or $M_\pi$ is observed except for the $a06m135$ ensemble,
    which we consider statistics limited.
}
  \label{fig:GAdata13}
\end{figure*}

The same data for $G_A(Q^2)$ and $\widetilde{G}_P(Q^2) \times
(Q^2+M_\pi^2)/(4M_N^2)$ from the 13 ensembles with the $\Ssim$
strategy are plotted in Fig.~\ref{fig:GAdata13}. Remarkably, they show
no significant variation with respect to the lattice spacing $a$ or
$M_\pi$ except for a $1\sigma$ lower values on the $a06m135$
ensemble, which we identify to be statistics limited.

In Appendix~\ref{sec:ff-tabs} we summarize why, with the methodology
for momentum insertion through the operator used in this work,
improving the lattice calculations ($M_\pi \to 135$~MeV, increasing
$M_\pi L > 4$, and reducing $a$) will increasingly give data at $Q^2 <
0.5$~GeV${}^2$. Even in this work, most of the data for $Q^2 >
0.7$~GeV${}^2$ comes from the $M_\pi \approx 310$~MeV ensembles.  If
the optimistic scenario presented by the current data, mild dependence
on $\{a,M_\pi\}$ as shown in Fig.~\ref{fig:GAdata13} and in
Ref.~\cite{Park:2021ypf}, holds then one will have confidence in the
final result for $G_A$ also for $0.5 \lesssim Q^2 \lesssim 1.5$~GeV${}^2$
even though its extraction in that region will be from data on $M_\pi
> 200$~MeV ensembles. To push precision to $Q^2 \sim 5$~GeV${}^2$
to meet the needs of the DUNE experiment will need further algorithmic
developments and much higher statistics data.

A detailed analysis of the extrapolation of these
$G_A$ to $Q^2 \to 0$ to get the axial charge $g_A$ is presented in Sec.~\ref{sec:aff-gA-rA}. The
marked improvement in satisfying the PCAC relation and the PPD
hypothesis shown by the $G_A$, $\GP$, and $G_P$ obtained with the $\Sfour$
and $\Ssim$ strategies (which include the $N\pi$ state) is 
discussed next in Sec.~\ref{ssec:ppd-pcac}.

\subsection{The PCAC relation and pion-pole dominance}
\label{ssec:ppd-pcac}
In this section, we evaluate how well the form factors from the three
strategies, tabulated in Appendix~\ref{sec:ff-tabs}, satisfy the PCAC relation, 
which in terms of the bare axial, $A_\mu(x)$, and
pseudoscalar, $P(x)$, currents is:\looseness-1
\begin{align}
  \partial_\mu A_\mu = 2\widehat{m} P \,,
  \label{eq:PCAC-op}
\end{align}
where the quark mass parameter $\widehat{m} \equiv Z_m m_{ud} Z_P Z_A^{-1}$ 
includes all the renormalization factors, and $m_{ud} = (m_u +
m_d)/2 =m_l$ is the light quark mass in the isospin symmetric limit.  Using
the decomposition in Eqs.~\eqref{eq:aff-a} and~\eqref{eq:aff-ps} of
GSME, the PCAC relation requires that the three form factors $G_A$,
$\GP$, and $G_P$ satisfy, up to discretization errors, the relation
\begin{align}
  2M_N G_A(Q^2) - \frac{Q^2}{2M_N}\GP(Q^2) = 2\widehat{m} G_P(Q^2) \,, 
\label{eq:PCAC2}
\end{align}
which we rewrite as 
\begin{align}
  R_1 + R_2 = 1 \,, \label{eq:PCAC}
\end{align}
with 
\begin{align}
  R_1 =&  \frac{Q^2}{4 M_N^2} \frac{\GP(Q^2)}{G_A(Q^2)}\,, \label{eq:R1}\\
  R_2 =& \frac{\widehat m}{M_N} \frac{G_P(Q^2)}{G_A(Q^2)} \,. \label{eq:R2}
\end{align}

The PPD  hypothesis relates  $\GP$ to $G_A$ as 
\begin{equation}
  R_3 \equiv  \frac{Q^2+M_\pi^2}{4 M_N^2} \frac{\GP(Q^2)}{G_A(Q^2)} = 1\,. 
\label{eq:PPD}
\end{equation}

\begin{figure}[!tb]  %F03
  \includegraphics[trim=0 24 0 0, clip, width=0.48\textwidth]{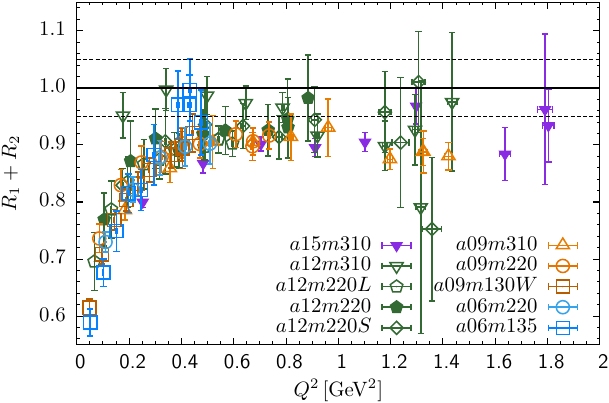}
  \\ %\vspace{1em}
  \includegraphics[trim=0 24 0 0, clip, width=0.48\textwidth]{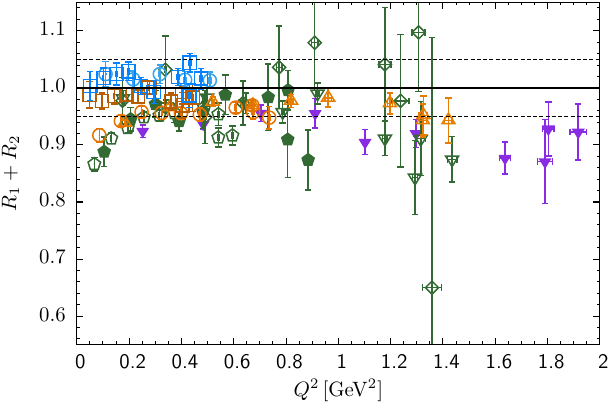}
  \\ %\vspace{1em}
  \includegraphics[trim=0 0 0 0, clip, width=0.48\textwidth]{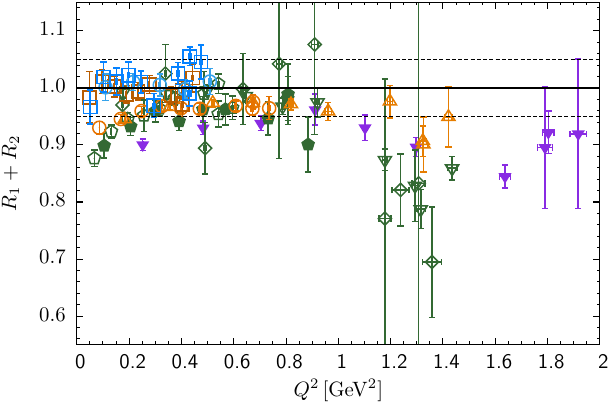}
  \caption{The data for $R_1+R_2$, which should equal unity to satisfy
    PCAC relation, is plotted versus $Q^2$ for analysis strategies
    $\Stwo$ (top), $\Sfour$ (middle), and $\Ssim$ (bottom). The PCAC
    relation, Eqs.\eqref{eq:PCAC-op} and~\eqref{eq:PCAC2}, requires
    $R_1+R_2=1$ up to discretization errors. The dashed lines give the
    $\pm 5\%$ deviation band.}
  \label{fig:PCAC}
\end{figure}

\begin{figure}[!tb] %F04
  \includegraphics[trim=0 24 0 0, clip, width=0.48\textwidth]{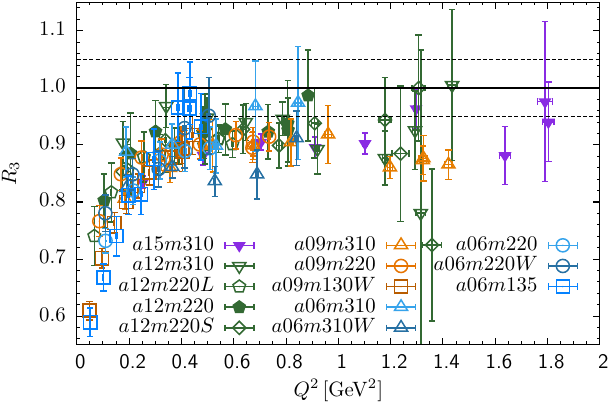}
  \\ %\vspace{1em}
  \includegraphics[trim=0 24 0 0, clip, width=0.48\textwidth]{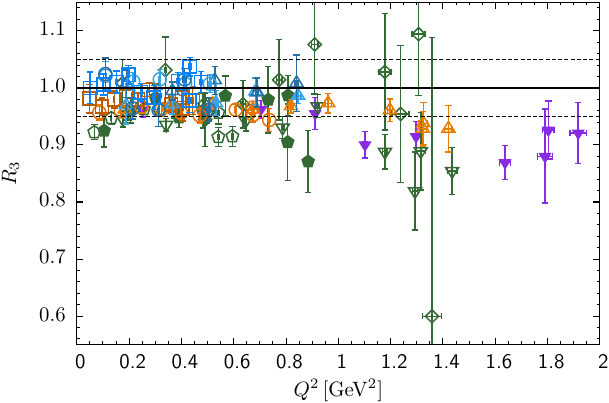}
  \\ %\vspace{1em}
  \includegraphics[trim=0 0 0 0, clip, width=0.48\textwidth]{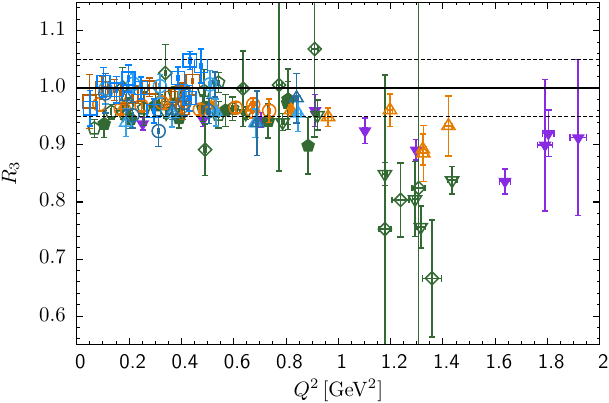}
  \caption{The ratio $R_3$, which should be unity for the pion-pole
    dominance hypothesis to be satisfied, is plotted versus $Q^2$ for
    analysis strategies $\Stwo$ (top), $\Sfour$ (middle), and $\Ssim$
    (bottom). The dashed lines mark the $\pm 5\%$ deviation band.}
  \label{fig:PPD}
\end{figure}

Tests of whether the form factors satisfy the PCAC ($R_1 + R_2 = 1$)
and PPD ($R_3 = 1$) relations are presented in Fig.~\ref{fig:PCAC} and
Fig.~\ref{fig:PPD}, respectively. Data with the $\Stwo$ strategy show
about $10\%$ deviation for both the PPD and PCAC relations for $Q^2 >
0.3\, \GeV^2$. Below it, the deviation grows to about $40\%$ at the
lowest $Q^2$ point on the two physical pion mass ensembles. See also 
the discussion in Appendix~\ref{sec:ff-tabs} on the differences in data
for the form factors obtained using the three ESC strategies. 

There is a very significant reduction in the deviations for both the
$\Sfour$ and $\Ssim$ strategies for $Q^2 < 0.3\, \GeV^2$. In fact,
except for three $M_\pi \approx 220$~MeV ensembles, data below $Q^2 =
1\, \GeV^2$ is essentially independent of $Q^2$ and the deviations
from unity and the variations between ensembles is in most cases
within about $5\%$, which can be due to possible
discretization errors. The differences between data with $\Ssim$ and
$\Sfour$ are much smaller. Also, the improvement in the PPD relation,
Eq.~\eqref{eq:PPD}, tracks that in PCAC, Eq.~\eqref{eq:PCAC}.

\begin{table}[tbh!]   %T02
  \centering
  \renewcommand{\arraystretch}{1.1}
  \begin{ruledtabular}
    \begin{tabular}{lcc}
      ID & $am_{ud}^\text{sea}$ & $a\widehat{m}^\text{2pt}$ \\ \hline
      $a15m310$  & 0.013   & - \\
      $a12m310$  & 0.0102  & 0.0121 \\
      $a12m220L$ & 0.00507 & - \\
      $a12m220$  & 0.00507 & - \\
      $a12m220S$ & 0.00507 & - \\
      $a09m310$  & 0.0074  & - \\
      $a09m220$  & 0.00363 & - \\
      $a09m130W$ & 0.0012  & 0.0015 \\
      %$a06m310$  & 0.0048 & - \\
      %$a06m310W$ & 0.0048 & - \\
      $a06m220$  & 0.0024 & 0.0028 \\
      %$a06m220W$ & 0.0024 & 0.0028 \\
      $a06m135$  & 0.00084 & 0.00088 
    \end{tabular}
  \end{ruledtabular}
  \caption{The HISQ sea quark mass is given in the second column. The quark mass $\widehat{m}$ is calculated from Eq.~\eqref{eq:mq}}
  \label{tab:mq}
\end{table}

We point out a caveat in our clover-on-HISQ calculation of the quark
mass $\widehat{m}$ used in Eq.~\eqref{eq:PCAC2}. For four ensembles, 
$a12m310$, $a09m130W$, $a06m220$, and $a06m135$ we have calculated 
$\widehat{m}^\text{2pt}$ using the following ratio of pion two-point correlators, 
\begin{align}
  2\widehat{m}^\text{2pt} =& \frac{\matrixe{\Omega}{\partial_\mu A_\mu(t)P(0)}{\Omega}}{\matrixe{\Omega}{P(t)P(0)}{\Omega}} \,. \label{eq:mq}
\end{align}
For the other ensembles, the data for these two-point functions were
not collected, so we use the HISQ sea quark mass $am_{ud}^\text{sea}$
for $\widehat{m}$ since for staggered fermions, in fact all lattice
fermions with chiral symmetry, $Z_m Z_P Z_A^{-1}=1$.  These quark
masses are given in Table~\ref{tab:mq} and we find that
$\widehat{m}^\text{2pt}$ is 5 -- 20\% larger than
$am_{ud}^\text{sea}$, which is not unexpected for our clover-on-HISQ
calculation.  Noting that $R_2 \approx 0.5 R_1$ (see Fig.~[15] in 
Ref.~\cite{Rajan:2017lxk}), such a 20\% systematic error would
increase $R_1 + R_2$ by about 7\%.  This would bring the data from the
physical mass ensembles, $a09m130W$ and $a06m135$, in better agreement
but would not alter our conclusion that form factors obtained with
$\Sfour$ and $\Ssim$ strategies show better agreement with the PCAC
relation compared to $\Stwo$. Also, $\widehat{m}$ does not
enter in the PPD relation, Eq.~\eqref{eq:PPD}, and the deviation from
unity of the PPD relation with $\Sfour$ and $\Ssim$ data is observed
to be smaller than seen in the PCAC relation as shown in Fig.~\ref{fig:PPD}.
Equally important, this caveat does not impact the
extraction of individual form factors or their subsequent analysis
since $\widehat{m}$ only enters in the test of how well the three form
form factors satisfy the PCAC relation, Eq~\eqref{eq:PCAC2}.

We further examine whether the deviation from unity in
Fig.~\ref{fig:PCAC} at small $Q^2$ is a discretization error. The
$\mathcal{O}(a)$ improvement affects only the axial current, $A_\mu
\to A_\mu + c_A a \partial_\mu P$, and adds to the left hand side in
Eq.~\eqref{eq:PCAC2} the term
$ - Q^2 a c_A G_P $, 
i.e., 
under improvement,  Eq.~\eqref{eq:PCAC2} can be written as 
\begin{equation}
  M_N \frac{G_A}{G_P} - \frac{Q^2}{4M_N}\frac{\GP}{G_P} = \widehat{m} + \frac{1}{2}a c_A Q^2 \,, 
\end{equation}
where the improvement coefficient, $c_A$, is typically
$\mathcal{O}(10^{-2})$ and negative.  Thus, this effect is expected to
be small for $Q^2 < 1$~GeV${}^2$, and will not change our
conclusions. On the other hand, effects due to possible mistuning of
the clover coefficient, $c_{SW}$, and $c_A$, and $O(a^2)$ corrections
are likely to increase with $Q^2$. Similarly, artifacts are expected to
increase with the quark mass since the improvement coefficient $b_m$
is not included

%% the shift absorbed in $\widehat{m}$ to $\widehat{m} + \frac{1}{2}a
%% c_A Q^2$. It vanishes as $Q^2 \to 0$ and can be considered part of
%% taking the continuum limit as discussed in Section~\ref{sec:aff-gA-rA}
%% for the axial form factor and in Section~\ref{sec:aff-gpx} for the
%% induced pseudoscalar form factor.

%

\begin{figure}[!tb]  %F05
  \includegraphics[trim=0 15 0 0, clip, width=0.48\columnwidth]{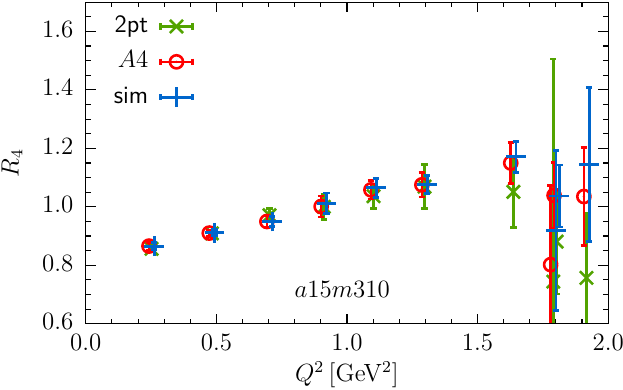}
  \includegraphics[trim=0 15 0 0, clip, width=0.48\columnwidth]{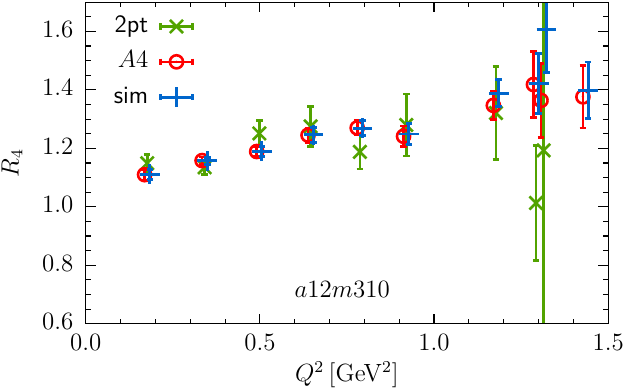}
  \\ \vspace{0.5em}
  \includegraphics[trim=0 15 0 0, clip, width=0.48\columnwidth]{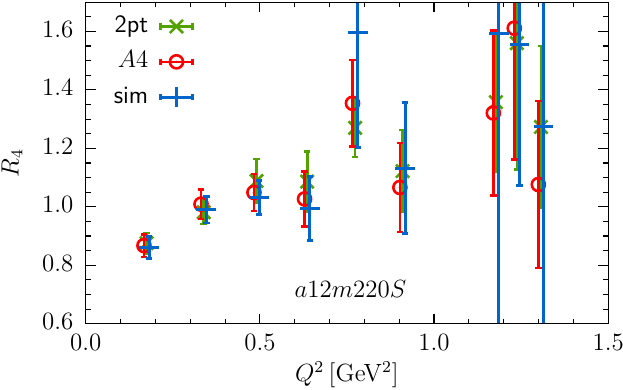}
  \includegraphics[trim=0 15 0 0, clip, width=0.48\columnwidth]{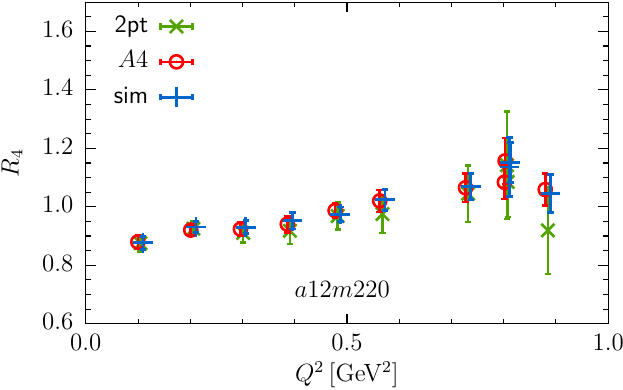}
  \\ \vspace{0.5em}
  \includegraphics[trim=0 15 0 0, clip, width=0.48\columnwidth]{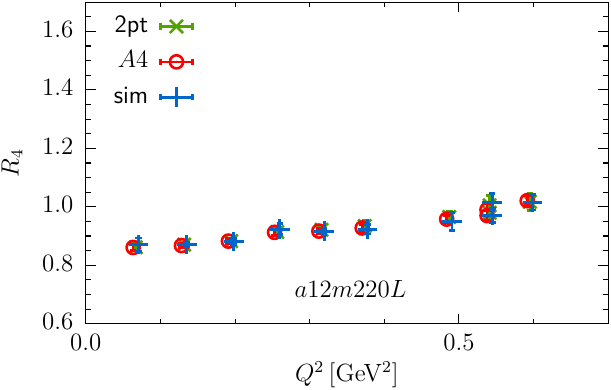}
  \includegraphics[trim=0 15 0 0, clip, width=0.48\columnwidth]{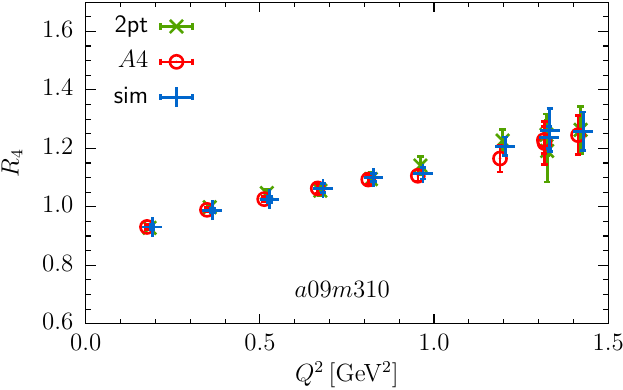}
  \\ \vspace{0.5em}
  \includegraphics[trim=0 15 0 0, clip, width=0.48\columnwidth]{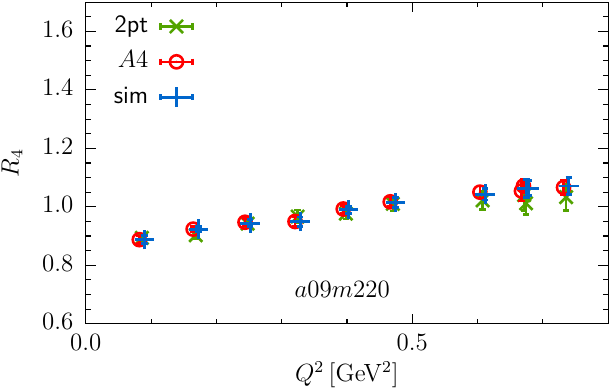}
  \includegraphics[trim=0 15 0 0, clip, width=0.48\columnwidth]{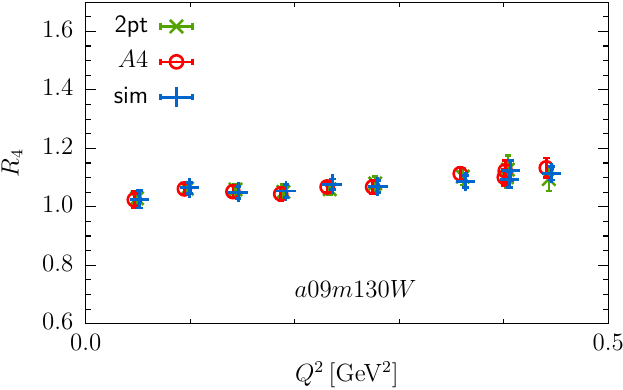}
  \\ \vspace{0.5em}
  \includegraphics[trim=0 0 0 0, clip, width=0.48\columnwidth]{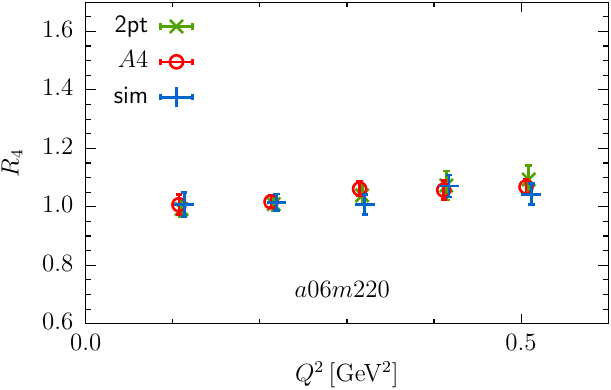}
  \includegraphics[trim=0 0 0 0, clip, width=0.48\columnwidth]{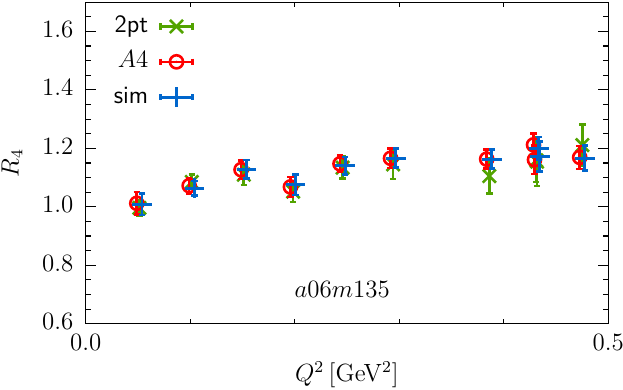}
  \caption{Results for the ratio $R_4 = (G_P / \widetilde{G}_P) \times
    (4\hat{m}M_N/M_\pi^2)$. For the pion-pole dominance hypothesis to
    be exact (derivable from the PCAC relation), $R_4 $ should be
    unity independent of $Q^2$. The data show an approximate linear
    increase with $Q^2$, which is consistent with the
    Goldberger-Trieman discrepancey as discussed in
    Ref.~\protect\cite{Bernard:2001rs,Park:2021ypf}. }
  \label{fig:R4-Scomp}
\end{figure}

The PPD relation (Eq.~\eqref{eq:PPD}) can be derived from PCAC (Eq.~\eqref{eq:PCAC2}) provided 
\begin{equation}
  R_4 \equiv  \frac{4\widehat{m}M_N}{M_\pi^2} \frac{G_P}{\widetilde{G}_P} =   1 \,.
\label{eq:R4} 
\end{equation}
In this case, $R_1+R_2 = 1$ would also imply $R_3= 1$.  In
Fig.~\ref{fig:R4-Scomp}, we compare $R_4$ from the three strategies
for all ensembles except $a06m220W$, $a06m310$, and $a06m310W$ where
$G_P$ is not available.  We note a roughly linear increase in $R_4$
with $Q^2$, which is consistent with the behavior observed in
Ref.~\cite{Park:2021ypf} and with the analysis of the
Goldberger-Trieman discrepancy using $\chi$PT in
Ref.~\cite{Bernard:2001rs}.  Lastly, we note that the data for $R_4$ from
all three strategies, $\Stwo$, $\Sfour$ and $\Ssim$, overlap implying
that the changes in $\GP$ and $G_P$, both of which have a pion pole,  between
different treatments of ESC ($\Stwo$ versus $\Sfour$ or $\Ssim$)
cancel in the ratio $R_4$
within our statistics.  This observation supports our hypothesis that 
the same excited states contribute to all five correlation functions.

\subsection{Excited States Spectrum}
\label{ssec:ESS}

\begin{figure*}[!tb] %F06
  \includegraphics[trim=0 0 0 0, clip, width=0.32\textwidth]{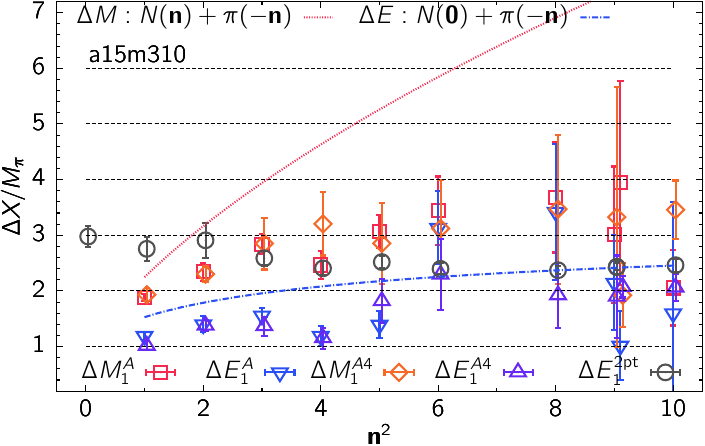}
  \hspace{0.5em}
  \includegraphics[trim=0 0 0 0, clip, width=0.32\textwidth]{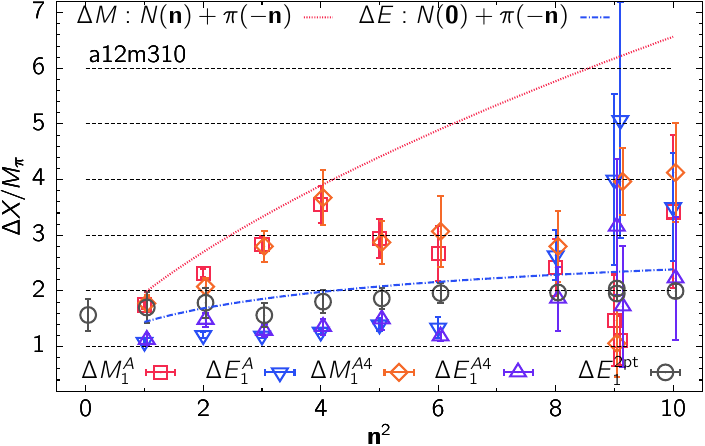}
  \phantom{
  \includegraphics[trim=0 0 0 0, clip, width=0.32\textwidth]{figs/man/spec_axial_a12m310}
  }
  \\ \vspace{0.5em}
  \includegraphics[trim=0 0 0 0, clip, width=0.32\textwidth]{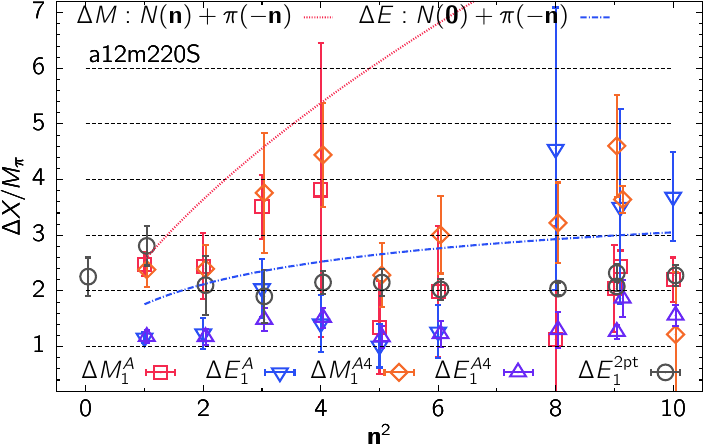}
  \hspace{0.5em}
  \includegraphics[trim=0 0 0 0, clip, width=0.32\textwidth]{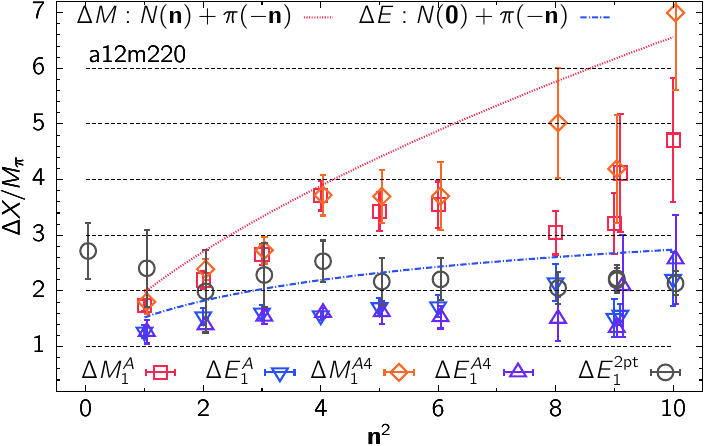}
  \hspace{0.5em}
  \includegraphics[trim=0 0 0 0, clip, width=0.32\textwidth]{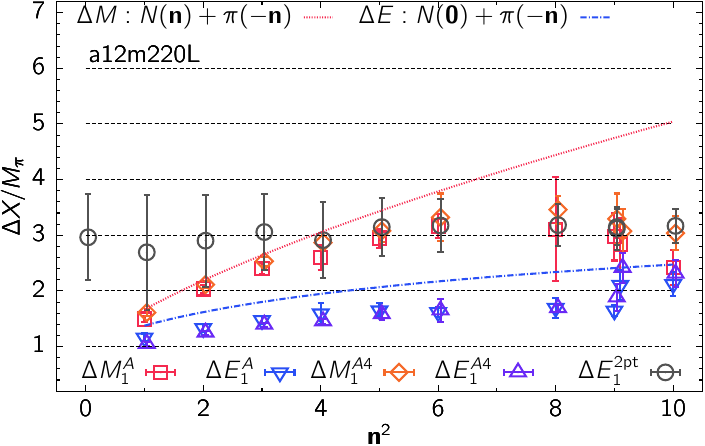}
  \\ \vspace{0.5em}
  \includegraphics[trim=0 0 0 0, clip, width=0.32\textwidth]{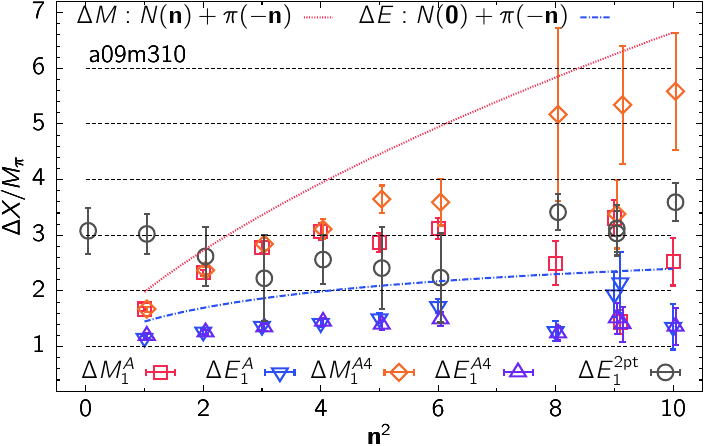}
  \hspace{0.5em}
  \includegraphics[trim=0 0 0 0, clip, width=0.32\textwidth]{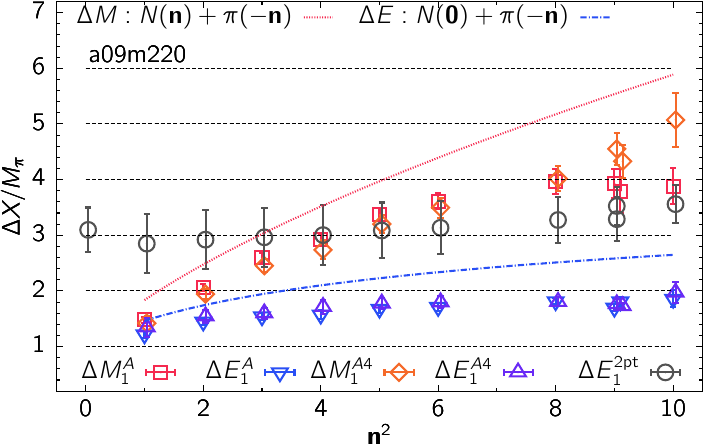}
  \hspace{0.5em}
  \includegraphics[trim=0 0 0 0, clip, width=0.32\textwidth]{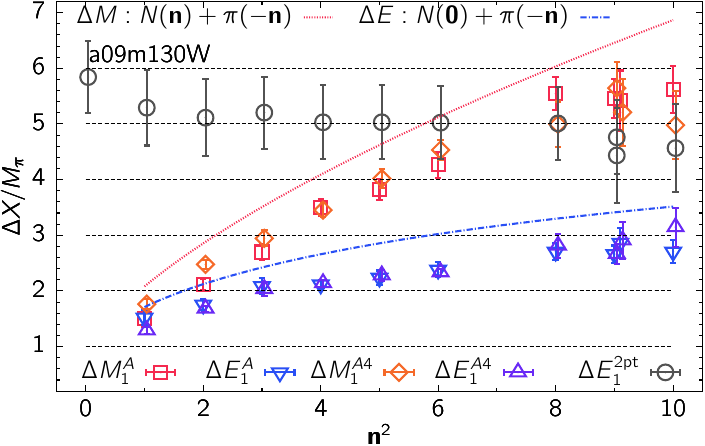}
  \\ \vspace{0.5em}
  \includegraphics[trim=0 0 0 0, clip, width=0.32\textwidth]{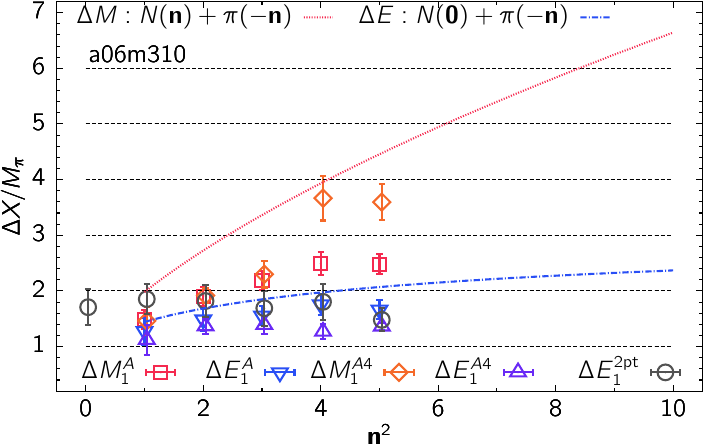}
  \hspace{0.5em}
  \includegraphics[trim=0 0 0 0, clip, width=0.32\textwidth]{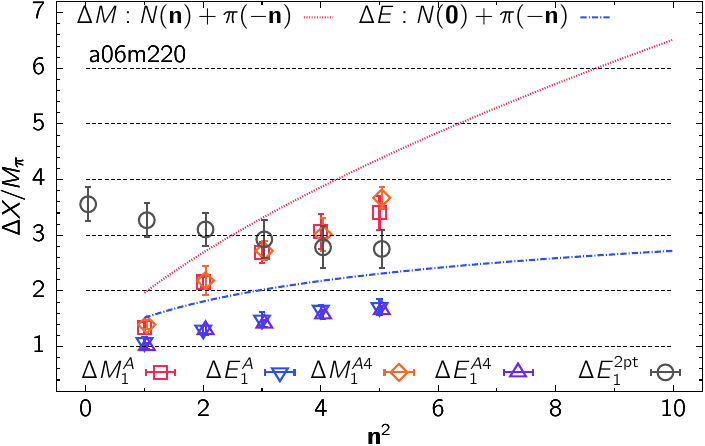}
  \phantom{
%  \hspace{0.5em}
  \includegraphics[trim=0 0 0 0, clip, width=0.32\textwidth]{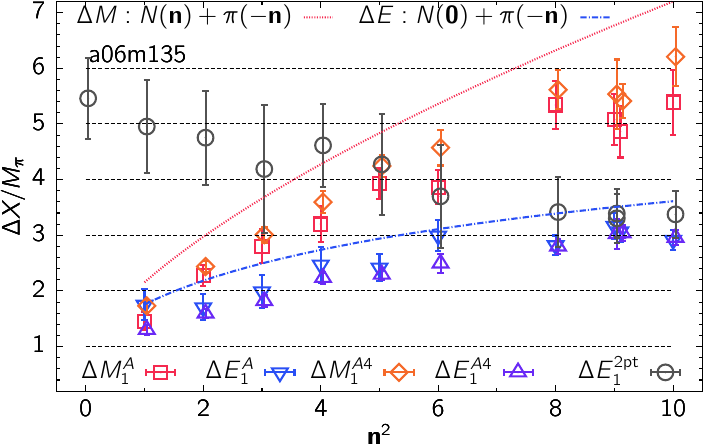}
  }
  \\ \vspace{0.5em}
  \includegraphics[trim=0 0 0 0, clip, width=0.32\textwidth]{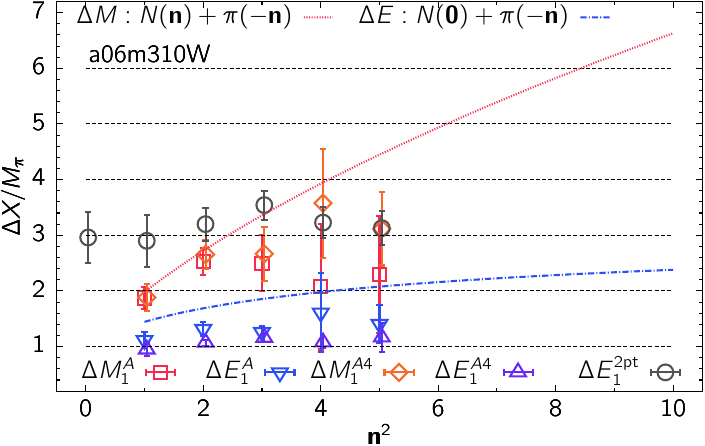}
  \hspace{0.5em}
  \includegraphics[trim=0 0 0 0, clip, width=0.32\textwidth]{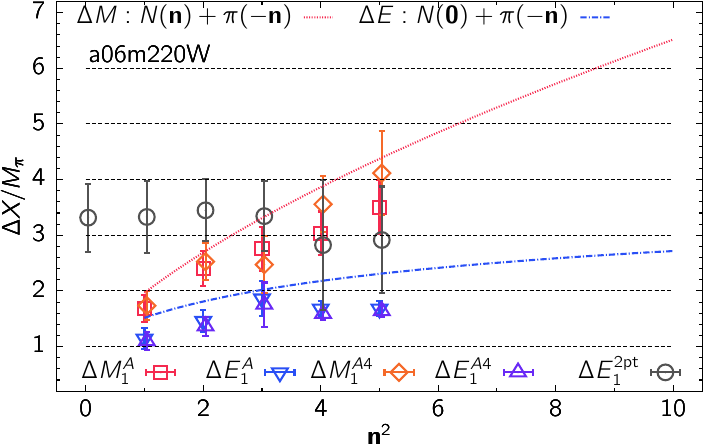}
  \hspace{0.5em}
  \includegraphics[trim=0 0 0 0, clip, width=0.32\textwidth]{figs/man/spec_axial_a06m135}
  \caption{Results for the energy (mass) gaps $\Delta E_1$ ($\Delta
    M_1$) for the first excited state extracted from (i) simultaneous
    fits to axial three-point correlators $C[A_\mu]$ and the
    pseudoscalar correlator $C_P$ ($\Ssim$ strategy and labeled
    $\Delta E_1^A$ and $\Delta M_1^A$), and (ii) from fits to the
    $C[A_4]$ correlator ($\Sfour$ strategy and labeled $\Delta
    E_1^{A4}$ and $\Delta M_1^{A4}$).  These mass gaps are compared
    with the first excited state energy $\Delta E_1^\text{2pt}$ from
    four-state fits to the nucleon two-point correlator. Note that the
    difference between them (black circles versus blue triangles), and consequently the difference between
    the form factors extracted, increases as $M_\pi \to 135$~MeV and ${\textbf n}^2 \to 0$ (equivalently $Q^2\to 0)$.
}
  \label{fig:esc-axial}
\end{figure*}

In Fig.~\ref{fig:esc-axial} we show data for the energy gaps, $\Delta
E_1$ and $\Delta M_1$ on the two sides of the operator insertion for
the various ensembles, including the two physical pion mass ones,
$a09m130W$ and $a06m135$.  The results for $\Delta E_1^{A}$ and
$\Delta M_1^{A}$, outputs of the simultaneous fits to all five
correlators (insertions of ${A_\mu}$ and $P$) at a given momentum
transfer $\bm{p}=2\pi\bm{n}/L$ overlap with the results $\Delta
E_1^{A4}$ and $\Delta M_1^{A4}$ obtained from fits to just $C_{A_4}$.  This
indicates that the energy gaps in the $\Ssim$ fits are
essentially controlled by $C_{A_4}$. The momentum dependence of the
data is consistent with the expectation that the relevant excited
states on the two sides are $N(\bm{n}) + \pi(-\bm{n})$ and
$N(\bm{0})+\pi(-\bm{n})$. This is based on the rough agreement between
the data and the corresponding noninteracting energies of these
states, $\Delta M_1$ and $\Delta E_1$, shown by the dashed red and 
blue lines, respectively, and consistent with the PPD hypothesis that
the currents inject a pion with momentum $\vec q$.

The data with open circles in Figure~\ref{fig:esc-axial} are the
energy gaps $\Delta E_1^\text{2pt}$ obtained from the nucleon
two-point correlators. These are roughly independent of momentum and
larger than those from $\Sfour$ or $\Ssim$ fits, especially for the
smaller $Q^2$ points.  The difference increases as $Q^2 \to 0$ and
$M_\pi \to 0$. This behavior is consistent with $\Delta
E_1^\text{2pt}$ corresponding to a mixture of radial and higher
multiparticle excitation whereas the energy of the intermediate excited states
identified by the $\Sfour$ and $\Ssim$ fits, $N(\bm{n}) +
\pi(-\bm{n})$ and $N(\bm{0})+\pi(-\bm{n})$, decreases with decreasing
$\bm n$ and $M_\pi$.  A word of caution when
making these identifications: it is very important to qualify that the
$\Delta E_1$ and $\Delta M_1$ from the two-state fits in $\Sfour$ and
$\Ssim$ strategies are effectively trying to account for all the
intermediate states that make significant contributions and not just
the lowest or the most intuitive ones. Given the size of the effect,
identifying and improving control over all the excited states that
make significant contribution to these correlation functions will be
key to obtaining, in the future, higher  precision results for the form
factors.

\clearpage

\subsection{Renormalization Constant $Z_A$}
\label{sec:renorm}

The renormalization constant $Z_A$ for the axial current needed for
the form factors $G_A$ and $\GP$ and the charges $g_A$, $g_P^\ast$ and
$g_{\pi NN}$ was determined nonperturbatively using the RI-sMOM
intermediate scheme in Ref.~\cite{Gupta:2018qil}. We use the results
given in Table~V there.

\section{Parameterization of $G_A(Q^2)$, and the extraction of $g_A$ and $\expv{r_A^2}$}  %S03
\label{sec:aff-gA-rA}

In this section, we present the analysis of $G_A(Q^2)$ without
including the values of the axial charge $g_A$ obtained from the
forward matrix element. Its extraction is discussed separately in
Sec.~\ref{sec:charges}.  This is done to keep the two extractions of
$g_A$---from the forward matrix element
(\eqref{eq:summary-charge-3RD}) and by extrapolating $G_A(Q^2)$ to
$Q^2 \to 0$ (Eqs.~\eqref{eq:gArAfinal})---separate. The 
final result, given in Table~\ref{tab:final_Comp}, is taken to be
the average of the two.

The axial-vector form factor $G_A(Q^2)$ can be parameterized, near $Q^2=0$,
by the axial charge $g_A$ and the axial charge radius squared
$\expv{r_A^2}$:
\begin{align}
  G_A(Q^2) &= g_A ( 1 - \frac{\expv{r_A^2}}{6} Q^2 + \cdots) \,,
\end{align}
where $g_A \equiv G_A(0)$ and
\begin{align}
  \expv{r_A^2} &\equiv -\frac{6}{g_A} \left. \frac{dG_A(Q^2)}{dQ^2} \right\vert_{Q^2=0} \,.
\end{align}
To extract these from lattice data obtained at $Q^2 \gsim 0.1$~GeV${}^2$, one parameterizes the $Q^2$
dependence of $G_A(Q^2)$. Among the various parameterizations, we study the dipole
ansatz and the model-independent $z$-expansion. 
The dipole ansatz 
\begin{align}
  G_A(Q^2) &= \frac{g_A}{(1+Q^2/\mathcal{M}_A^2)^2} \,.
  \label{eq:dip}
\end{align}
has two free parameters, the axial charge $g_A$ and the axial mass
$\mathcal{M}_A$.  
The $z$-expansion is the series 
\begin{align}
  G_A(Q^2) &= \sum_{k=0}^\infty a_k\,z^k\,,\;
  \label{eq:zexp}
\end{align}
in terms of the variable
\begin{align}
  z &= \frac{\sqrt{t_c+Q^2} - \sqrt{t_c+t_0}}{\sqrt{t_c+Q^2} + \sqrt{t_c+t_0}} \quad {\rm with}\ \; t_c \equiv 9M_\pi^2
  \label{eq:def-z}
\end{align}
that maps the kinematically allowed analytical region $Q^2 \geq 0$ to
that within a unit circle, $\abs{z} <
1$~\cite{Bhattacharya:2011ah}. The parameter $t_0$ is discussed
later. For sufficiently small $z$, fits with the first few terms
should suffice.  In practice, to stabilize the fits we impose the
condition $\abs{a_k} \leq 5$ for all $z^{k \ge 1}$
truncations~\cite{Bhattacharya:2011ah}.  
With increasingly precise data over a sufficiently large range of
$Q^2$, our goal is to demonstrate that a data-driven choice can be
made between the various parameterizations.

In the data presented here, the statistical signal is good for
momentum transfer with ${\bm n}^2 \le 6$ but often poor in the four
points with $ 8 \le {\bm n}^2 \le 10$.  To test the stability of the
dipole and $z^k$ fits with such few points, we compare the output of
the fits to the lowest six versus all ten $Q^2$ points on nine
ensembles where data on all ten $Q^2$ values exist.  Observing
consistency, the final results are taken from fits to six (five in 4
cases) points, i.e., to data up to the $Q^2|_{\rm max}^{n^2 \le 6}$
given in Table~\ref{tab:ens}. This implies that results for form
factors presented for $Q^2 \gtrsim 0.5$~GeV${}^2$ come mainly from the
$M_\pi \sim 310$~MeV ensembles, and, a priori, could have an large
systematic uncertainty. In practice, however, the observed weak
dependence of form factors on $M_\pi$ (see Fig.~\ref{fig:GAdata13} and
similar result from a clover-on-clover calculation presented in
Ref.~\cite{Park:2021ypf}) suggests that the uncertainty is likely
within the quoted errors. Readers should, nevertheless, keep in mind
that results at $Q^2 \gtrsim 0.5$~GeV${}^2$ are based
mainly on data from the $M_\pi \approx 310$~MeV ensembles.\looseness-1

Estimates of $G_A$ from the dipole fit to data from the three ESC
strategies are consistent, however, on six ensembles these dipole fits
to $\Ssim$ and $\Sfour$ data have poor $p$-values. Our evaluation of
the failure is that the dipole ansatz does not have enough parameters
to capture the change in the curvature over the range of $Q^2$
studied. A consequence is that estimates for $g_A$ and $\expv{r_A^2}$
are smaller than those from $z^2$ fits to the same $\Ssim$ and
$\Sfour$ data. Since agreement with PCAC is essential and $\Ssim$ data
do the best job, while the dipole fits are poor, we rule out the
dipole ansatz. Henceforth, our final results are obtained using the
$z$-expansion, and the dipole results are given only for comparison.

In the $z$-expansion fits, the free parameter $t_0$ in
Eq.~\eqref{eq:def-z} is used to adjust the maximum value of $z$ within
the unit circle $|z| \le 1$.  We take $t_0=0.4, 0.2, 0.12$ for
$M_\pi \approx 310, 220, 130\,\MeV$ ensembles, which gives
$\abs{z} \lesssim 0.2$. We have
checked that using $t_0=0$ does not change the fits or the values
significantly.

\begin{figure}[!tbh]   %F13 cont
  \centering
  \includegraphics[trim=0 25 0 0, clip, width=0.75\columnwidth]{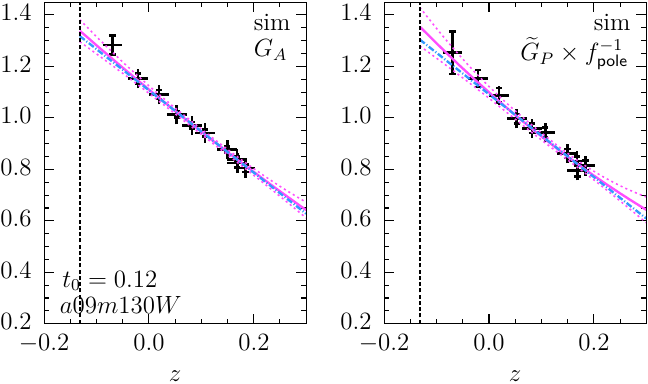}
  \\ %\vspace{0.5em}
  \includegraphics[trim=0 25 0 0, clip, width=0.75\columnwidth]{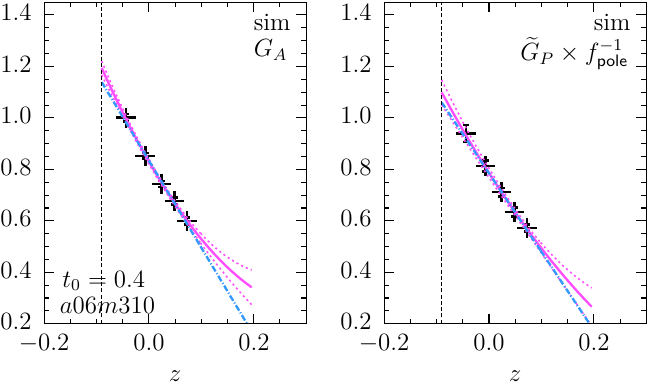}
  \\ %\vspace{0.5em}
  \includegraphics[trim=0 25 0 0, clip, width=0.75\columnwidth]{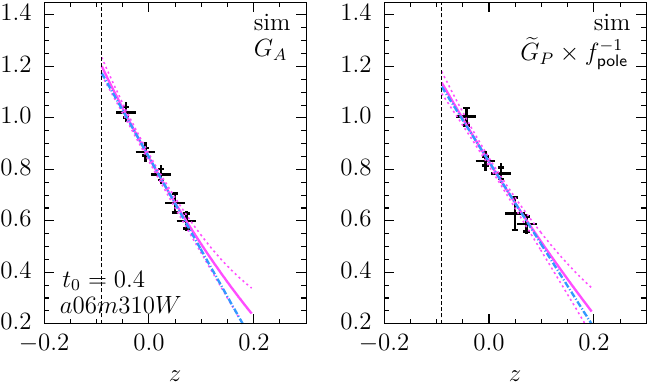}
  \\ %\vspace{0.5em}
  \includegraphics[trim=0 25 0 0, clip, width=0.75\columnwidth]{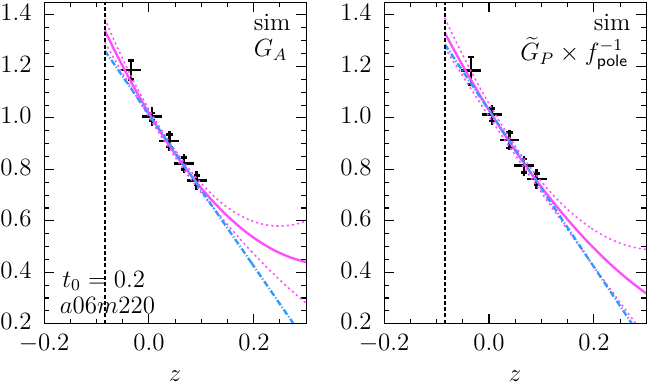}
  \\ %\vspace{0.5em}
  \includegraphics[trim=0 25 0 0, clip, width=0.75\columnwidth]{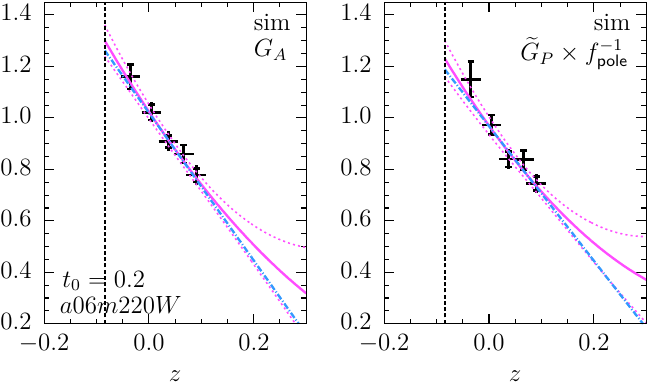}
  \\ %\vspace{0.5em}
  \includegraphics[trim=0 0 0 0, clip, width=0.75\columnwidth]{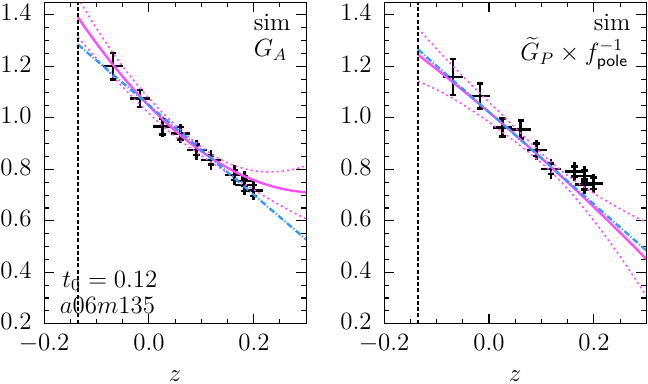}
  \caption{Comparison of $G_A$ and $\widetilde{F}_P \equiv \GP \times f^{-1}_{\rm pole}$ from strategy $\Ssim$ for
    $a\approx 0.15, 0.12\, 0.09, 0.06\, \fm$ lattices. The $z^2$ (magenta lines with 
    error band) is compared to the $z^1$ fit (blue dash
    dot). The four largest  $z$ points are
    excluded from the fits to the $M_\pi \approx 310\,\MeV$ and $a12m220S$ data. The vertical black dotted line 
    corresponds to $Q^2=0$.     The value of $t_0$ and ensemble name are given in the labels. }
  \label{fig:GA-GP-zexp-ALL-simA}
\end{figure}

To ensure that the form factors satisfy the expected $1/Q^4$
perturbative behavior in the limit $Q^2 \to \infty$, sum rules can be
imposed as was done in Ref.~\cite{Rajan:2017lxk}.  However, to obtain the
behavior near $Q^2=0$ from six or ten data points with $Q^2_\text{max}
\approx 1\, \GeV^2$, we choose to make fits without the sum
rules~\cite{Jang:2019jkn}, i.e., to not increase the weight of the
larger error high $Q^2$ points by imposing the sum rules. The $z^1$
and $z^2$ fits to $G_A(Q^2)$ from the $\Ssim$ strategy are shown in
Fig.~\ref{fig:GA-GP-zexp-ALL-simA}. The resulting bare axial charge
$g_A\equiv G_A(0)$ and the charge radius squared $\expv{r_A^2}$ from
the $z^{2}$ fits are shown in Fig.~\ref{fig:gA-rAsq-comp}, and the 
data are summarized in Table~\ref{tab:Qsqfit-gA-1} in
Appendix~\ref{sec:EnsembleResults}. From these data and the
$z$-expansion fits, we conclude the following:

\begin{figure*}[!tb]  %F07
  \includegraphics[width=0.95\columnwidth]{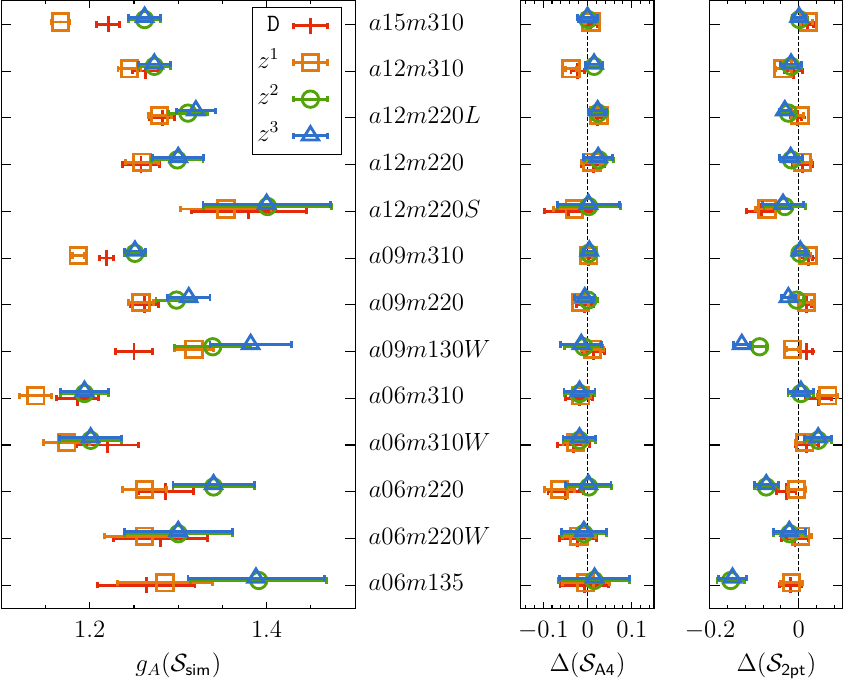} \hspace{1.0cm}
  \includegraphics[width=0.95\columnwidth]{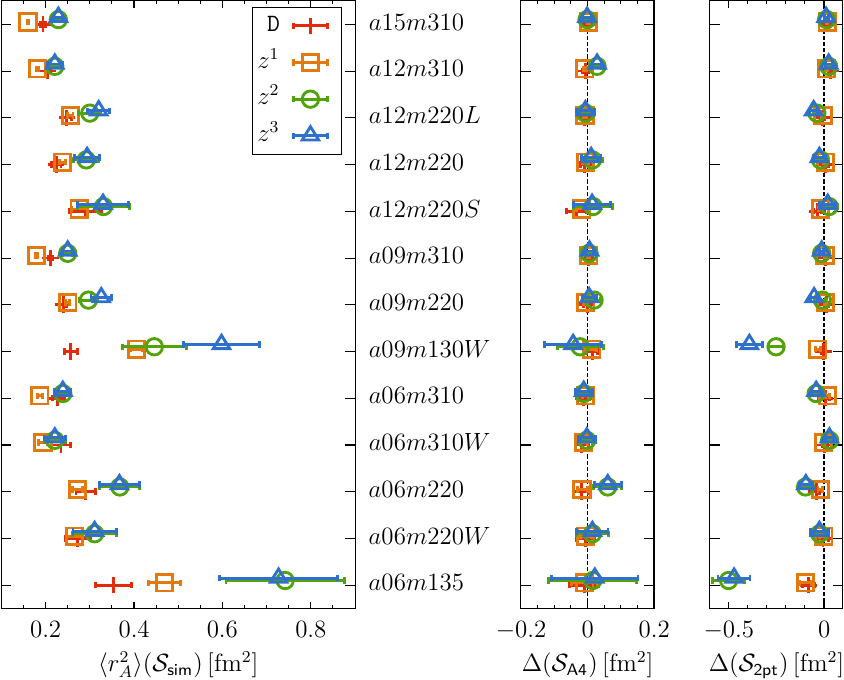}
  \caption{(Left) Results for bare $g_A$ from the strategy $\Ssim$ and the
    differences $\Delta(S_{A4})=g_A|_{\Sfour} - g_A|_{\Ssim}$ and
    $\Delta(\Stwo)=g_A|_{\Stwo} - g_A|_{\Ssim}$. To facilitate
    visualization of the spread, the errors plotted for $\Delta(X)$
    are those in $g_A (X)$.  Results are shown for the dipole fit 
    labeled ``D'' and  $z^{1,2,3}$-truncations.
    (Right) Analogous results for $\expv{r_A^2}$.  }
  \label{fig:gA-rAsq-comp}
\end{figure*}

\begin{itemize}
\item
There is agreement in results between $z^2$ and $z^3$ fits in all
cases. To account for the small curvature observed in the data shown
in Fig.~\ref{fig:GA-GP-zexp-ALL-simA} and yet avoid
overparameterization, evaluated using the Akaika Information Criteria
(AIC)~\cite{1100705}, we will present final results with the $z^2$
truncation.
\item
The errors in the data from the two physical mass ensembles $a09m130W$
and especially $a06m135$ are large and underscore the need for higher statistics. 
\item
Results for both $g_A$ and $\expv{r_A^2}$ from both $\Sfour$ and
$\Ssim$ analyses overlap and increase in value as $M_\pi \to 135$~MeV. This
increase is correlated with the increasing ESC of the $N \pi$ state. 
\end{itemize}
We take the final results from the $\Ssim$ strategy in which a
simultaneous fit is made to all five correlators and the form factors
come closest to satisfying the PCAC relation as shown in
Fig.~\ref{fig:PCAC}. This is a 2-state fit and we find that stable 3-state fits
require higher statistics. Thus, with the current data,
we do not have a reliable way of estimating the systematic uncertainty 
associated with possible residual ESC.

The analysis of $g_A$ obtained from the forward matrix elements is postponed
to Sec.~\ref{sec:charges}.

\subsection{Extrapolation of $g_A$ and $\expv{r_A^2}$ to the Physical Point}
\label{ssec:aff-extrap}

Extrapolation of the renormalized axial charge $g_A$ and the 
axial charge radius squared $\expv{r_A^2}$ 
to the physical point ($a\to 0$, $M_\pi \to
135\,\MeV$, $L \to \infty$) is performed using a simultaneous 
CCFV fit keeping only the lowest order corrections in the ansatz 
\begin{align}
  Y = b^Y_0 + b^Y_1 a + b^Y_2 M_\pi^2 + b^Y_3 M_\pi^2 \exp{(-M_\pi L)} \,,
  \label{eq:ccfv}
\end{align}
where $Y = \expv{r_A^2}$ or $g_A$ and $\{b^Y_i\}$ denote the
corresponding set of fit parameters.  The discretization artifacts are
taken to start at $O(a)$ since the clover action used is only tadpole
improved and the axial and pseudoscalar currents are
unimproved~\cite{Bhattacharya:2015wna}.  Similarly, only the lowest
order term in the chiral expansion is kept to avoid
over-parameterization as data at only three values of the pion mass
have been simulated.

We have performed four CCFV fits to (i) the full set of thirteen ensembles
(13-pt); and three ``12-pt'' fits that exclude (ii) the coarsest
lattice point $a15m310$, (iii) the smallest volume point $a12m220S$
that also has large errors, and (iv) the point $a06m135$ that  has
large statistical errors and shows the largest difference from the
other 12 points.  The three 12-pt fits are used to estimate
systematics due to discretization and finite volume effects, and the
impact of the $a06m135$ point. Results of the 13-point CCFV
extrapolation for $g_A$ and $\expv{r_A^2}$ are summarized in
Table~\ref{tab:gA} for six cases: the three strategies used for
removing ESC, $\Ssim$, $\Sfour$, and $\Stwo$, and the two $Q^2$
parameterizations, $z^2$ and dipole.  The parameters of the 13-point
CCFV extrapolation of the $z^2$ fit to the $\Ssim$ data, used to
get the final central values, are given in
Table~\ref{tab:ccfv-rA-13pt-Ssim} in Appendix~\ref{sec:ccfv-summary}
for both $g_A$ and $\expv{r_A^2}$. Results for all the other cases can
be constructed using the data for the form factors given in
Tables~\ref{tab:ff-a15m310}--\ref{tab:ff-a06m135} in
Appendix~\ref{sec:ff-tabs}. 

\begin{figure*}[!tbh]  %F08
  \centering
\subfigure{
 \includegraphics[width=0.96\textwidth]{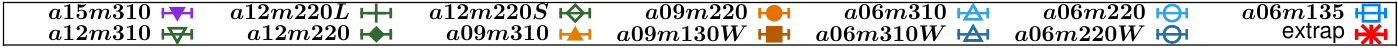}
}
\subfigure{
  \includegraphics[trim=0 0 14 60, clip, width=0.32\textwidth]{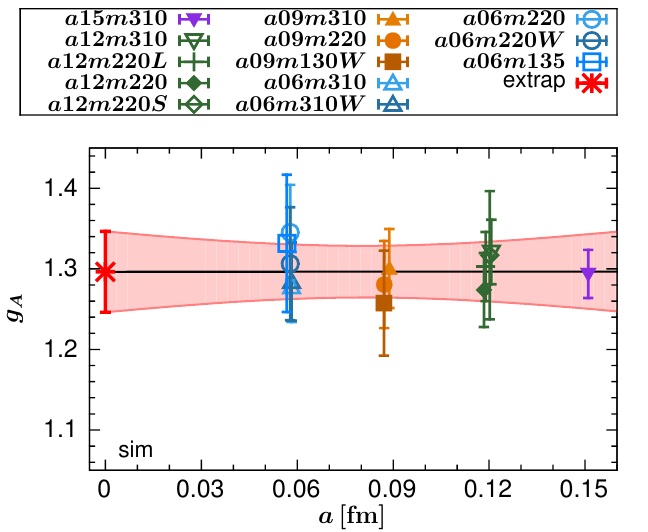}
  \hfill
  \includegraphics[trim=0 0 14 60, clip, width=0.32\textwidth]{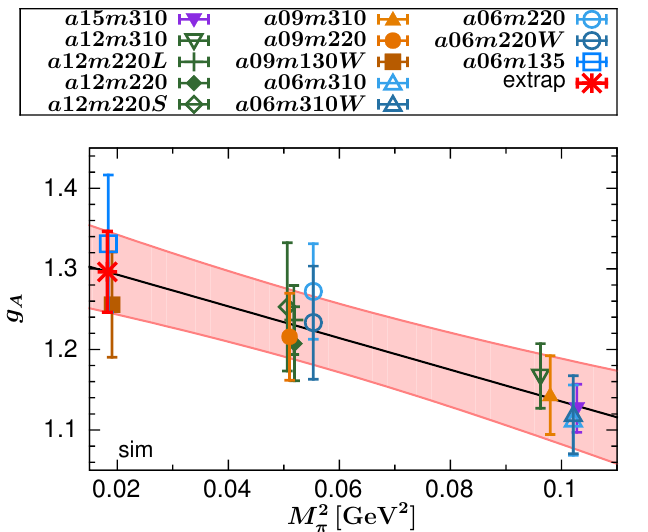}
  \hfill
  \includegraphics[trim=0 0 14 60, clip, width=0.32\textwidth]{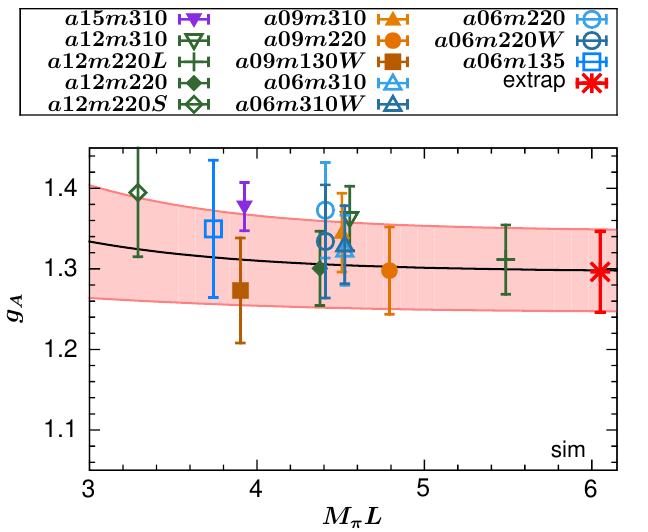}
}
  \caption{The axial charge $g_A$ given by the 13-pt CCFV fit to
    $\Ssim$ data using the $z^2$ fit to $G_A(Q^2\neq 0)$. The pink
    band in each panel gives the result of the CCFV fit
    (Eq.~\eqref{eq:ccfv}) versus the x-axis variable with the other two
    variables set to their physical values.  The data points in each
    panel have been shifted in the other two variables using the same
    CCFV fit, however, the size of errors are not changed.  The final
    result at the physical point is shown by the red cross.}
  \label{fig:gA-z2-simA}
\end{figure*}

\begin{figure*}[!tbh]    %F09
  \centering
\subfigure{
 \includegraphics[width=0.96\textwidth]{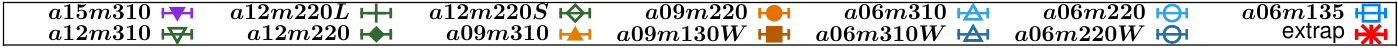}
}
%\subfigure{
  \includegraphics[trim=0 0 14 60, clip, width=0.32\textwidth]{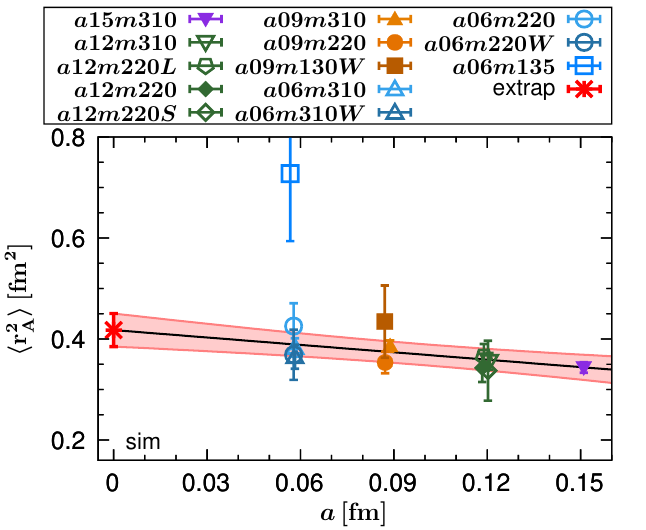}
  \hfill
  \includegraphics[trim=0 0 14 60, clip, width=0.32\textwidth]{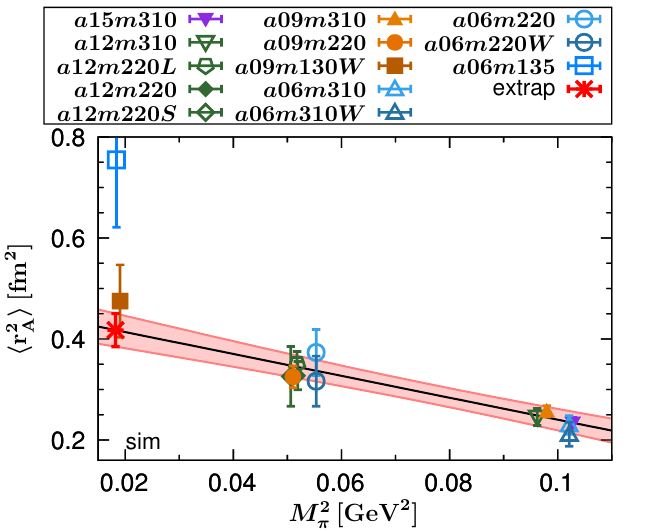}
  \hfill
  \includegraphics[trim=0 0 14 60, clip, width=0.32\textwidth]{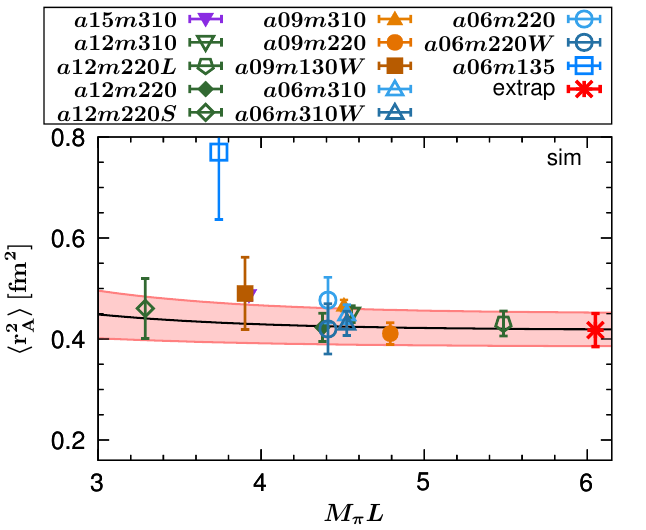}
\\
  \includegraphics[trim=0 0 14 60, clip, width=0.32\textwidth]{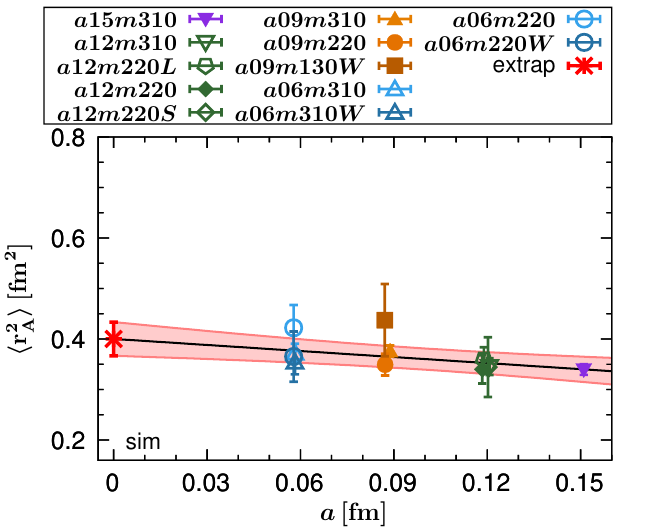}
  \hfill
  \includegraphics[trim=0 0 14 60, clip, width=0.32\textwidth]{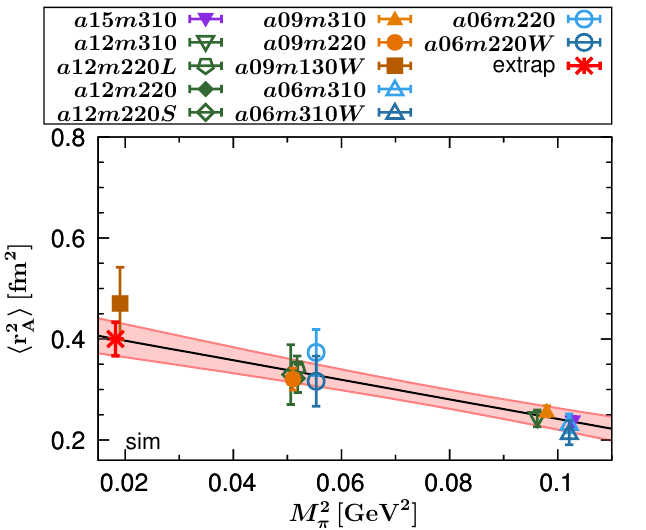}
  \hfill
  \includegraphics[trim=0 0 14 60, clip, width=0.32\textwidth]{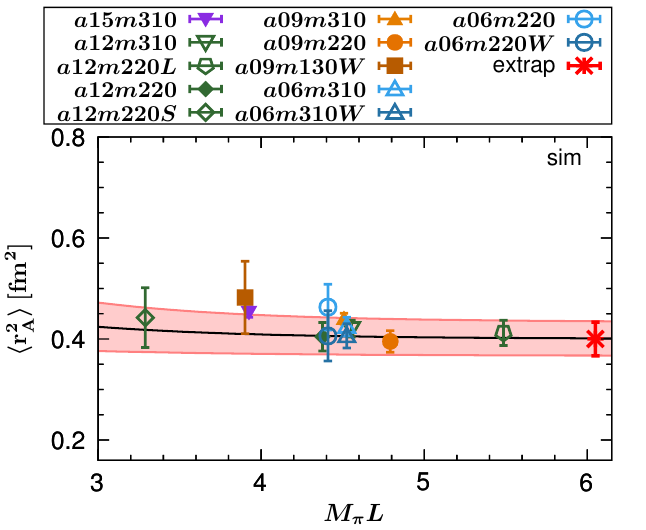}
%}
  \caption{(Top panels) The axial charge radius squared $\expv{r_A^2}$
    given by the 13-pt CCFV fit to data obtained using the $z^2$ fit
    to $G_A(Q^2\neq 0)$ with the $\Ssim$ strategy. (Bottom panels) The
    12-pt fit without the $a06m135$ point (open blue square in the three 
    panels on the top) that has large errors.  The pink band
    in each panel gives the result of the CCFV fit (Eq.~\eqref{eq:ccfv})
    versus the x-axis variable with the other two variables set to
    their physical values.  The data points in each panel have been
    shifted using the same CCFV fit, however, the size of errors are not changed.
    The final result at the physical point is shown by the red cross.
  }
  \label{fig:rA-z2-simA}
\end{figure*}

\subsubsection{$g_A$}
\label{ssec:aff-gA}

The result for $g_A$, taken from the 13-point CCFV extrapolation of 
$z^2$ fits to the $\Ssim$ data, shown in
Fig.~\ref{fig:gA-z2-simA}, is \looseness-1
\begin{align}
  % from the full CCFV fits
  g_A =& 1.296 (50)_\text{stat}\, (13)(11) \CL
      =& 1.296 (50)_\text{stat}\, (17)_\text{sys}\,, \qquad [z^2]  \,.
\label{eq:gA-z} 
\end{align}
The first error is the total analysis uncertainty given by the overall bootstrap 
process, and the next two
are additional systematic uncertainties: (i) the difference between
using $z^2$ and $z^3$ fits and (ii) the difference of this central
value from the average of the three 12-point CCFV fits. The two
systematics are added in quadrature to get the total systematic error
given in the second line in Eq.~\eqref{eq:gA-z}.  In
Sec.~\ref{sec:charges}, this result is compared with an independent
analysis of $g_A$ obtained from the forward matrix element, i.e., from
the zero momentum correlator, $C_{A_3}(\bm{p=0})$, as defined in
Eq.~\eqref{eq:Ai-ff-decomp}.

\begin{table}[!tb]   %T03
  % full CCFV fits
  \centering
  \renewcommand{\arraystretch}{1.1}
  \begin{ruledtabular}
    \begin{tabular}{cll}
      $g_A$ & \multicolumn{1}{c}{$z^2$} & \multicolumn{1}{c}{dipole} \\ \hline
      $\Ssim$  & 1.296(50)(13)(11) & 1.239(43)(-)(39) \\
      $\Sfour$ & 1.281(51)(11)(21) & 1.204(44)(-)(21) \\
      $\Stwo$  & 1.213(39)(02)(-)  & 1.228(37)(-)(-)  \\
\hline
      $\expv{r_A^2}$ & \multicolumn{1}{c}{$z^2$} & \multicolumn{1}{c}{dipole} \\ \hline
      $\Ssim$  & 0.418(33)(29)(18) & 0.305(13)(-)(06) \\
      $\Sfour$ & 0.428(31)(21)(19) & 0.305(15)(-)(06) \\
      $\Stwo$  & 0.282(27)(16)(-)  & 0.275(14)(-)(-) 
    \end{tabular}
  \end{ruledtabular}
  \caption{$g_A$ and $\expv{r_A^2}$ from the 13-point CCFV
    fit. Results are given for the $z^2$ and dipole fits to
    $G_A(Q^2\neq 0)$, and for the three strategies used to control
    ESC. In each case, in addition to the central value and the total
    analysis error, the two systematic errors are the difference
    between the $z^2$ and $z^3$ estimates, and the difference from the
    12-pt CCFV fits explained in the Sec.~\protect\ref{ssec:aff-extrap}.}
  \label{tab:gA}
\end{table}

\subsubsection{$\expv{r_A^2}$}
\label{ssec:aff-rA}

The CCFV fit to $\expv{r_A^2}$, obtained from the $\Ssim$ data with $z^2$ fit, is 
shown in Fig.~\ref{fig:rA-z2-simA} (top panels). It gives 
\begin{align}
  % from the full CCFV fits
  \expv{r_A^2} =& 0.418 (33)_\text{stat} (29)(18)\,\fm^2 \CL
               =& 0.418 (33)_\text{stat} (34)_\text{sys}\,\fm^2 \,,\qquad [z^2] 
\label{eq:rA-z} 
\end{align}
with the errors derived in the same way as for $g_A$. 

For both $g_A$ and $\expv{r_A^2}$, the largest dependence in the CCFV
fit is on $M_\pi^2$ for the $\Ssim$ and $\Sfour$ strategies. This is a
consequence of the increasing influence of the $N \pi$ state as its mass gap 
decreases well below the N(1440) as $M_\pi \to 135$~MeV. In contrast, the
$\Stwo$ data, which do not include the $N \pi $ state in the analysis,
show mild dependence on all three variables $\{a, M_\pi, M_\pi L\}$.

Estimates from the 12-pt CCFV fit excluding the $a06m135$ point, shown
in the bottom panels of Fig.~\ref{fig:rA-z2-simA}, are consistent with
the 13-point results. This is expected since the errors in the $a06m135$
point are large.  Clearly, to further improve the estimates of both
$g_A$ and $\expv{r_A^2}$ requires much higher statistics data at small
$Q^2$ on the physical pion-mass ensembles.

%%%%%%%%%%%%%%%%%%%%%%%%%%%%%%%%%%%%%%%%%%%%%%%%%%%%%%%%%%%%%%%%%%%%%%%%

\begin{figure}[!tbh]  %F10
  \includegraphics[trim=0 0 0 0, clip, width=0.95\columnwidth]{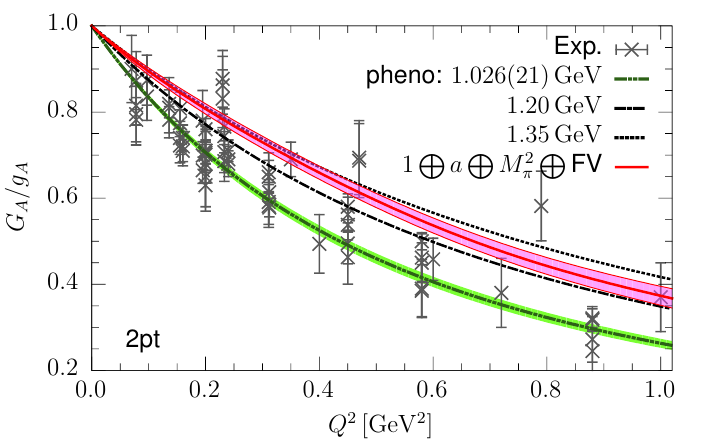}
  \\
  \includegraphics[trim=0 0 0 0, clip, width=0.95\columnwidth]{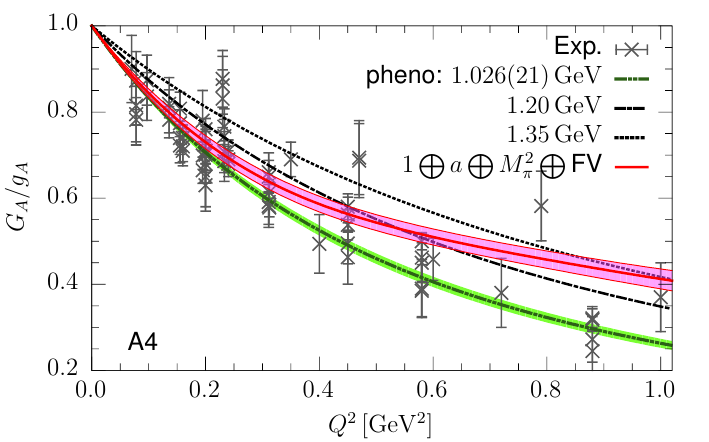}
  \\
  \includegraphics[trim=0 0 0 0, clip, width=0.95\columnwidth]{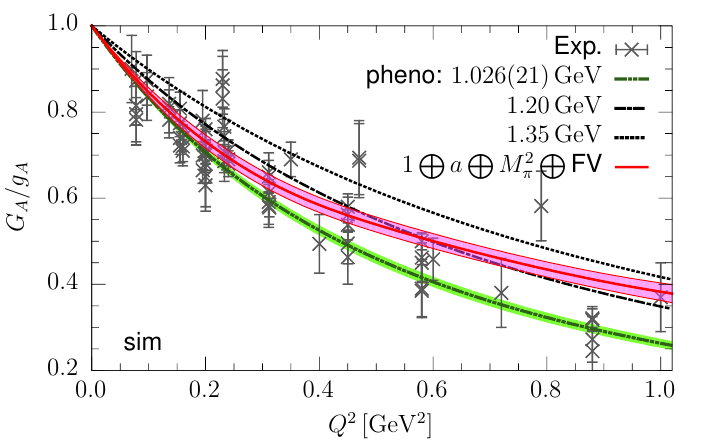}
  \caption{Results for $G_A/g_A$ at the physical point for the three
    strategies $\Stwo$, $\Sfour$ and $\Ssim$ (labeled ``2pt'',
    ``A4'', and ``sim'', respectively) used to control the
    excited-state contamination. The three step process used to get
    these results shown by the pink band is described in the text.  In
    each case, the error band represents the full analysis error for
    that strategy but with the value at $Q^2=0$ fixed to unity. The
    label $1 \oplus a \oplus M_\pi^2 \oplus FV$ specifies that all 4
    terms in the CCFV ansatz, Eq.~\eqref{eq:ccfv}, were kept. The
    experimental $\nu$-deuterium data (gray crosses labeled Exp.)
    were provided by Ulf Meissner and the dipole result
    $M_A=1.026(21)$~GeV is taken from
    Ref.~\protect\cite{Bernard:2001rs}. This and the two other dipole
    fit with $M_A = 1.20$ and $ 1.35$~GeV are shown only for
    comparison.}
  \label{fig:GA-CCFV-Qsq}
\end{figure}

\begin{figure*}[!tbh]  %F11
  \includegraphics[trim=0 0 0 0, clip, width=0.95\textwidth]{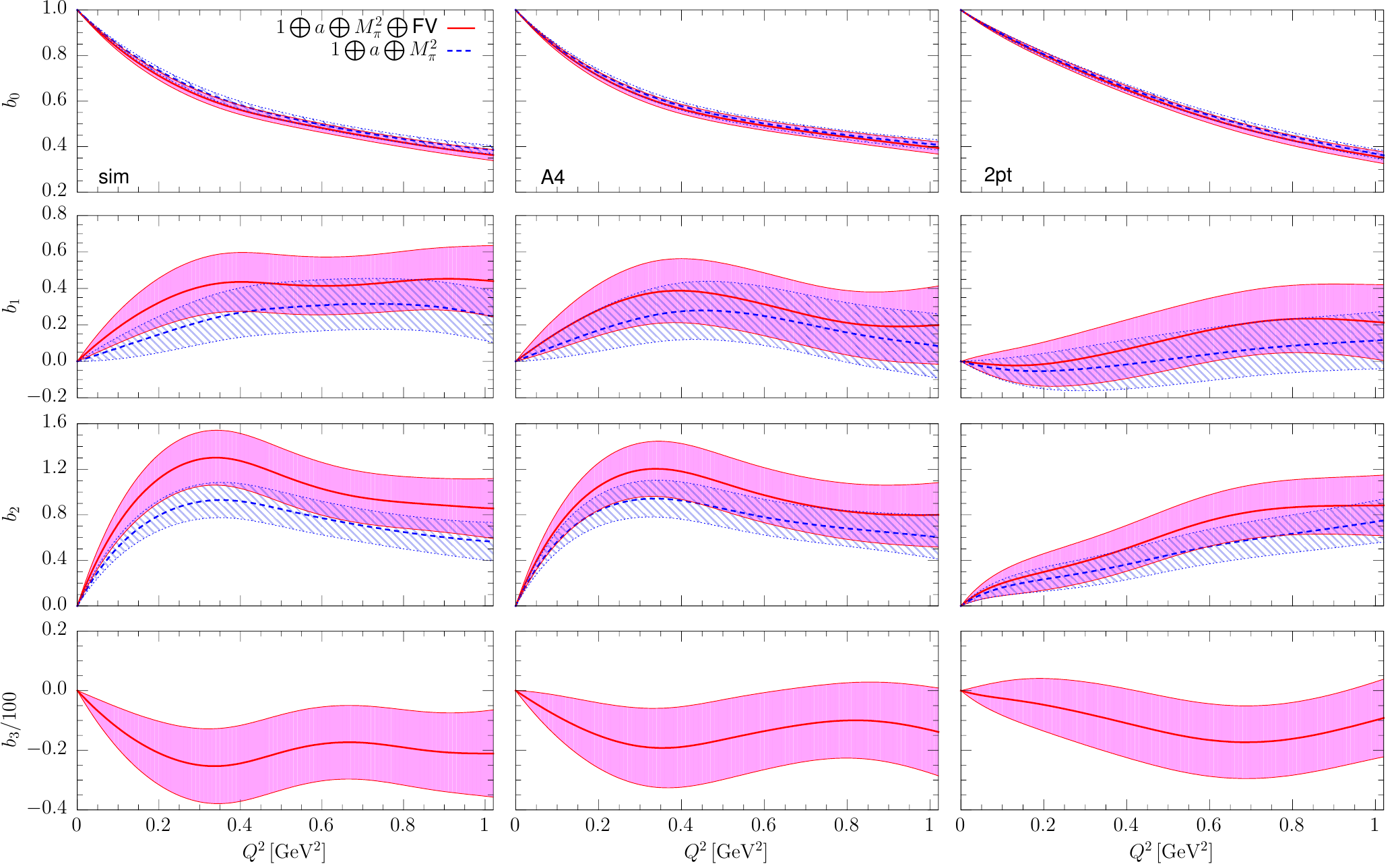}
  \caption{The fit coefficients $b_i$, $i=0,1,2,3$, defined in
    Eq.~\eqref{eq:ccfv}, for the CCFV extrapolation of the axial form
    factor $G_A(Q^2)/g_A$ obtained with strategy $\Ssim$ (left), $\Sfour$
    (middle) and $\Stwo$ (right) and fit with the $z^2$
    truncation. The extrapolated $G_A(Q^2)/g_A$ with $\Ssim$ is shown in
    Fig.~\ref{fig:GA-CCFV-Qsq}. The hatched blue curves correspond to 
    $b_{0,1,2}$ obtained neglecting the finite-volume term in the analysis.}
  \label{fig:GA-CCFV-Qsq-coeff-all3}
\end{figure*}

%%%%%%%%%%%%%%%%%%%%%%%%%%%%%%%%%%%%%%%%%%%%%%%%%%%%%%%%%%%%%%%%%%%%%%%%

\subsection{$G_A(Q^2)$ at the Physical Point}
\label{ssec:aff-GA}

The $Q^2$ dependence of the axial form factor up to $1\,\GeV^2$,
obtained at the physical point, is shown in Fig.~\ref{fig:GA-CCFV-Qsq}
for the three strategies $\Stwo$, $\Sfour$, and $\Ssim$. 
The pink band in these figures was obtained using the following three
step process starting with the renormalized lattice data for
$G_A(Q^2)/g_A$, which on each of the thirteen ensembles are at
different discrete values of $Q^2$.  First the data on each ensemble
were fit using the $z^2$-ansatz (see Eq.~\eqref{eq:zexp}) and the
result is taken to specify $G_A(Q^2)/g_A$ for $0 < Q^2 \leq
1\,\GeV^2$. Second, we chose a set of eleven $Q^2$ values evenly
distributed over this range, and for the thirteen data points at each
of these $Q^2$ values carry out a 13-point CCFV extrapolation using
Eq.~\eqref{eq:ccfv}. The result was taken to be the physical point
value of $G_A/g_A $ at that $Q^2$.  In each of these CCFV fits, the
thirteen points from the thirteen ensembles are uncorrelated as these
are independent calculations.  Lastly, these eleven extrapolated
points were fit by the $z^2$ ansatz to obtain the final
parameterization valid in the interval $ 0 \leq Q^2 \leq 1.0$
GeV${}^2$ and shown by the pink band in
Fig.~\ref{fig:GA-CCFV-Qsq}. The errors in the original lattice data
were fully propagated through this three step process carried
out within a single bootstrap setup. This gives the central value and error. 
Possible uncertainty due to incomplete removal of ESC or due to using
only the leading order CCFV fit ansatz is to be estimated separately. 

Figure~\ref{fig:GA-CCFV-Qsq} also shows the experimental bubble
chamber data and the dipole ansatz with $M_A=1.026(21)$~GeV extracted
from it (green band)~\cite{Bernard:2001rs}.  A recent $z$-expansion analysis of the
$\nu$-deuterium data~\cite{Meyer:2016oeg} finds a $\approx$10X larger
uncertainty. In our analysis, only the $\Stwo$ data are 
roughly consistent with a dipole ansatz with $M_A \approx
1.30$~GeV, however, the three form factors extracted using $\Stwo$ fail to
satisfy the PCAC relation. We, therefore, reemphasize that the dipole
curves with $M_A= 1.026$, $1.2$ and $ 1.35$~GeV are shown only for 
comparison.\looseness-1

The data from the $\Sfour$ and $\Ssim$ strategies are consistent, and
show a more rapid fall until $Q^2 \approx 0.3\,\GeV^2$, and give
results roughly consistent with the dipole values $M_A=1.026(21)$~GeV
(and $\expv{r_A^2} = 0.444(28)$~fm${}^2$)) in
Ref.~\cite{Bernard:2001rs}. They then level out falling more slowly,
however, note that for $Q^2 > 0.5 \,\GeV^2$ our results are mainly
from the heavier $M_\pi \approx 310$~MeV ensembles.  On the heavier
$M_\pi$ ensembles, the mass gaps in the $N \pi$ analyses (blue
triangles and red squares in Fig.~\ref{fig:esc-axial}) increase rapidly towards those
from the ${{\mathcal{S}_\text{2pt}}}$ fit (black
circles). Consequently, as shown in Fig.~\ref{fig:GA-CCFV-Qsq}, the $G_A/g_A$ from the
${{\mathcal{S}_\text{sim}}}$ analysis moves towards the
${{\mathcal{S}_\text{2pt}}}$ result for $Q^2 \gtrsim 0.5$~GeV${}^2$.

To obtain data for $Q^2 > 0.5 \,\GeV^2$ on
physical pion mass ensembles with $M_\pi L > 4$ requires simulations
at much larger values of $\textbf q$ where statistical and
discretization errors are large with the methodology used in this
work. A more promising method for generating data at large $Q^2$ is
momentum smearing~\cite{Bali:2016lva}.  Also, when including points
with larger $Q^2$, the $z$-expansion fits with and without sum-rules
should be compared since it is not known, a priori, when the expected
$1/Q^4$ asymptotic behavior becomes significant. Alternately, as shown
in Ref.~\cite{Park:2021ypf}, one can analyze the data using a Pad\'e
parameterization. Our fits to the final $G_A$ using the Pad\'e ansatz
$g_A/(1 + b_1 Q^2 + b_2 Q^4)$, which has the $1/Q^4$ behavior built in,
gave estimates consistent with Eqs.~\ref{eq:gA-z} and~\ref{eq:rA-z}.

The coefficients $b_i$ in the CCFV fit using Eq.~\eqref{eq:ccfv} are
shown in Fig.~\ref{fig:GA-CCFV-Qsq-coeff-all3} for the three
strategies $\Stwo$, $\Sfour$, and $\Ssim$.  The coefficients
$b_1(Q^2)$ and $b_2(Q^2)$ are similar within errors for $\Ssim$ and
$\Sfour$, significantly different from zero, and qualitatively
different from the case $\Stwo$. The $b_i$ for $\Ssim$ and $\Sfour$
show a change in behavior at $Q^2 \approx 0.3$~GeV${}^2$, coincident
with the region where the curvature in $G_A$ changes as shown in
Fig.~\ref{fig:GA-CCFV-Qsq}.  This could be due to the fact that most
of the raw data controlling the parameterization at $Q^2 \gtrsim
0.3$~GeV${}^2$ comes from the $M_\pi \approx 220$ and $310$~MeV
ensembles. On the other hand, support for the parameterization comes
from the observation that the data, plotted in
Fig.~\ref{fig:GAdata13}, do not show a significant variation versus
$\{a, M_\pi\}$.  Cutting out the data with $Q^2 >
0.3$~GeV${}^2$ to see if fits change, unfortunately eliminates most of the $M_\pi \approx 220$ and
$310$~MeV ensembles---they do not have enough points to perform even a
$z^2$ fit.  We are, therefore, not able to resolve the reason for the
change in parameterization around $Q^2 \approx 0.3$~GeV${}^2$.\looseness-1

\begin{figure*}[!tbh]  %F12
  \includegraphics[trim=0 0 0 0, clip, width=1.4\columnwidth]{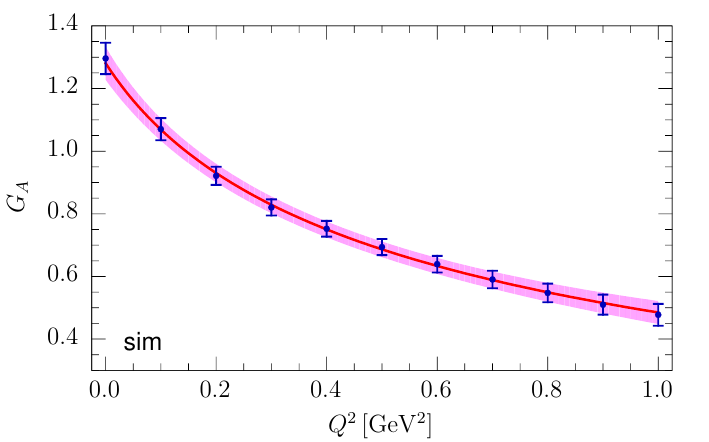}
  \caption{The final estimate of $G_A(Q^2)$ at the physical
    point. The eleven fiducial points used to make the fit are shown
    in blue and their errors are obtained from the one overall bootstrap
    analysis covering the three step process described in the text. The 
    parameterization versus $z$ is given in Eq.~\protect\eqref{eq:GAz2fit}.  }
  \label{fig:GAz2fit}
\end{figure*}
%%%%%%%%%%%%%%%%%%%%%%%%%%%%%%%%%%%%%%%%%

To provide our best parameterization of $G_A(Q^2)$ for phenomenology, we
repeated the above 3-step procedure using the
$\Ssim$ strategy data. Again, the data after extrapolation to the continuum
limit (at $Q^2 = 0, 0.1, 0.2, . . . , 1.0 $ GeV${}^2$) were fit with a
$z^2$ ansatz, $t_c = 9M_\pi^2$ with $M_\pi=135$~MeV, and $t_0 =
0.25$~GeV${}^2$. The result, shown in Fig.~\ref{fig:GAz2fit}, has the
parameterization
\begin{align}
G_A(Q^2) &= a_0 + a_1\ z + a_2 \ z^2  \CL
         &= 0.876(28) -1.669(99) z + 0.483(498) z^2   \,,
\label{eq:GAz2fit}
\end{align}
with the correlation matrix:
\begin{equation}
\begin{blockarray}{cccc}
&a_0&a_1&a_2\\
\begin{block}{l(rrr)}
a_{0\ \ }  &  1.0        &  -0.45170  &  -0.02966  \\
a_{1\ \ }  &   -0.45170  &  1.0       &  -0.24394  \\
a_{2\ \ }  &  -0.02966   & -0.24394   &   1.0       \\
\end{block} 
\end{blockarray}
\label{eq:GAz2Cov}
\end{equation}
This fit gives 
\begin{align}
g_A &= 1.281(53) \,,   \CL
\expv{r_A^2} &=  0.498(56)\ {\rm fm}^2   \,,
\label{eq:gAz2final}
\end{align}
which are consistent with the estimates in Eqs.~\eqref{eq:gA-z} and
~\eqref{eq:rA-z}, {\it albeit} with a $\expv{r_A^2}$ larger by
roughly $ 1\sigma$.

We also carried out this final $z^2$ fit setting $t_0 = 0$ in the
definition of $z$. The results are 
\begin{align}
g_A &= 1.282(54)   \CL
\expv{r_A^2} &=  0.505(66)\ {\rm fm}^2   \,.
\label{eq:gAz2t00}
\end{align}
While consistent, the coefficient $c_2$ in this fit is essentially undetermined. We, therefore, 
choose the results given in Eq.~\eqref{eq:gAz2final}. 

For our final results from the analysis of $G_A$, we take the average weighted by the ``stat'' errors of values given in
Eqs.~\eqref{eq:gA-z},~\eqref{eq:rA-z} and~\eqref{eq:gAz2final} to get
\begin{align}
g_A &= 1.289(53)_\text{stat} (17)_\text{sys}   \CL
\expv{r_A^2} &=  0.439(56)_\text{stat} (34)_\text{sys}\ {\rm fm}^2   \,.
\label{eq:gArAfinal}
\end{align}
For errors, we take the larger of the ``stat'' error and keep the ``sys'' errors 
given in Eqs.~\eqref{eq:gA-z} and~\eqref{eq:rA-z}. 

We now return to two related issues that arise because the majority of the data
used to get the parameterization in Eq.~\eqref{eq:GAz2fit} are at 
$Q^2 \lesssim 0.5$~GeV${}^2$ as shown in
Fig.~\ref{fig:GA-Scomp-all}. The first is 
whether this parameterization is compatible with $G_A(Q^2=\infty) = 0$ since
the sum rule constraints needed to build in the perturbative $1/Q^4$
behavior have not been imposed?  And the second is---how reliable is this $G_A$
for $Q^2 > 0.5$~GeV${}^2$ since most of the data in this region are
from the $M_\pi \approx 310$~MeV ensembles?  Regarding the first issue, 
the parameterization in Eq.~\eqref{eq:GAz2fit} at $z=1$ ($\Rightarrow Q^2=\infty$) gives 
\begin{align}
G_A &= -0.31(48) \CL
\frac{d G_A(z)}{dz} &= -0.70(98)    \,.
\label{eq:GAtest}
\end{align}
These are consistent with zero within one sigma. For the second issue,
data in Fig.~\ref{fig:GAdata13} show that $G_A(Q^2)$ extracted on 12
ensembles show little dependence on $\{a, M_\pi\}$. The one exception,
$a06m135$, where the data lie about one sigma lower, has already been
identified as statistics limited. As stated before, if future data
continue to show little dependence on $a$ and $M_\pi$, then even data
from the $M_\pi \approx 310$~MeV ensembles would provide a good
approximation to the continuum $G_A$ and increase confidence in the
result for $Q^2 > 0.5$~GeV${}^2$.

Lastly, we made a $z^2$ fit to the same continuum extrapolated data but
imposed a prior, $G_A(Q^2 = \infty)=0$ with width $0.3$ based on the
value in Eq.~\eqref{eq:GAtest}. The result is
\begin{align}
G_A(Q^2) &= 0.872(28) -1.705(109) z + 0.767(102) z^2   \,,
\end{align}
\label{eq:GAz2prior}
%
\iffalse
\begin{equation}
\begin{blockarray}{cccc}
&a_0&a_1&a_2\\
\begin{block}{l(rrr)}
a_{0\ \ }  &  1.0        &  -0.35062  &  0.23426  \\
a_{1\ \ }  &  -0.35062   &  1.0       &  -0.54468  \\
a_{2\ \ }  &  0.23426    & -0.54468   &  1.0       \\
\end{block}
\end{blockarray}
\label{eq:GAz2Cov}
\end{equation}
with $g_A = 1.300(46)$ and $\expv{r_A^2} = 0.533(24)$. 
\fi
%
%
\noindent with $G_A=-0.07(10)$ and $d G_A/d z = -0.17(17)$ at $z=1$.
The main difference from the result in Eq.~\eqref{eq:GAz2fit} is the
tightening of the estimate of the $z^2$ term. Overall, in all these
fits, the coefficients $a_0$ and $a_1$, defined in
Eq.~\eqref{eq:GAz2fit} are stable, whereas higher precision data are
needed to improve $a_2$.

\section{Couplings $g_P^\ast$ and $g_{\pi NN}$ from the Induced Pseudoscalar Form Factor}  %S04
\label{sec:aff-gpx}

The induced pseudoscalar coupling $g_P^\ast$ is defined as 
\begin{align}
  g_P^\ast \equiv \frac{m_\mu}{2M_N}\GP(Q^{\ast 2}) \,,
  \label{eq:gPstar}
\end{align}
where $m_\mu$ is the muon mass and $Q^{\ast 2} = 0.88 m_\mu^2$ is the energy scale of muon capture. Similarly, 
the pion-nucleon coupling $g_{\pi NN}$ is obtained from the residue of
$\GP(Q^2)$ at the pion pole, i.e.,  through the relation
\begin{align}
g_{\pi NN} &\equiv \;  \lim_{Q^2 \to -M_\pi^2} \frac{M_\pi^2 + Q^2}{4M_N F_\pi} {\widetilde G}_P(Q^2) \CL
           &=    \frac{  \FP(-M_\pi^2) M_N }{F_\pi}  \,,
\label{eq:gAGT} 
\end{align}
where $F_\pi=92.9$~MeV is the pion decay constant. The function $\FP$ defined as 
\begin{align}
  \widetilde{F}_P(Q^2) \equiv \frac{Q^2+M_\pi^2}{4M_N^2}  \GP(Q^2) \,. 
  \label{eq:GP2FP}
\end{align}
is $\GP$ without the pion pole and should equal $G_A$ if PPD were
exact. This requires $R_3 \equiv \FP/G_A $, plotted in
Fig.~\ref{fig:PPD}, to be unity. Deviations for $\Stwo$ are
significant, while those for $\Ssim$ and $\Sfour$ are one within the
size of discretization errors and/or violations of PPD expected.

To extract $g_P^\ast$ and $g_{\pi NN}$ from the lattice data, a
parameterization of the $Q^2$ behavior of $\GP$ and $\FP$ was carried
out. A comparison of the $z^1$ and $z^2$ fits to $G_A$ and $\FP$ from
the $\Ssim$ strategy is shown in Fig.~\ref{fig:GA-GP-zexp-ALL-simA}
for the thirteen ensembles.  Results from $z^2$ and $z^3$ fits are
consistent, indicating convergence, while $z^1$ fits miss the small
curvature seen. To avoid over parameterization, we again take the $z^2$
results as the central values.

\subsection{Parameterization of $\GP(Q^2)$ and $\FP$}
\label{ssec:GP-param}

Based on the analysis of PCAC (see Figs.~\ref{fig:PCAC}), we 
focus on the $\GP(Q^2)$ data from the $\Ssim$ and $\Sfour$ strategies
and again give the $\Stwo$ results only for comparison.

We consider two ways to parameterize $\GP(Q^2)$, both of which build in the 
pion-pole dominance hypothesis. The first is the expansion
\begin{align}
  \GP(Q^2) &= \frac{c_0}{Q^2+M_\pi^2} + c_1 + c_2 Q^2 \,,
  \label{eq:expandGP}
\end{align}
where the $c_i$ $(i=0,1,2)$ are fit parameters, and we have kept 
as many terms in the polynomial as can be resolved by the data. Results for
$g_P^\ast$, $g_{\pi NN}F_\pi$ and $g_{\pi NN}F_\pi/M_N$ using this fit
(labeled PD) to the $\Ssim$ data are given in
Table~\ref{tab:Qsqfit-gPstar-PD} along with the $\chi^2/\text{DOF}$
and $p$-value for the fits.  

In the second way, we treat $\FP(Q^2)$ as an analytic function that can be
fit using either the dipole ansatz (with free parameters $\FP(0)$ and
$\widetilde{\mathcal{M}}_P$) or the $z$-expansion, Eq.~\eqref{eq:zexp}, 
with $z$ again defined by Eq.~\eqref{eq:def-z}.  Results for
$g_P^\ast$, $g_{\pi NN}F_\pi$ and $g_{\pi NN}F_\pi/M_N$, from $z^2$
fits to $\widetilde{F}_P(Q^2)$ obtained with the $\Ssim$ strategy are
given in Table~\ref{tab:Qsqfit-gPstar-2} and agree with those in 
Table~\ref{tab:Qsqfit-gPstar-PD}. 

\begin{figure*}[!h]  %F14
 \includegraphics[width=0.95\columnwidth]{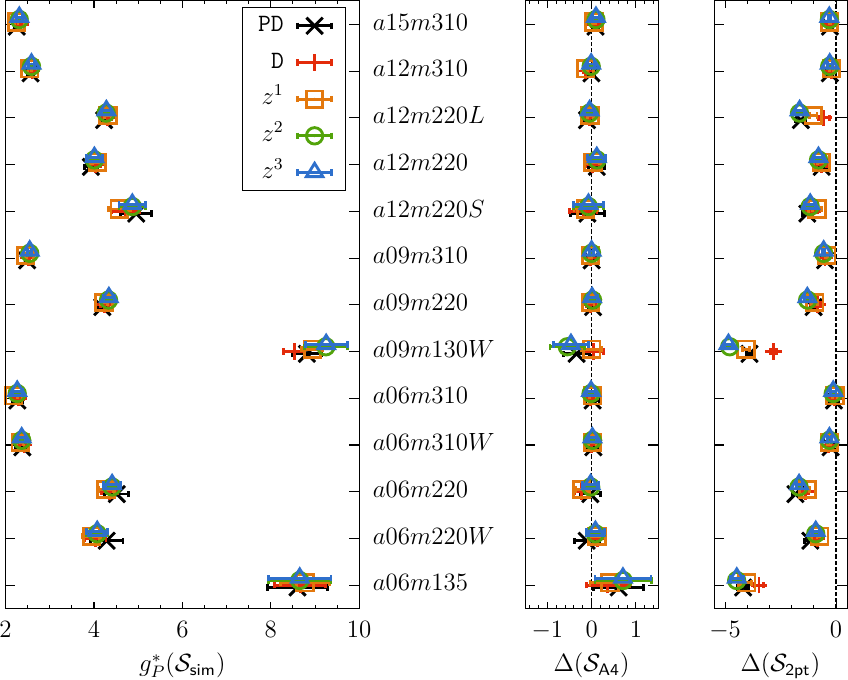} \hspace{1.0cm}
 \includegraphics[width=0.95\columnwidth]{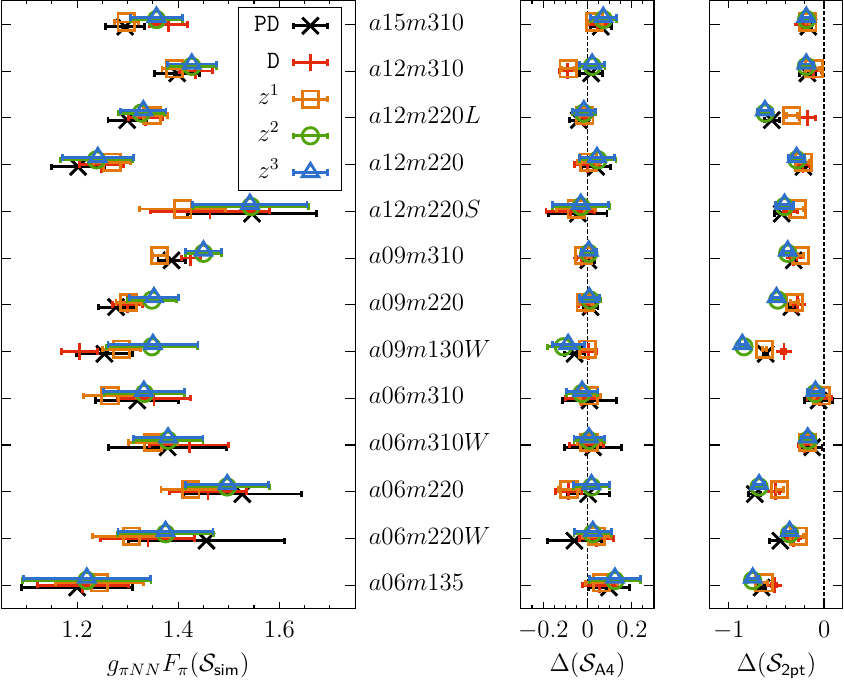}
\vspace{-0.2cm}
  \caption{(Left) Results for the bare $g_P^\ast$ from the strategy $\Ssim$ and
    the differences $\Delta(X)=g_P^\ast(X) - g_P^\ast(\Ssim)$. To
    facilitate visualization of the spread, the errors plotted for
    $\Delta(X)$ are those in $g_P^\ast(X)$.  The fits used to
    parameterize the $Q^2$ behavior are labeled ``PD'' defined in 
    Eq.~\eqref{eq:expandGP}; ``D'' for the dipole fit, and $z^k$ for
    various truncations of the $z$-expansion.
    (Right) Results for bare values of $g_{\pi NN}F_\pi$ obtained with the
    strategy $\Ssim$ and the differences $\Delta(X)=g_{\pi NN}F_\pi(X)
    - g_{\pi NN}F_\pi(\Ssim)$.  
\label{fig:gPstar-gpiNNFpi-comp}}
\end{figure*}

Results for the bare $g_P^\ast$ from the $\Ssim$ strategy for 
five $Q^2$ parameterizations of $\GP$ are shown in
Figure~\ref{fig:gPstar-gpiNNFpi-comp} (left) along with their
differences from results obtained using the $\Sfour$ and $\Stwo$
strategies. The analogous results for unrenormalized $g_{\pi NN}F_\pi$
are shown in Fig.~\ref{fig:gPstar-gpiNNFpi-comp} (right).

\subsection{Extrapolation of $g_P^\ast$ and $g_{\pi NN}$ to the Physical Point}
\label{sec:extrap-gpstar}

\subsubsection{$g_P^\ast$}
\label{sec:gPstar}

Renormalized $g_P^\ast$ is extrapolated to the physical point in two
ways.  In the first method $2 m_\mu M_N \FP(Q^{\ast\,2})$ is
extrapolated using the CCFV fit function given in Eq.~\eqref{eq:ccfv}
and multiplied by the the pion-pole factor at the physical point:
\begin{align}
  g_P^\ast = \left.2 m_\mu M_N \FP(Q^{\ast\,2})\right\vert_\text{extrap} \times \left.\frac{1}{Q^{\ast\,2} + M_\pi^2}\right\vert_\text{phys} \,.
\end{align}
In the second method, extrapolation of $g_P^\ast$ is carried out by
adding a pion-pole term, $b_4^{g_P^\ast} / (Q^{\ast\,2}+M_\pi^2)$,
to the CCFV fit function in Eq.~\eqref{eq:ccfv}.  The two methods give
consistent estimates and their unweighted average is used to get the
final results summarized in Table~\ref{tab:gPstar} for each of the
three strategies, $\Ssim$, $\Sfour$, and $\Stwo$.

The error obtained from the overall analysis is quoted as the first ``stat'' uncertainty.  The
systematical errors associated with truncation of the $z$-expansion, and the
largest difference of the central value from the three 12-pt CCFV fits
are quoted as the second and third errors.  The difference between the
two extrapolation  methods described above is quoted as the fourth error. 
For the final result, we take the $\Ssim$ with $z^2$ fits value:
\begin{align}
  g_P^\ast =& 9.03(47)_\text{stat}(01)(32)(27) \,, \quad [z^2]\,, \label{eq:gPstar-z} \CL
           =& 9.03(47)_\text{stat}(42)_\text{sys} \,.
\end{align}
In the second line, the three systematic errors are
combined in quadrature.  The 13-pt CCFV fit to the $\Ssim$ data 
on each ensemble fit with $z^2$, is shown in Fig.~\ref{fig:CCFV-4-z2-simA}. 

\begin{table*}[!tbh]   %T04
  % full CCFV fits
  \centering
  \renewcommand{\arraystretch}{1.1}
  \begin{ruledtabular}
    \begin{tabular}{clll}
      $g_P^\ast$ & \multicolumn{1}{c}{$z^2$} & \multicolumn{1}{c}{dipole} & \multicolumn{1}{c}{PD} \\ \hline
      $\Ssim$    & 9.03(47)(01)(32)(27) & 8.61(39)(-)(19)(23) & 8.92(45)(-)(38)(33) \\
      $\Sfour$   & 8.92(44)(04)(20)(23) & 8.70(37)(-)(17)(15) & 8.94(43)(-)(28)(33) \\
      $\Stwo$    & 4.50(26)(02)(-)(22)  & 5.36(25)(-)(-)(12)  & 4.73(27)(-)(-)(10)
    \end{tabular}
  \end{ruledtabular}
  \caption{$g_P^\ast$ from the $z^2$-expansion, dipole, and pion-pole
    dominance (PD) fits. The first column gives the strategy used for extracting the
    matrix elements. In each value, the first error is the total analysis error and the 
    rest are systematic errors explained in the text.}
  \label{tab:gPstar}
\end{table*}

\begin{table*}[!tbh]   %T05
  % full CCFV fits with average of extrapolation type I and II
  \centering
  \renewcommand{\arraystretch}{1.1}
  \begin{ruledtabular}
    \begin{tabular}{clll}
               & \multicolumn{1}{c}{$z^2$} & \multicolumn{1}{c}{dipole} & \multicolumn{1}{c}{PD} \\ \hline
      $\Ssim$  & 14.14(81)(01)(77)(35) & 13.03(67)(-)(41)(28) & 13.80(81)(-)(99)(33) \\
      $\Sfour$ & 13.77(79)(07)(33)(29) & 13.06(64)(-)(38)(26) & 13.90(79)(-)(57)(31)   \\
      $\Stwo$  & \phantom{0}5.76(57)(00)(-)(10)&  \phantom{0}7.57(46)(-)(-)(09)  &  \phantom{0}6.24(57)(-)(-)(05)   
    \end{tabular}
  \end{ruledtabular}
  \caption{Results for $g_{\pi NN}$ from the $z^2$, dipole, and pion-pole
    dominance (PD) fits.  The first column gives the ESC strategy used
    to extract the matrix elements. The first error is statistical
    and the rest are systematic as explained in the text.}
  \label{tab:gpiNN}
\end{table*}

\subsubsection{$g_{\pi NN}$}
\label{sec:gpiNN}

\begin{figure*}[!h]   %F15
  \centering
  \hspace{2em} \includegraphics[width=0.98\textwidth]{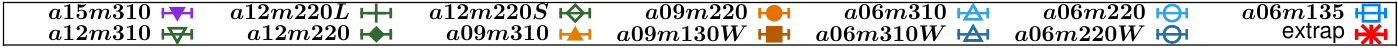} 
\vspace{-0.06cm}
  \includegraphics[trim=0 0 14 60, clip, width=0.32\textwidth]{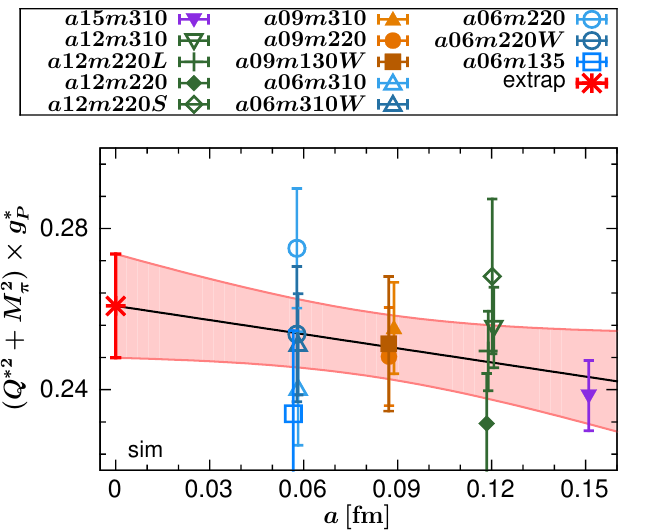}
  \hfill
  \includegraphics[trim=0 0 14 60, clip, width=0.32\textwidth]{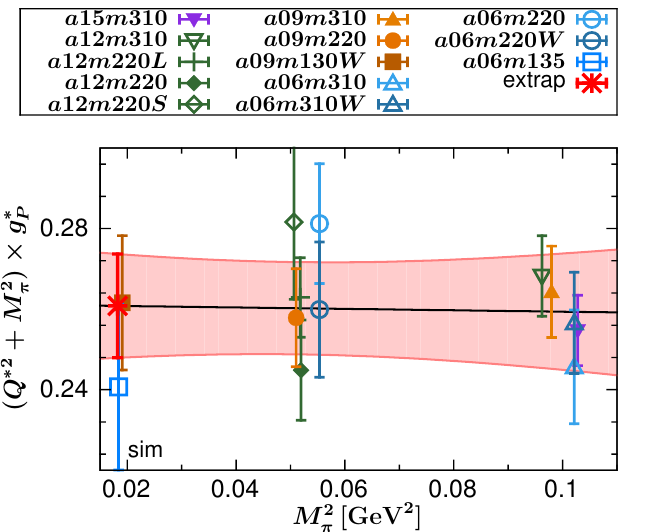}
  \hfill
  \includegraphics[trim=0 0 14 60, clip, width=0.32\textwidth]{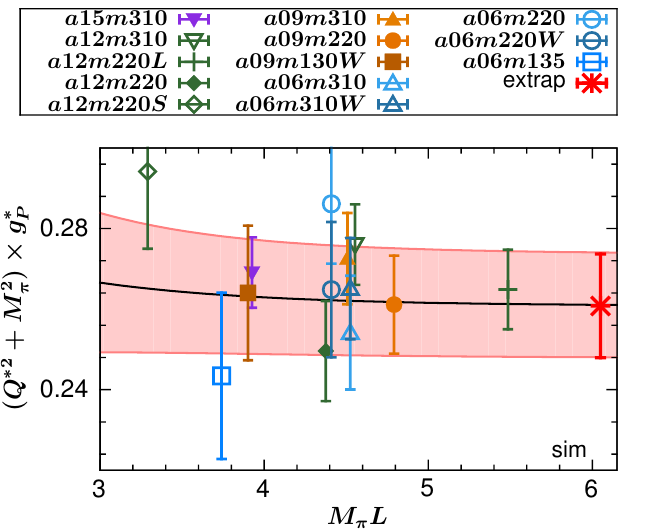}
\vspace{-0.01cm}
  \includegraphics[trim=0 0 14 60, clip, width=0.32\textwidth]{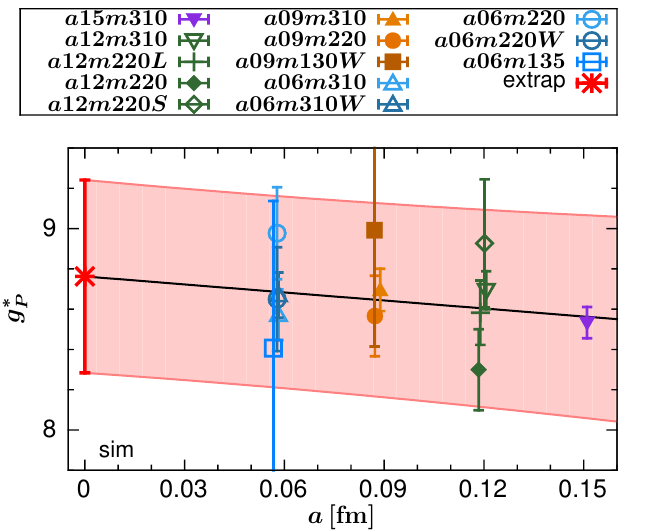}
  \hfill
  \includegraphics[trim=0 0 14 60, clip, width=0.32\textwidth]{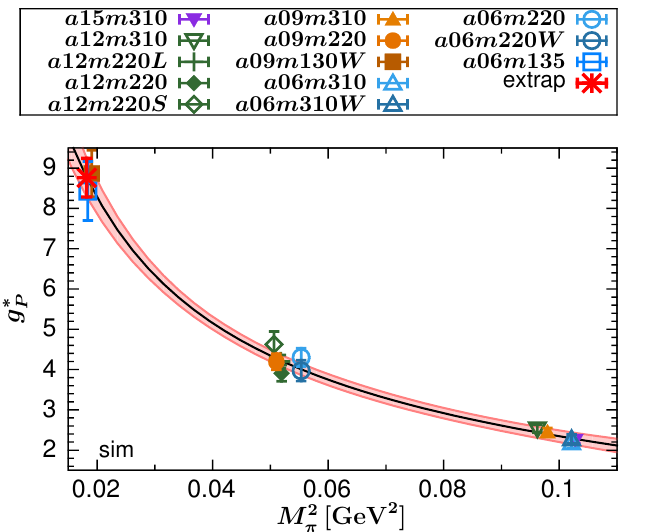}
  \hfill
  \includegraphics[trim=0 0 14 60, clip, width=0.32\textwidth]{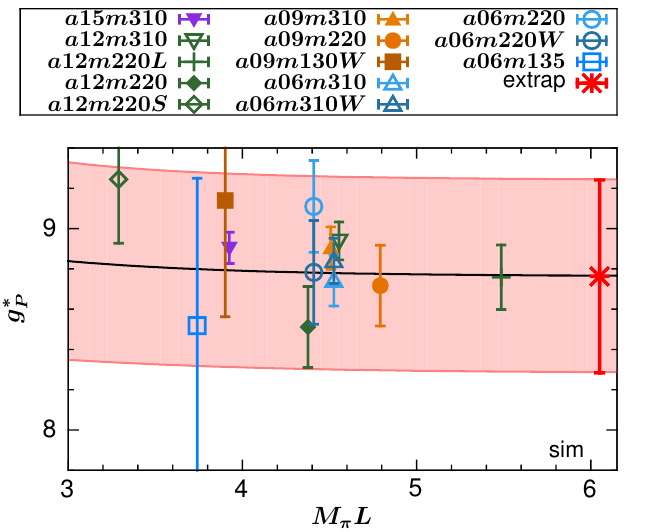}
\vspace{-0.06cm}
  \includegraphics[trim=0 0 14 60, clip, width=0.32\textwidth]{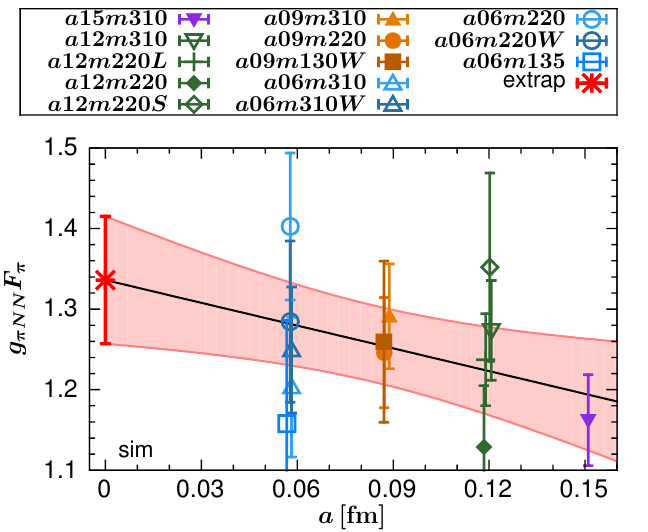}
  \hfill
  \includegraphics[trim=0 0 14 60, clip, width=0.32\textwidth]{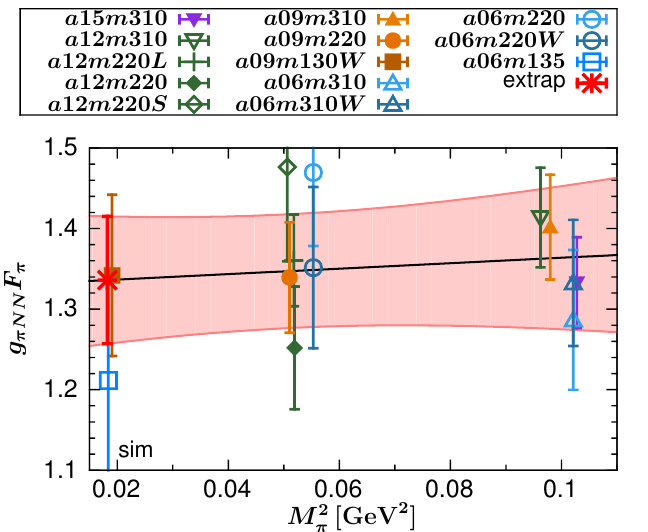}
  \hfill
  \includegraphics[trim=0 0 14 60, clip, width=0.32\textwidth]{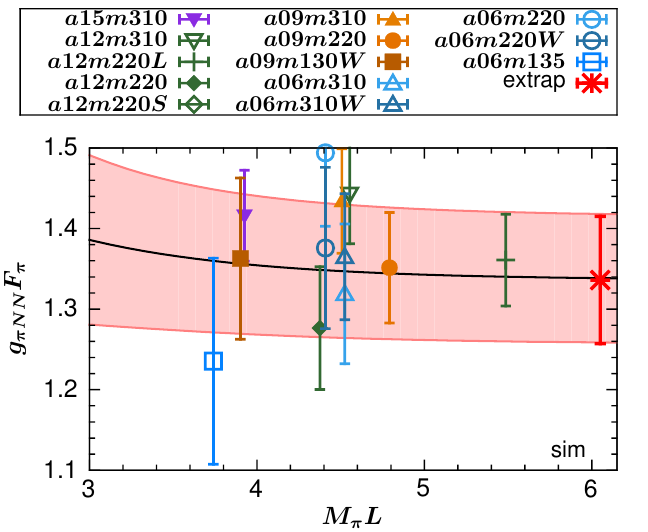}
  \caption{The chiral-continuum-finite-volume extrapolation of the
    $(Q^{\ast 2}+M_\pi^2)\times g_P^\ast$ (top row), $g_P^\ast$
    (middle row), and $g_{\pi NN}F_\pi$ (bottom row) data using
    the 13-point fit. In each case the data were obtained using the
    $z^2$ parameterization of $\FP(Q^2\neq 0)$ with strategy $\Ssim$. The black solid line
    with the pink error band represents a hyperplane obtained by taking
    the physical limit of all CCFV fit variables except the one shown
    on the $x$-axis. The plotted data points have been shifted by
    using the fit coefficients, while the errors are unchanged.}
  \label{fig:CCFV-4-z2-simA}
\end{figure*}

The CCFV extrapolation to obtain $g_{\pi N N }$ is carried out using
Eq.~\eqref{eq:ccfv} for (i) the product $g_{\pi NN}F_\pi = M_N
\FP(-M_\pi^2)$, and the result, in the continuum, divided by $F_\pi =
92.9\,\MeV$; and (ii) $\FP(-M_\pi^2)$ and the result multiplied by
$M_N (=939\,\MeV)/F_\pi (= 92.9\,\MeV)$.  It turns out that these two
extrapolations have different systematics: the slopes with respect to
$M_\pi^2$ of $g_{\pi NN} F_\pi$ and $\FP(-M_\pi^2)$ are different as
shown in Fig.~\ref{fig:CCFV-4-z2-simA}. The two estimates are,
however, consistent and we take their average to get the $g_{\pi NN}$
for the nine cases summarized in Table~\ref{tab:gpiNN}: the three
strategies and the three types of $Q^2$ fits.

The central value %
\begin{align}
  % full CCFV fits
  g_{\pi NN} =& 14.14(81)_\text{stat}(1)(77)(35) \CL
             =& 14.14(81)_\text{stat}(85)_\text{sys} \,. \qquad\quad [z^2] \label{eq:gpiNN-z}
\end{align}
is taken from the $\Ssim$ data with $z^2$ fits and the errors 
are estimated as for $g_P^\ast$.

\subsection{$\FP(Q^2)$ at the Physical Point}

\begin{figure*}[!tbh]  %F16
  \includegraphics[width=0.98\textwidth]{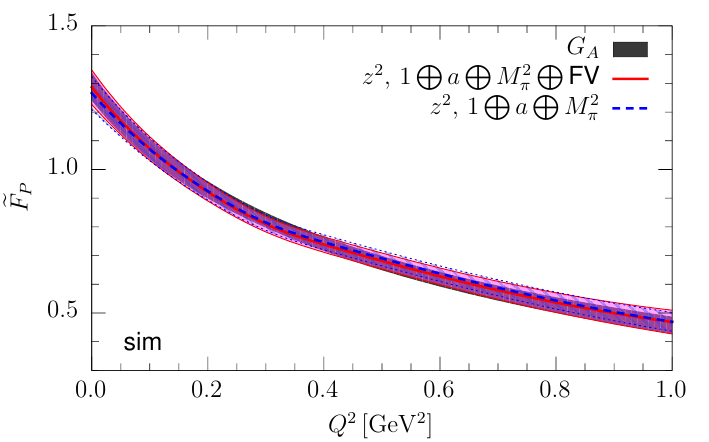}
\vspace{-2pt}
  \caption{The close overlap in the physical point results for
    $\FP(Q^2)$, defined in Eq.~\protect\eqref{eq:GP2FP}, with
    $G_A(Q^2)$ (black lines) reproduced from
    Fig.~\protect\ref{fig:GAz2fit}. Both were obtained using the three
    step process described in Sec.~\protect\ref{ssec:aff-GA} and the
    $\Ssim$ strategy data. The full CCFV fits to the $\FP$ data are
    shown with the solid red line and pink error band and the fit
    without the FV term with dashed blue line and hatched error
    band. These error bands show only the central analysis
    uncertainty.}
  \label{fig:FP-CCFV-Qsq}
\end{figure*}

\begin{figure*}[!tbh]  %F17
  \includegraphics[width=0.98\columnwidth]{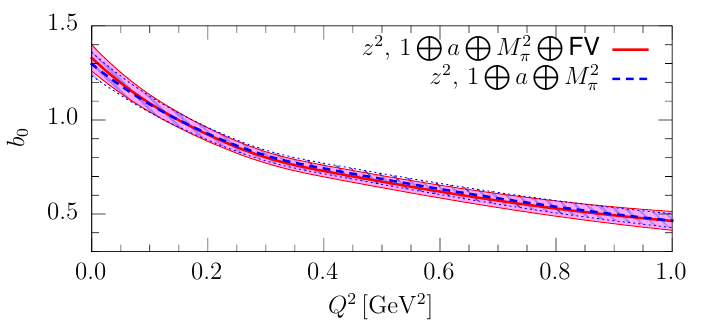}
  \includegraphics[width=0.98\columnwidth]{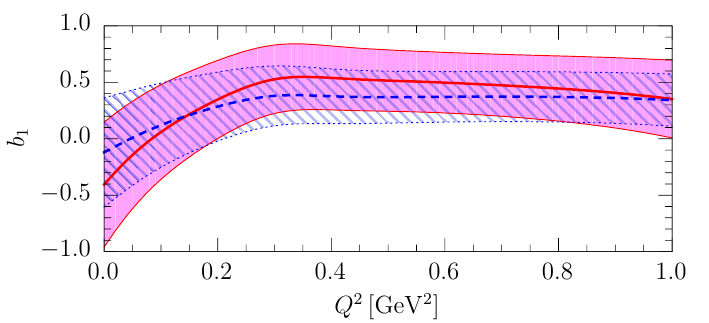} \\
  \includegraphics[width=0.98\columnwidth]{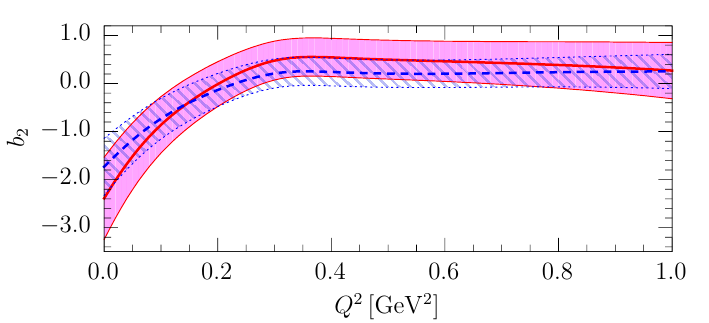}
  \includegraphics[width=0.98\columnwidth]{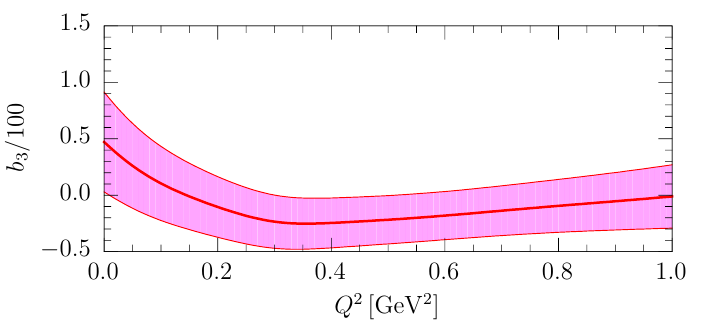}
\vspace{-2pt}
  \caption{The behavior of the four CCFV fit coefficients, $b_0, b_1,
    b_2$ and $ b_3$ defined in Eq.~\protect\eqref{eq:ccfv} versus
    $Q^2$ obtained in the extrapolation of $\FP(Q^2)$ to the physical
    point. The rest is the same as in
    Fig.~\protect\ref{fig:FP-CCFV-Qsq}.}
  \label{fig:FP-CCFV-params}
\end{figure*}

The $\FP(Q^2)$ at the physical point was obtained following the same
three step procedure used for extrapolating $G_A(Q^2)$ that is
described in Sec.~\ref{ssec:aff-GA}.  This $\FP(Q^2)$ is compared with
the $G_A(Q^2)$ already shown in Fig.~\ref{fig:GAz2fit} in
Fig.~\ref{fig:FP-CCFV-Qsq}. If PPD were exact, then $\FP$ should equal
$G_A$. The overlap of the two bands turns out to be surprisingly good
over the whole $Q^2$ interval.  Results for the four fit parameters,
$b_i(Q^2)$, versus $Q^2$ obtained in the CCFV extrapolation process
are shown up to $1\,\GeV^2$ in Fig~\ref{fig:FP-CCFV-params} for data
obtained with the $\Ssim$ strategy.

Similar to $G_A(Q^2)$, the two physical mass ensembles impact the
coefficients $b_i(Q^2)$ shown in Fig.~\ref{fig:FP-CCFV-params} only
for $Q^2 \lesssim 0.4\,\GeV^2$.  The plots show some pion mass
dependence for $Q^2 < 0.2\,\GeV^2$, i.e., $b_2(Q^2) \neq 0$. The
coefficients for the lattice spacing dependence, $b_1(Q^2)$, and for
finite volume, $b_3(Q^2)$, have large uncertainty. Also, neglecting
the finite volume term does not change $b_1(Q^2)$ and $b_2(Q^2) $
significantly. Overall, the shape of these coefficients versus $Q^2$
is somewhat different from those for $G_A$ shown in
Fig.~\ref{fig:GA-CCFV-Qsq-coeff-all3}.

The $z^2$ fit to the physical point $\FP$, shown in
Fig.~\ref{fig:FP-CCFV-Qsq}, with $t_c = 9M_\pi^2$ 
and $t_0 = 0.25$~GeV${}^2$ has the parameterization
\begin{align}
\FP(Q^2) &= a_0 + a_1\ z + a_2 \ z^2  \CL
         &= 0.868(30) -1.702(136) z + 0.587(601) z^2   \,,
\label{eq:FPz2fit}
\end{align}
with the correlation matrix:
\begin{equation}
\begin{blockarray}{cccc}
&a_0&a_1&a_2\\
\begin{block}{l(rrr)}
a_{0\ \ }  &  1.0        &  -0.45085  &  -0.05106  \\
a_{1\ \ }  &   -0.45085  &  1.0       &  -0.23890  \\
a_{2\ \ }  &  -0.05106   & -0.23890   &   1.0       \\
\end{block} 
\end{blockarray}
\label{eq:FPz2Cov}
\end{equation}
The agreement, within errors, with the parameterization of $G_A(Q^2)$
given in Eqs.~\eqref{eq:GAz2fit} and~\eqref{eq:GAz2Cov} is very
good. This is not unexpected based on the overlap between the two
shown in Fig.~\ref{fig:FP-CCFV-Qsq}, nevertheless, one should keep in
mind the $Q^2$ dependence shown in Fig.~\ref{fig:R4-Scomp}, and the
Goldberger-Trieman discrepancey~\cite{Bernard:2001rs,Park:2021ypf}.

%--------------------
% GA and GP versus z
%--------------------

\begin{figure}[!h]   %F13
  \centering
  \includegraphics[trim=0 25 0 0, clip, width=0.75\columnwidth]{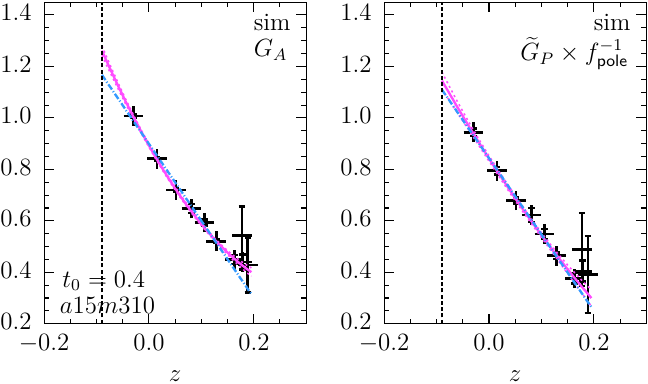}
  \\ %\vspace{0.5em}
  \includegraphics[trim=0 25 0 0, clip, width=0.75\columnwidth]{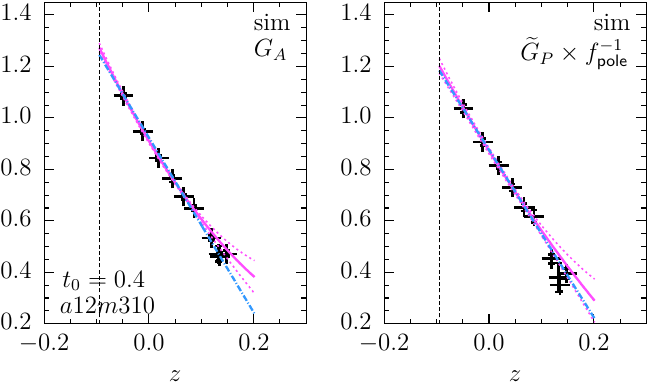}
  \\ %\vspace{0.5em}
  \includegraphics[trim=0 25 0 0, clip, width=0.75\columnwidth]{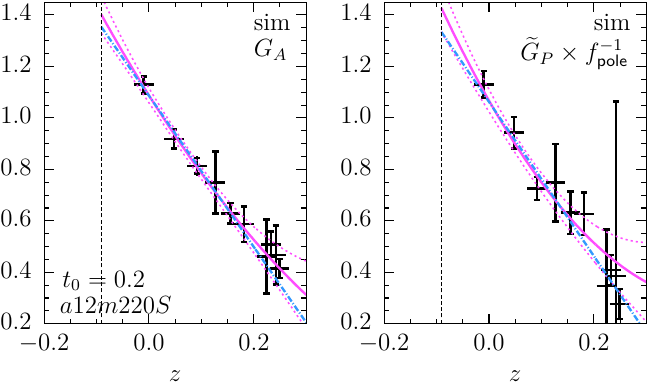}
  \\ %\vspace{0.5em}
  \includegraphics[trim=0 25 0 0, clip, width=0.75\columnwidth]{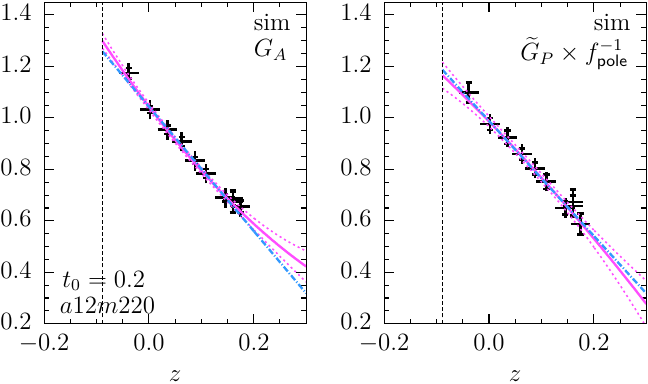}
  \\ %\vspace{0.5em}
  \includegraphics[trim=0 25 0 0, clip, width=0.75\columnwidth]{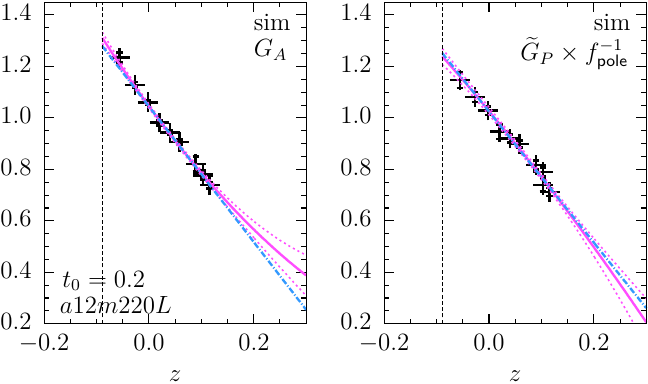}
  \\ %\vspace{0.5em}
  \includegraphics[trim=0 25 0 0, clip, width=0.75\columnwidth]{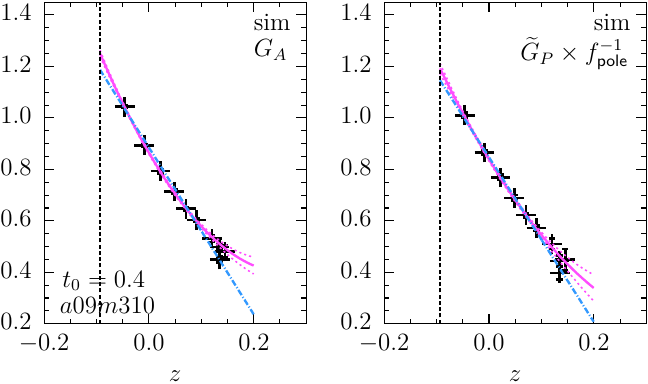}
  \\ %\vspace{0.5em}
  \includegraphics[trim=0 0 0 0, clip, width=0.75\columnwidth]{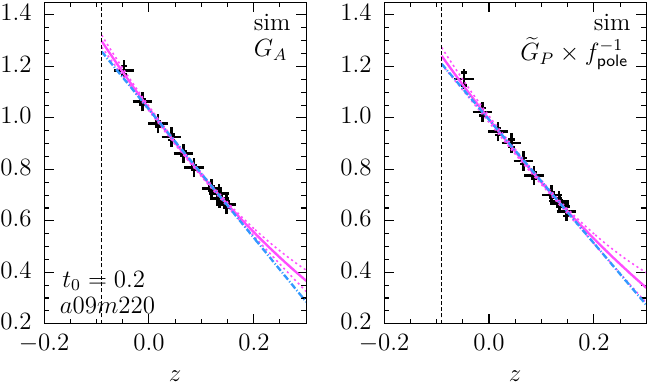}
  \\ %\vspace{0.5em}
  \label{fig:GA-GP-zexp-a15-a12-simA}
\end{figure}

%
%--------------------------
% gPstar, gpiNN, gAGT CCFV
%--------------------------

\clearpage

%------------------------------------------
% GA, GP, GPS CCFV coefficients versus Q^2
%------------------------------------------

% \clearpage

\section{Nucleon Charges from forward matrix elements} %S05
\label{sec:charges}

The spectral decomposition of the forward, $\bm q = 0$, three-point function
truncated at three states, $\ket{i}$ with $i=0,1,2$, can be written as
%
%\begin{widetext}
\begin{align}
  C_\Gamma^\text{3pt}&(t;\tau) =
  \sum_{i,j=0} \abs{A_j}\abs{A_i} \matrixe{j}{\mathcal{O}_\Gamma}{i} e^{-M_i t -M_j(\tau-t)} \CL
  =& \abs{A_0}^2 g_\Gamma e^{-M_0 \tau} \times \Big[ 
     1 + r_1^2 b_{11} e^{-\Delta M_1 \tau} \CL
   & + r_2^2 b_{22} e^{-(\Delta M_1+\Delta M_2) \tau} \CL
   & + 2 r_1 b_{01} e^{-\Delta M_1 \tau/2} \cosh(\Delta M_1 t_s) \CL
   &  + 2 r_2 b_{02} e^{-(\Delta M_1+\Delta M_2) \tau/2} \cosh\left\{(\Delta M_1+\Delta M_2) t_s\right\} \CL
   & + 2 r_1 r_2 b_{12} e^{-(2\Delta M_1+\Delta M_2) \tau/2} \cosh(\Delta M_2 t_s)
     \Big] \CL 
   & + \cdots \,,
  \label{eq:charge-s3}
\end{align}
%\end{widetext}
%
where $t_s \equiv t-\tau/2$, $\matrixe{0}{\mathcal{O}_\Gamma}{0}$ is
the bare charge $g_\Gamma$, the transition matrix elements are $b_{ij}
\equiv
\matrixe{i}{\mathcal{O}_\Gamma}{j}/\matrixe{0}{\mathcal{O}_\Gamma}{0}$,
the ratios of amplitudes are $r_i = \abs{A_i}/\abs{A_0}$, and the
successive mass gaps are $\Delta M_i \equiv M_i - M_{i-1}$. The
prefactors in terms involving the excited states are products such as
$r_2^2 b_{22}$. These products are simply parameters in the fits and
are not used subsequently. Thus the ratios $r_i$ for the excited states 
are, by themselves, not needed.

\begin{figure}[!t]  %F18
  \includegraphics[trim=10 100 10 100, clip, width=0.98\columnwidth]{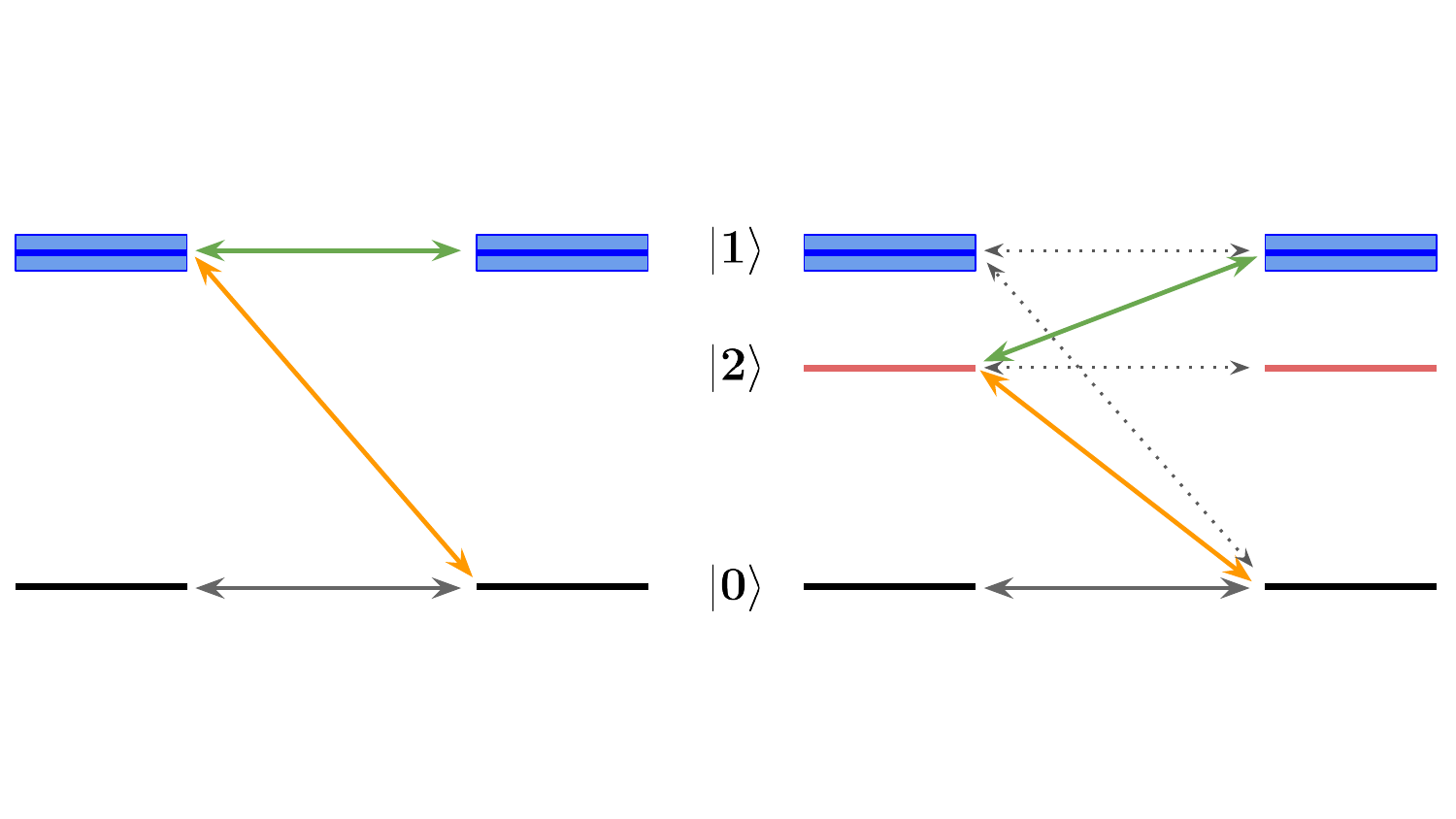}
  \caption{A pictorial representation of the standard 2-state fit
    (left) and the 3-RD fit (right). In the 3-RD fit, the $M_0$ and
    $M_1$ are taken from the nucleon two-point correlator fit but
    $\Delta M_2$ is determined from the fit to the three-point
    correlator. Negative values for $\Delta M_2$ in
    Table~\ref{tab:charge-esc-params} indicate that $\ket{2}$ lies
    below $\ket{1}$. Both fits include only two transitions,
    $\ket{0}\to\ket{1}$ (yellow) and $\ket{1}\to\ket{1}$ (green) in
    the 2-state fit and $\ket{0}\to\ket{2}$ (yellow) and
    $\ket{1}\to\ket{2}$ (green) in the 3-RD fit. The
    transitions turned off with respect to the full 3-state ansatz
    given in Eq.~\eqref{eq:charge-s3} are represented by dashed
    lines. }
  \label{fig:3-RD-fit}
\end{figure}

To remove excited states contributions, we made three kinds
of fits to the 3-point functions using Eq.~\eqref{eq:charge-s3}:
\begin{itemize}
\item
$3^\ast$: This is a 3-state fit with $b_{22}$ set to zero. The four
  parameters $A_0$, $M_0$, $M_1$, $M_2$ are taken from four-state
  fits to the two-point function, leaving only $\matrixe{0}{\mathcal{O}_\Gamma}{0}$, and products 
  such as $r_1^2 b_{11}$ as free
  parameters.  This strategy (along with its two-state version) was
  used to get the results presented in Ref.~\cite{Gupta:2018qil} that are 
  reproduced in Eq.~\eqref{eq:summary-charge-2018PRD}.
\item
3-RD: This is a 3-state fit with $b_{01}$, $b_{11}$ and $b_{22}$ set
to zero, otherwise the fits become unstable. The three parameters
$A_0$, $M_0$, $M_1$ are again taken from four-state fits to the
two-point function. The value of the second mass gap, $\Delta M_2$, is
left as a free parameter in the fit. The sign of $\Delta M_2$ for a given
charge determines whether $\ket{1}$ lies above or below $\ket{2}$ as
shown pictorially in Fig.~\ref{fig:3-RD-fit}.
\item
3-RD-$N\pi$: In this fit, $M_1$ is fixed to the noninteracting energy
of the $ (N(\bm{n})\pi(-\bm{n}))$ state with $\bm{n}=(1,0,0))$. For the
value of $M_2 $, we use a Bayesian prior with a narrow width centered
about the first excited state mass determined from the two-point
correlator as given in Table~\ref{tab:charge-Npi-hard-params} 
in Appendix~\ref{sec:gASTdata}.
\end{itemize}
We also tried two-state fits with $\Delta M_1$ left as a free
parameter. For the axial charge, we found large fluctuations in
$\Delta M_1$ between the jackknife samples leading to unreliable
values. So we do not present these estimates.

Results for unrenormalized isovector nucleon charges, $g_A$, $g_T$,
and $g_S$, using the $3^\ast$, $3$-RD, and the 3-RD-$N\pi$ fits are
given in Table~\ref{tab:charge}, and the other parameters of the
$3$-RD fits are given in Table~\ref{tab:charge-esc-params} 
in Appendix~\ref{sec:gASTdata}. 

The final renormalized charges are presented in the
$\overline{\mathrm{MS}}$ scheme at $2\,\GeV$. We carry out the
renormalization using the RI-sMOM intermediate scheme as described in
Ref.~\cite{Gupta:2018qil}.  To understand systematics, we use three
methods: (i) $g_X = Z_X g_X^\text{(bare)}$, where $X=A,T,S$; and (ii)
$g_X = Z_X/Z_V \times g_X^\text{(bare)}/g_V^\text{(bare)}$ with the
relation $Z_A g_V =1$. Empirically, some of the
systematics cancel in each ratio in the second method, giving slightly smaller
overall errors.  In method three, we take the average of the first two within
the jackknife process, and use it to get the final estimates. The renormalization
factors $Z_X$ and $Z_X/Z_V$ used in this study are given in Table~V in
Ref.~\cite{Gupta:2018qil}.

We use the same leading order CCFV ansatz, given in
Eq.~\eqref{eq:ccfv}, for extrapolating results to the physical point for all three
strategies: $3^\ast$, $3$-RD, and 3-RD-$N\pi$.

Results from the $3^\ast$ (or 2-state for $g_S$) analysis, have already been
published in Ref.~\cite{Gupta:2018qil}, and reproduced here to facilitate 
comparison.\footnote{The statistics in the a06m135 and a12m310
  ensembles have been increased, however, the changes in the estimates
  are insignificant, so we continue to quote the results from Ref.~\cite{Gupta:2018qil}.}
These are: 
\begin{align}
  \label{eq:summary-charge-2018PRD}
  g_A =& 1.218(25)_\text{stat}(30)_\text{sys} \qquad (3^\ast\text{-state,~\cite{Gupta:2018qil}}) \CL
  g_T =& 0.989(32)_\text{stat}(10)_\text{sys} \qquad (3^\ast\text{-state,~\cite{Gupta:2018qil}}) \\
  g_S =& 1.022(80)_\text{stat}(60)_\text{sys} \qquad (2\text{-state,~\cite{Gupta:2018qil}}) \,. \nonumber
\end{align}

\begin{figure*}[!tbh] % F19
  \includegraphics[trim=0 0 0 0, clip, width=0.32\textwidth]{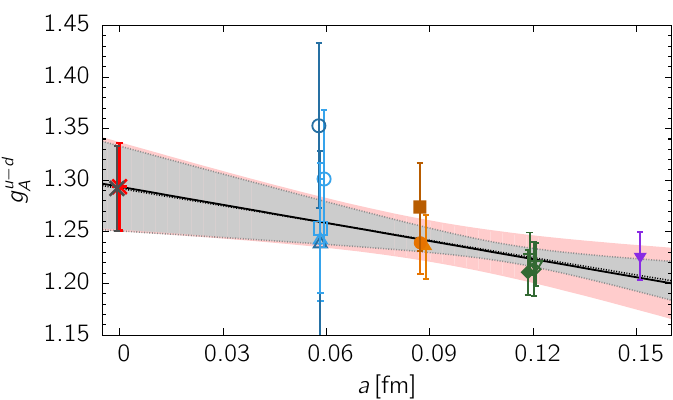}
  \hspace{0.5em}
  \includegraphics[trim=0 0 0 0, clip, width=0.32\textwidth]{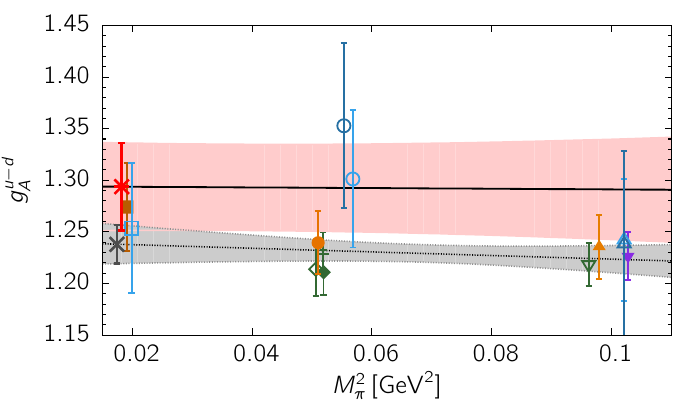}
  \hspace{0.5em}
  \includegraphics[trim=0 0 0 0, clip, width=0.32\textwidth]{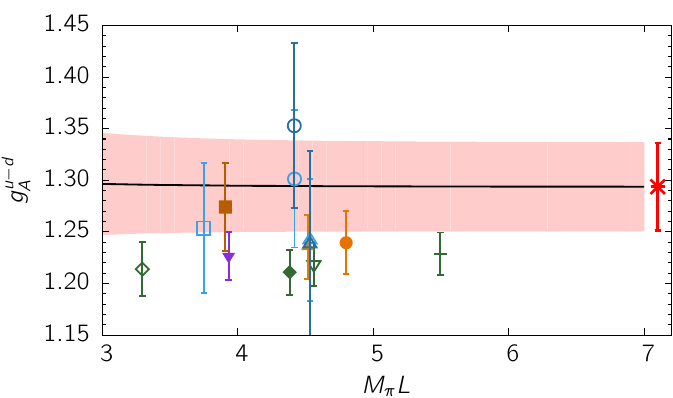}
  \\ \vspace{0.5em}
  \includegraphics[trim=0 0 0 0, clip, width=0.32\textwidth]{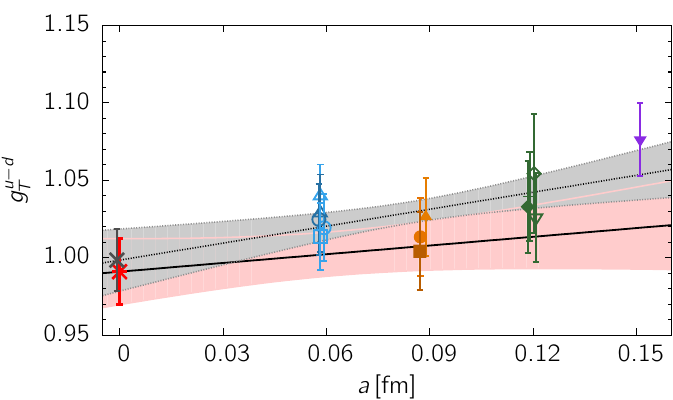}
  \hspace{0.5em}
  \includegraphics[trim=0 0 0 0, clip, width=0.32\textwidth]{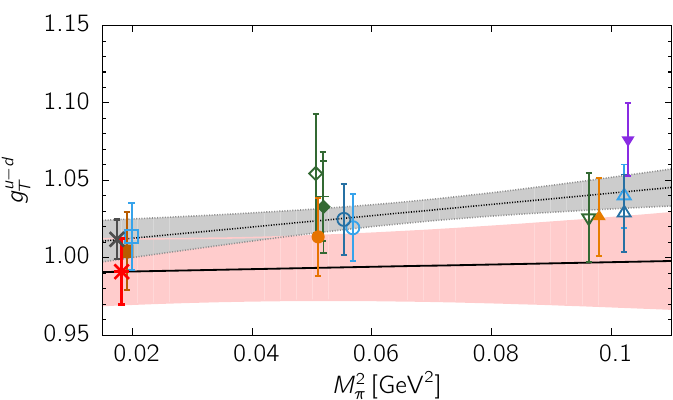}
  \hspace{0.5em}
  \includegraphics[trim=0 0 0 0, clip, width=0.32\textwidth]{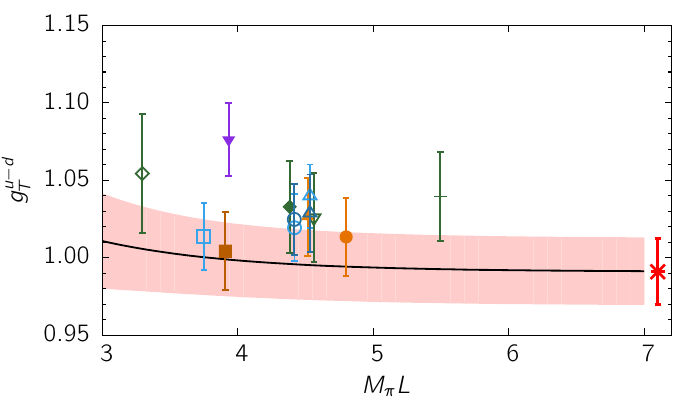}
  \\ \vspace{0.5em}
  \includegraphics[trim=0 0 0 0, clip, width=0.32\textwidth]{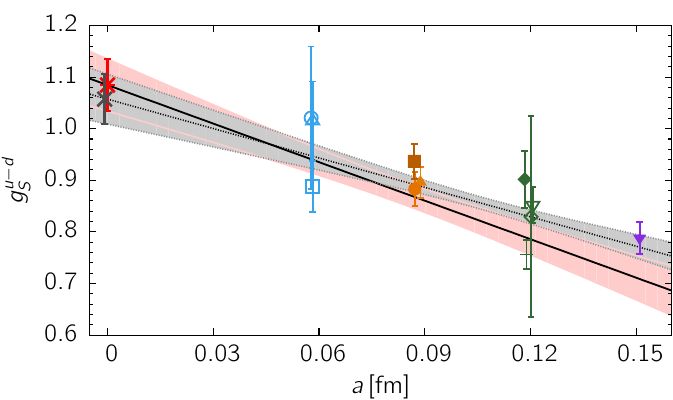}
  \hspace{0.5em}
  \includegraphics[trim=0 0 0 0, clip, width=0.32\textwidth]{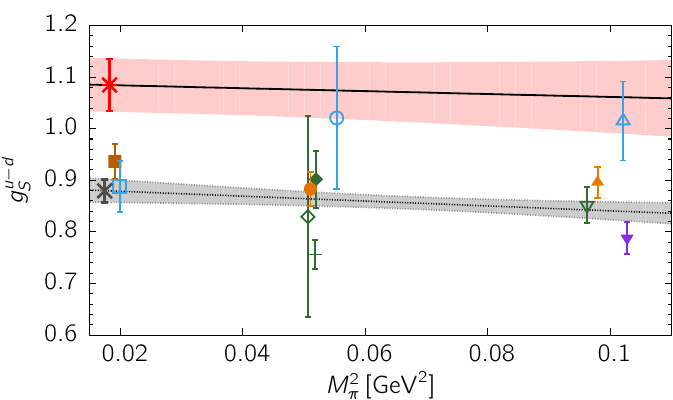}
  \hspace{0.5em}
  \includegraphics[trim=0 0 0 0, clip, width=0.32\textwidth]{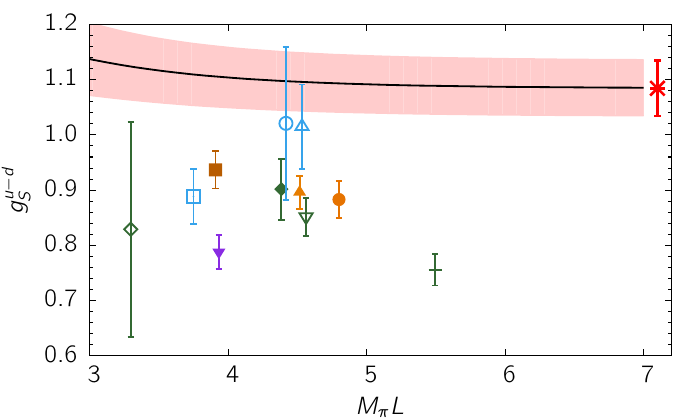}
%  \\ \vspace{0.5em}
  \caption{The simultaneous chiral-continuum-finite-volume (CCFV) fit
    to the axial $g_A$ (top, 13-point), tensor $g_T$ (middle,
    13-point), and scalar $g_S$ (bottom, 11-point) charges.
    The data are extracted using the 3-RD fit described in the text
    and are the average over the two renormalization methods $Z_X
    \times g_X^\text{(bare)}$ and $Z_X/Z_V \times
    g_X^\text{(bare)}/g_V^\text{(bare)}$ where the $g_V$ is the vector
    charge.  In each panel, the pink bend with black solid line
    represents the full CCFV fit.  In the left (middle) panels, the
    gray band shows the fit to the date keeping only the $a$
    ($M_\pi^2$) dependent term in Eq.~\protect\eqref{eq:ccfv}. The
    value at the physical point is marked by the red star. The data in each
    panel have not been shifted to the physical point in the other
    two fit variables.}
  \label{fig:charges-3RD-ccfv-13pt}
\end{figure*}

We now focus on the 3-RD analysis and make three CCFV extrapolation 
with the following cuts on the data
\begin{itemize}
\item
``13-point'' CCFV fit uses all thirteen points. 
\item
``11-point'' CCFV fit: This fit excludes the $a06m310W$ and
  $a06m220W$ points obtained with larger smearing radius for sources
  used to calculate the quark propagators~\cite{Gupta:2018qil}. Larger
  smearing radius reduces the ESC at smaller values of $\tau$ but
  gives larger statistical errors at the values of $\tau$ used in our
  excited-state fits to get the $\tau \to \infty$ values as discussed
  in Ref.~\cite{Yoon:2016dij}. Also, we expect significant
  correlations between the two pairs, ($a06m310$, $a06m310W$) and
  ($a06m220$, $a06m220W$), since they use the same gauge
  configurations.  Comparing the two pairs, the results for the three
  charges agree except for $g_S$ between $a06m220$ and $a06m220W$,
  which can be accounted for by the larger statistical errors,
  especially in the $a06m220W$ data. Consequently, the
  ``11-point'' CCFV fit is used to get the central value for
  $g_S$, which is shown in Fig.~\ref{fig:charges-3RD-ccfv-13pt}.
\item
``10-point'' CCFV fit excludes $a15m310$, $a06m310W$ and $a06m220W$
  points.  Since the variation with the lattice spacing is the
  dominant systematic, removing the $a15m310$ point (coarsest lattice
  with $a \sim 0.15$~fm) aims to provide a handle on higher order,
  $\mathcal{O}(a^2)$, corrections neglected in Eq.~\eqref{eq:ccfv}.
\end{itemize}
Results from these three CCFV extrapolations, different truncations of the CCFV
ansatz, and the three renormalization methods are given in
Tables~\ref{tab:ccfv-gA-3RD},~\ref{tab:ccfv-gT-3RD},
and~\ref{tab:ccfv-gS-3RD} in Appendix~\ref{sec:gASTdata} and used to
assess the various systematics.

The central values are taken from the ``13-point fit'' for $g_A$ and
$g_T$ and the ``11-pt fit'' for $g_S$ with the 3-RD data
renormalized using the third (average of the first two) method.  Note
that we find a systematic shift of $\approx 0.03$, 0.02 and $0.03$
between the first two renormalization methods for the three charges,
$g_A$, $g_T$ and $g_S$, respectively.

These CCFV fits are shown in Fig.~\ref{fig:charges-3RD-ccfv-13pt}.
Each panel in a given row shows the fit result versus one of the three
variables with the other two set to their physical point values.  In
the left two panels, we show two fits: (i) using the full ansatz given
in Eq.~\eqref{eq:ccfv} (pink band), and (ii) assuming there is
dependence only on the x-axis variable (gray band). For example, in
the left panels the gray band corresponds to a fit with $b_2^{g_X} =
b_3^{g_X} = 0$. The data show that the discretization errors are the
dominant systematic, i.e., there is an almost complete overlap of the
two fits (pink and gray bands) for $g_A$ and a significant overlap for
$g_S$ and $g_T$. The variation with $a$ over the range $0 < a \leq
0.15$~fm is about 10\%, 5\% and 30\% for $g_A$, $g_T$, and $g_S$,
respectively. The large variation with $a$ in $g_S$ is similar to that 
found in the clover-on-clover calculation~\cite{Park:2021ypf}.

The final results of the 3-RD analysis are: 
\begin{align}
  g_A =& 1.294(42)_\text{stat}(18)_\text{CCFV}  (16)_\text{Z} \qquad (3\text{-RD}) \CL
  g_T =& 0.991(21)_\text{stat}(04)_\text{CCFV}  (09)_\text{Z} \qquad (3\text{-RD}) \CL
  g_S =& 1.085(50)_\text{stat}(102)_\text{CCFV} (13)_\text{Z} \qquad (3\text{-RD}) \,. 
  \label{eq:summary-charge-3RD}
\end{align}
The first error quoted
(labeled stat) is the total uncertainty from the central analysis.  The
second error is an estimate of the uncertainty in the CCFV
extrapolation.  For $g_A$ and $g_T$, this is taken to be the average
of the differences |11-pt -- 13-pt| and
|10-pt -- 13-pt|. For $g_S$, it is the difference |10-pt
-- 11-pt|. The third error is half the difference in estimates
between the first two renormalization methods.

The $g_A$ from the 3-RD fit is in good agreement with the result
obtained from the extrapolation of the axial form factor $G_A(Q^2)$ to
$Q^2 = 0$ given in Eq.~\eqref{eq:gA-z}. It is also consistent
with the experimental value $g_A=1.2766(20)$ but has much larger errors.
The difference between the $3^\ast$ (PNDME~\cite{Gupta:2018qil}) value reproduced in
Eq.~\eqref{eq:summary-charge-2018PRD}, and the 3-RD  is due to different 
 excited state energies used in the fits to the spectral
decomposition in Eq.~\ref{eq:charge-s3}. The data in Table~\ref{tab:charge-esc-params} show that
the fit parameter $M_2$ when left free satisfies $M_\pi \lesssim M_2-M_0
\lesssim 3M_\pi$ for all but the $a\approx 0.12$~fm lattices.  In
~\cite{Jang:2019vkm}, we showed evidence that the
$N(\bm{p}_1)+\pi(-\bm{p}_1)$ with $\bm{p}_1 = (1,0,0)2\, \pi/La$ state
makes a significant contribution on the zero momentum side of the
operator insertion in the calculation of the form factors, and the
noninteracting energy of this state is $M_{N\pi} - M_0 \approx
2M_\pi$. In short, the $M_2$ output by the 3-RD fit has a mass lower
than $M_1$ obtained from the two-point correlator and broadly
consistent with the hypothesis that the $N\pi$ states contribute. We again 
caution the reader that these excited state masses should
only be regarded as effective fit parameters that encapsulate the
effect of the full tower of $N(\bm{p})+\pi(-\bm{p})$ states with
momenta $\bm{p}=(2\pi/L)\bm{n}$ as well as other multihadron and
radial excitations that can contribute. 

For $g_S$ and $g_T$, the 3-RD fit reduces to a 2-state fit if $\Delta
M_2 (= M_2 - M_1) = 0$, i.e., $M_2 \simeq M_1$. This is the case for
many of the ensembles as shown in Table~\ref{tab:charge-esc-params}.
Results given in Eq.~\eqref{eq:summary-charge-3RD} are consistent with
those in Eq.~\eqref{eq:summary-charge-2018PRD} indicating that
sensitivity to excited state energies input into the analysis is small.

Based on the 3-RD fits, which indicate that the data for $g_A$ prefer
a low-mass excited state with $\Delta M \approx 2M_\pi$, the
3-RD-$N\pi$ fit defined above, with the mass gaps summarized in
Table~\ref{tab:charge-Npi-hard-params}, were performed. Charges from
this fit are compared with the 3-RD and the 3$^\ast$-state fits (or the 2-state
fit for the $g_S$) in Table~\ref{tab:charge} for the thirteen
ensembles.  The $p$-value for many of the 3-RD-$N\pi$ fits are low. To
stabilize the 3-RD-$N\pi$ fits, we increased the width of the priors
for $M_2$, however, this still did not lead to stable fits for several
ensembles. The 3-RD-$N\pi$  results are, therefore, not included in the final 
analysis. 

Results for $g_{A,S,T}$ in
Eq.~\eqref{eq:summary-charge-3RD} are compared with those from 
other collaborations in Table~\ref{tab:final_Comp}. This comparison 
complements the latest FLAG review that considered results published prior to
2021~\cite{FlavourLatticeAveragingGroupFLAG:2021npn,FlavourLatticeAveragingGroup:2019iem}.
Overall, results for $g_A$ and $g_T$ from all calculations are consistent within five
percent and for $g_S$ at ten percent. Their precision will continue
to improve steadily as higher statistics data are generated at additional $\{a,M_\pi\}$ points 
with $M_\pi L \gtrsim 4$. 

Our conclusion is that, with current statistics, fits for the axial
charge are more stable with input of $M_0$ and $M_1$ from the 4-state
fit to the 2-point function and letting the 3-point function determine
$M_2$ (corresponding roughly to the $N \pi$ state), i.e., the 3-RD
fit.  In future works with higher statistics, we expect results from
3-RD and 3-RD-$N\pi$ fits to come together as the same three states
provide the dominant contributions. Only the
details of their inclusion are different.

\begin{table*}[!b]   %T06
  % full CCFV fits
  \centering
  \renewcommand{\arraystretch}{1.1}
  \begin{ruledtabular}
    \begin{tabular}{ccccccc}
Collaboration  &      $g_A$ &  $g_S$  & $g_T$  & $\expv{r_A^2}$~fm${}^2$ & $g_P^\ast$   & $g_{\pi NN}$ \\ \hline
PNDME 23  & 1.292(53)(24) & 1.085(50)(103) & 0.991(21)(10)  &  0.439(56)(34)  & 9.03(47)(42)  &  14.14(81)(85) \\
RQCD 19/23& $1.284^{28}_{27}$ & $1.11^{14}_{16}$ & $0.984^{19}_{29}$  & 0.449(88) &  8.68(45) & 12.93(80) \\
ETMC 20/22& 1.283(22)     &  1.35(17) & 0.924(54)  &  0.343(42)(16)  &     &  \\
NME  21   & 1.32(6)(5)    & 1.06(9)(7)  & 0.97(3)(2)  &   0.428(53)(30) &  7.9(7)(9)  & 12.4.(1.2) \\
Mainz 22  & 1.225(39)(25) & $1.13(11)({}^7_6)$    & $0.965(38)({}^{13}_{41})$  &  0.370(63)(16)  &   &  \\
PACS 22   & 1.288(14)(9)  &  0.927(83)(22) & 1.036(6)(20)  &   &    &  \\
    \end{tabular}
  \end{ruledtabular}
  \caption{Comparison of $g_{A,S,T}$, $\expv{r_A^2}$, $g_P^\ast$ and
    $g_{\pi NN}$ from recent calculations labeled as: PNDME 23 is this
    work, RQCD~\protect\cite{RQCD:2019jai} (here we list values
    obtained with the $!z^{4+3}$ fit, and take $g_{A,S,T}$ from their
    recent work~\protect\cite{Bali:2023sdi}),
    ETMC~\protect\cite{Alexandrou:2020okk,Alexandrou:2022dtc},
    NME~\protect\cite{Park:2021ypf},
    Mainz~\protect\cite{Djukanovic:2022wru}, and
    PACS~\protect\cite{Tsuji:2022ric}. All results for $g_{A,S,T}$ are
    in the $\overline{\rm MS}$ scheme at scale 2~GeV.  For
    completeness, we also give results for $g_{S,T}$ from the Mainz
    collaboration~\protect\cite{Harris:2019bih} and from the ETMC
    collaboration~\protect\cite{Alexandrou:2019brg,Alexandrou:2022dtc}. These and earlier
    results are reviewed by the
    FLAG~\cite{FlavourLatticeAveragingGroupFLAG:2021npn,FlavourLatticeAveragingGroup:2019iem}. }
  \label{tab:final_Comp}
\end{table*}

%\clearpage
\section{Conclusions and Comparison with Previous Calculations}   %S06
\label{sec:con}

We have presented results for the axial, $G_A(Q^2)$, the induced
pseudoscalar, $\widetilde{G}_P(Q^2)$, and the pseudoscalar,
${G}_P(Q^2)$, form factors of nucleons using thirteen ensembles of
2+1+1-flavors of HISQ ensembles generated by the MILC
collaboration~\cite{Bazavov2012:PhysRevD.87.054505}. A large part of
the focus of this work is on understanding the nature of the excited
states that contribute significantly to the relevant correlation
functions and removing their contributions.  The analysis presented
here strengthens the case for including multihadron excited states,
such as $N \pi$, made in Ref.~\cite{Jang:2019vkm}.  Our data driven
analysis strategy, labeled $\Ssim$, identifies the contributions from
$N \pi$ state in the extraction of the GSME. The three form factors
obtained including $N \pi$ state satisfy the PCAC relation to within
10\% as opposed to a $\sim 50\%$ deviation without them.  For the
final results, we therefore choose the data obtained with the $\Ssim$
strategy for removing ESC, parameterize the $Q^2$ behavior using the
model-independent $z^2$ fit; and extrapolate the data to the physical
point using a simultaneous chiral-continuum-finite-volume (CCFV) ansatz
including the leading order corrections given in
Eq.~\eqref{eq:ccfv}. For errors, we quote two estimates: the first
labeled ``stat'' is the total error obtained from the analysis used to
produce the central value, and the second, labeled ``sys'', includes
the various systematic uncertainties combined in quadrature as discussed
in the appropriate sections.

Our final results are:
\begin{itemize}
\item
The axial charge is $g_A = 1.292 (53)_\text{stat}\,
(24)_\text{sys}$. This is the unweighted average of the value from the
extrapolation of $G_A(Q^2)$ to $Q^2=0$ (Eqs.~\eqref{eq:gArAfinal}) and from
the forward matrix element (Eq.~\eqref{eq:summary-charge-3RD}).  The
``stat'' and ``sys'' errors quoted are the larger of those from the
two determinations.  This result is consistent with the experimental value
but has much larger errors.
\item
The scalar charge $g_S = 1.085 (50)_\text{stat}\, (103)_\text{sys}$ and 
the tensor $g_T = 0.991 (21)_\text{stat}\, (10)_\text{sys}$ are given in 
Eq.~\eqref{eq:summary-charge-3RD}. 
\item
The extraction of the axial charge radius squared is discussed in
Sec.~\ref{ssec:aff-GA}, and the result from Eq.~\eqref{eq:gArAfinal} is
$\expv{r_A^2} = 0.439 (56)_\text{stat} (34)_\text{sys}$~fm${}^2$.
\item
The extraction of the induced pseudoscalar charge is discussed in
Sec.~\ref{sec:gPstar} with the result 
$g_P^\ast = 9.03(47)_\text{stat}(42)_\text{sys} $ given in Eq.~\eqref{eq:gPstar-z}. 
\item
The pion-nucleon coupling is discussed in Sec.~\ref{sec:gpiNN} with the result 
$g_{\pi NN} = 14.14(81)_\text{stat}(85)_\text{sys}$ given in Eq.~\eqref{eq:gpiNN-z}.
\item
Our procedure for obtaining the axial form factor, $G_A(Q^2)$, in the
continuum limit is discussed in Sec.~\ref{ssec:aff-GA}.  The final
parameterization is given in Eq.~\eqref{eq:GAz2fit}, the covariance
matrix of the fit in Eq.~\eqref{eq:GAz2Cov}, and the corresponding
values of $g_A = 1.281(53)$ and $\expv{r_A^2} = 0.498 (56)$~fm${}^2$
in Eq.~\eqref{eq:gAz2final}. The overall final values from the
analysis of $G_A$ are given in Eq.~\eqref{eq:gArAfinal}.
\end{itemize}

\begin{figure}[!th]  %F20
  \includegraphics[trim=14 0 20 0, clip, width=1.0\columnwidth]{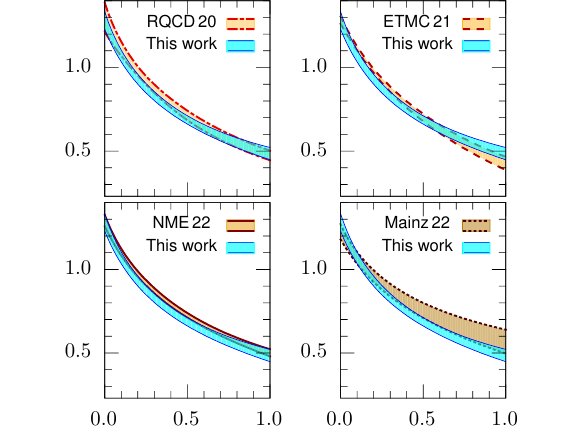}
  \includegraphics[width=1.0\columnwidth]{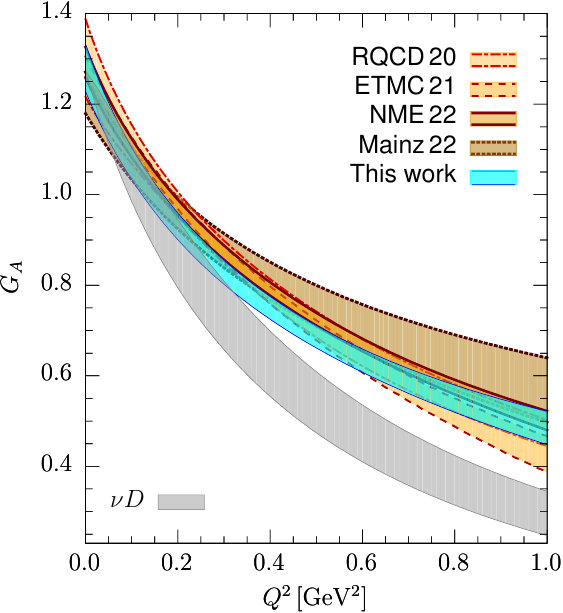}
  \caption{A comparison of the isovector axial form factor $G_A(Q^2)$ at the physical point 
    obtained using a $z$-expansion fit by the RQCD~\cite{RQCD:2019jai}
    (light faun band), ETMC~\cite{Alexandrou:2020okk} (light tan  band),
    NME~\cite{Park:2021ypf} (tan band), 
    Mainz~\cite{Djukanovic:2022wru} (brown band) collaborations and
    this work (turquoise band).  The $G_A$ extracted using the $\nu$-deuterium
    bubble chamber scattering experiments data~\cite{Meyer:2016oeg} is
    shown by the gray band and labeled $\nu D$ in the lower panel. }
  \label{fig:GA_comparison}
\end{figure}

A comparison of lattice results from various collaborations for all
the above quantities was presented recently in
Ref.~\cite{Park:2021ypf}. The charges $g_{A,S,T}$ have also been
reviewed by
FLAG~\cite{FlavourLatticeAveragingGroupFLAG:2021npn,FlavourLatticeAveragingGroup:2019iem}.
Since then, new results have been presented in
Refs.~\cite{Alexandrou:2021wzv,Djukanovic:2022wru,Tsuji:2022ric,Alexandrou:2022dtc}.  The
full list of relevant publications that have included $N \pi$ states
in the analysis of ESC and checked whether form-factors satisfy the
PCAC relation
are~\cite{Jang:2019vkm,RQCD:2019jai,Park:2021ypf,Alexandrou:2020okk,Djukanovic:2022wru,Tsuji:2022ric}.
We first summarize the results and the important features in each of
these calculations, and then show a comparison of $G_A(Q^2)$ obtained
by the various collaborations in Fig.~\ref{fig:GA_comparison}.
Results for the charges are compared in Table~\ref{tab:final_Comp}. 

The observation that the form factors extracted using the spectrum
from the nucleon 2-point function fail to satisfy the PCAC relation
Eq.~\eqref{eq:PCAC} was made in Ref.~\cite{Rajan:2017lxk}. The
possible cause, enhanced contributions of multihadron ($N\pi$)
excited states in the axial channel was proposed by
B\"ar~\cite{Bar:2018xyi} using a $\chi PT$ analysis. This was
confirmed using the data for the three-point function with the
insertion of the $A_4$ current in Ref.~\cite{Jang:2019vkm}. This
data-driven analysis, including only the lowest $N \pi$ excited state,
found that the ESC  to the
$\GP(Q^2)$ and $G_P(Q^2)$ form factors were about 35\%, while that in
$G_A(Q^2)$ could be $O(5\%)$ as the latter is affected only at
one-loop in $\chi PT$~\cite{Bar:2018akl,Bar:2018xyi}.  The  smallest $Q^2$ data 
in Tables~\ref{tab:ff-a09m130W} and~\ref{tab:ff-a06m135} in Appendix~\ref{sec:ff-tabs} for the two physical mass ensembles 
$a09m130W$ and $a06m135$ show $\sim 5\%$, $\sim 45\%$ and $\sim 45\%$ difference between 
the $\Stwo$ and $\Ssim$ values. A $\sim 5\%$ level of effect in $G_A(Q^2)$ is also 
consistent with what is observed in the axial charge $g_A$ extracted
from the forward matrix element as shown in Table~\ref{tab:charge} in Appendix~\ref{sec:gASTdata}.  

A brief comparison of our results with other lattice calculations published 
in~\cite{RQCD:2019jai,Park:2021ypf,Alexandrou:2020okk,Djukanovic:2022wru,Tsuji:2022ric,Alexandrou:2022dtc} is as follows.

The RQCD collaboration~\cite{RQCD:2019jai} has extracted $G_A(Q^2)$
from a two-state fit to thirty-six 2+1-flavor Wilson-clover ensembles
generated by the coordinated lattice simulations (CLS)
collaboration. The $\GP(Q^2)$ and $G_P(Q^2)$ are, on the other hand,
extracted using a 3-state fit in which the first excited state
energies are fixed to be the noninteracting energies of the lowest $N
\pi$ state and the second excited state energies are taken to be the
first excited state (values higher than $N(1440)$) given by fits to
the 2-point nucleon correlators.  While their form factors satisfy the
PCAC relation, they are equally well fit by the dipole ansatz and
$z^{4+3}$ (i.e., $z^3$ with sum rule constraints). The axial charge
$g_A = 1.302(86)$ from $z$-expansion (fit labeled $!z^{4+3}$) is
larger than $g_A = 1.229(30)$ from dipole (labeled $!2P$) with the
latter agreeing with that from the forward matrix element. The
corresponding difference in $\expv{r_A^2} $ is $0.449(88)$ versus
$0.272(33)$~fm${}^2$. Results for $g_P^\ast$ (8.68(45) versus
8.30(24)) and for $g_{\pi NN}$ (12.93(80) versus 14.78(1.81)) are
consistent.  They have recently~\cite{Bali:2023sdi} updated their
results for $g_{A,S,T}$ based on the analysis of 47 ensembles.  We
quote these new values in Table~\ref{tab:final_Comp}.

The preferred estimates from the ETM
collaboration~\cite{Alexandrou:2020okk} are from a single 2+1+1-flavor
physical mass $64^3 \times 128$ ensemble at $a \approx 0.8$~fm.  For
the analysis of $G_A(Q^2)$, they take excited-state energies from the
2-point function and find $\expv{r_A^2} = 0.343 (42) (16)$~fm${}^2$.
Their result for $g_A=1.283(22)$ is obtained from the forward matrix
element extracted without including possible contamination from $N\pi$
states.  When results from the direct calculations of $\GP(Q^2)$ and
$G_P(Q^2)$ are used, the three form factors show large deviations from
the PCAC relation which they attribute partially to large
discretization errors in their  twisted mass
formulation~\cite{Alexandrou:2022yrr}. Consequently, they quote final
estimates of $\GP(Q^2)$ derived from $G_A(Q^2)$ using the pion-pole
dominance hypothesis, i.e., the quoted $\GP$ is not independently determined.

The NME collaboration~\cite{Park:2021ypf} analyzed seven ensembles
generated with 2+1-flavors of Wilson-clover fermions. They make a
simultaneous fit to all five correlation functions with insertion of
the axial, $A_\mu$, and pseudoscalar, $P$, currents, i.e., same as the
$\Ssim$ strategy used in this work. The $A_4$ correlator provides the
dominant contribution to fixing the excited-state energies which turn
out to be close to the lowest $N\pi$ states as also discussed in this
paper and in Ref.~\cite{Jang:2019vkm}. The resulting form factors
satisfy the PCAC relation to within ten percent. Observing only a small 
dependence of $G_A(Q^2)$ on $a$ and $M_\pi$, they provide a
continuum parameterization of $G_A(Q^2)$ neglecting these effects, and 
thus underestimate the uncertainty. This $G_A(Q^2)$ is reproduced in
Fig.~\ref{fig:GA_comparison}. The value of the axial charge without
including $N \pi $ state is $g_A=1.242(46)(42)$ and including it 
gives $1.32(6)(5)$. Their other results are $\expv{r_A^2} = 0.428 (53)
(30)$~fm${}^2$, $g_P^\ast=7.9(7)(9)$ and $g_{\pi NN} = 12.4.(1.2)$.

The Mainz Collaboration~\cite{Djukanovic:2022wru} analyze fourteen
2+1-flavor Wilson-clover ensembles also generated by the coordinated
lattice simulations (CLS) collaboration. They obtain a
parameterization of $G_A(Q^2)$ in the continuum from a single combined
fit---summation method for dealing with ESC and the $z^2$ fit for the
$Q^2$ behavior. This result is shown in Fig.~\ref{fig:GA_comparison}
and from it they get $g_A = 1.225(39)(25)$ and $\expv{r_A^2} = 0.370
(63) (16)$~fm${}^2$.

The PACS collaboration~\cite{Tsuji:2022ric} has analyzed one physical pion mass ensemble
with a large volume ($128^4$, i.e., $(10.9{\rm fm})^4)$ at $a =
0.085$~fm and get $g_A = 1.288(14)(9)$. Remarkably, they find that
sources with exponential smearing for the generation of quark
propagators, in contrast to Gaussian smearing used by all other
calculations, leads to essentially no excited-state effects.  The
limitation of this calculation is only 20 configurations, each
separated by 10 molecular dynamics trajectories, were analyzed. Most
likely, this total of 200 trajectories represents less than one unit
of autocorrelation time.  Consequently, the errors are likely
underestimated.  Since no parameterization of $G_A(Q^2)$ was  presented, 
we do not include their results in the comparison.

Phenomenologically, the most important quantity needed for the
analysis of neutrino oscillation experiments is $G_A(Q^2)$, and we
show a comparison of results from various lattice collaborations in
Fig.~\ref{fig:GA_comparison} along with the extraction from the
$\nu$-deuterium bubble chamber scattering
experiments~\cite{Meyer:2016oeg}. In all cases, except ETM, the data
are extrapolated to the physical point and then fit using a truncated
$z$-expansion. The bands in Fig.~\ref{fig:GA_comparison} overlap
indicating that the lattice results are consistent within one sigma
and the envelope of the bands suggests a roughly 10\% uncertainty
throughout the range $0 < Q^2 < 1.0$ GeV${}^2$.  The other significant
observation is that the lattice results fall slower than the
phenomenological extraction (the $\nu N$ band) for $Q^2 \gsim 0.3$
GeV${}^2$.

When comparing these lattice data, it is important to
note that the various collaborations handle various
systematics differently.  These systematic effects will become clearer
and the analysis more robust as the precision of the data
increases. Similarly, recent calculations including $N \pi$
states in a variational basis of interpolating
operators~\cite{Barca:2022uhi} is a step forward in 
providing further insight into the excited states contributions 
and developing better methods for removing them.

What has become clear is that $N \pi$ states need to be included in
the analysis for the three form factors, and satisfying the PCAC
relation~\eqref{eq:PCAC} is an essential and necessary condition.  The
questions that remain for higher precision are how many multihadron
states need to be kept in the spectral decomposition for a given
precision and the size of their contributions.  The roughly $10\%$
spread in the lattice results compared in Fig.~\ref{fig:GA_comparison}
will be reduced with much higher statistics data that will be
available over the next few years, and better analyses of excited
states contributions.

\section{Acknowledgement}

We thank the MILC collaboration for providing the 2+1+1-flavor HISQ
lattices used in this study. The calculations used the Chroma software
suite~\cite{Edwards:2004sx}.  This research used resources at (i) the
National Energy Research Scientific Computing Center, a DOE Office of
Science User Facility supported by the Office of Science of the
U.S.\ Department of Energy under Contract No.\ DE-AC02-05CH11231; (ii)
the Oak Ridge Leadership Computing Facility, which is a DOE Office of
Science User Facility supported under Contract DE-AC05-00OR22725, through 
awards under the ALCC program project LGT107 and INCITE award
HEP133; (iii) the USQCD collaboration, which is funded by the Office
of Science of the U.S.\ Department of Energy; and (iv) Institutional
Computing at Los Alamos National Laboratory. T.~Bhattacharya and
R.~Gupta were partly supported by the U.S.\ Department of Energy,
Office of Science, Office of High Energy Physics under Contract
No.\ DE-AC52-06NA25396. T.~Bhattacharya, R.~Gupta, and B.~Yoon were
partly supported by the LANL LDRD program. 

%-----------
% reference
%-----------
%\clearpage
\bibliographystyle{apsrev} %%% physical review
\bibliography{refs} %%% refs.bib file
\clearpage

%************************************************/
\appendix

\onecolumngrid
\section{Determining the nucleon spectrum from $ C^\text{2pt}(t)$}    %A0A
\label{sec:meff-gen}
\vspace{0.01cm}
\twocolumngrid

\begin{figure}[!tbh]  %F21
  \centering
%  \begin{subfigure}[t]
  \centering
  \includegraphics[width=0.48\textwidth]{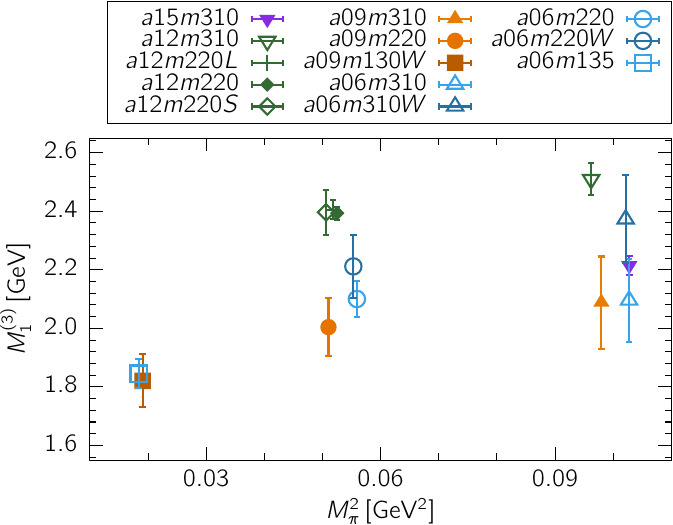} 
%  \\ %\vspace{0.5em}
%  \end{subfigure}
%  \begin{subfigure}[t]{width=0.35\textwidth}
  \centering
  \includegraphics[width=0.48\textwidth]{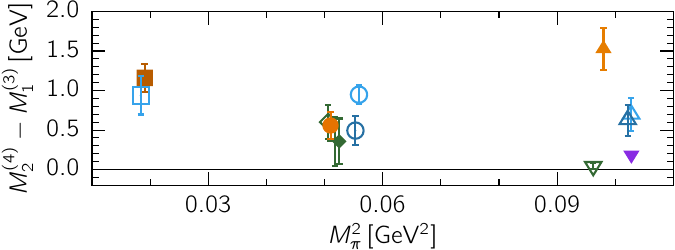}
  \\ %\vspace{0.5em}
  \includegraphics[width=0.48\textwidth]{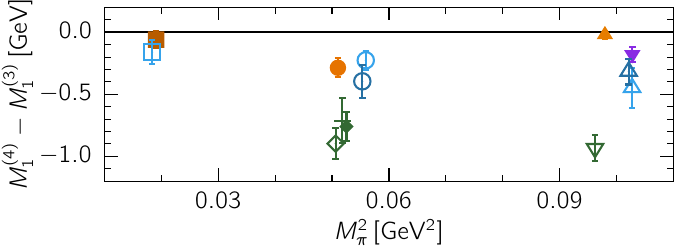} 
  \\ %\vspace{0.5em}
  \includegraphics[width=0.48\textwidth]{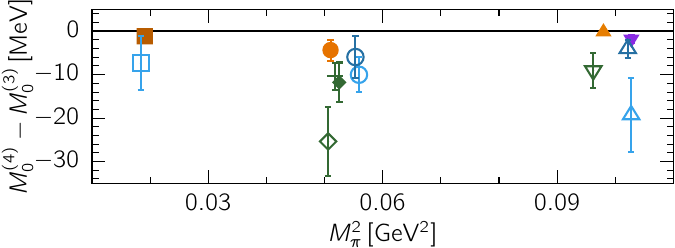} 
%  \end{subfigure}
  \caption{(Top) The first excited state mass, $M_1^{(3)}$, from the
    frequentist 3-state (or 2-state) fit. The mass differences
    $M_2^{(4)}-M_1^{(3)}$ and $M_1^{(4)}-M_1^{(3)}$ are shown in the
    second and third panels. The difference in the ground
    state mass, $M_0^{(4)} - M_0^{(3)}$, is given in the bottom
    panel. }
  \label{fig:comp-M1}
\end{figure}

\begin{figure}[!tbh]  %F22
  \centering
  \includegraphics[width=0.98\columnwidth]{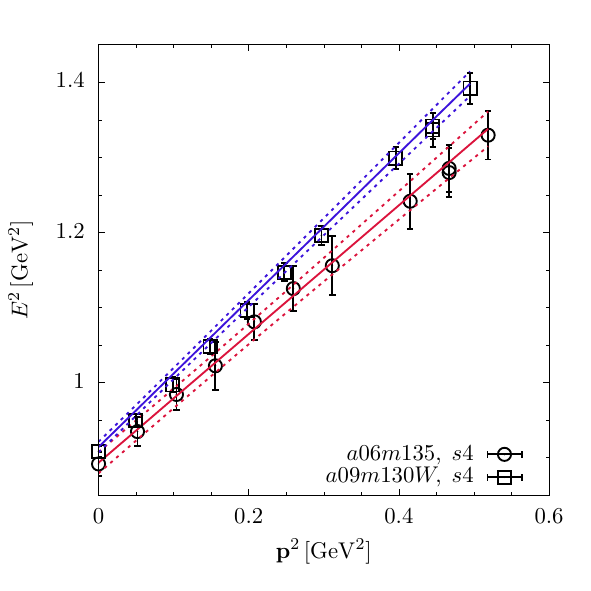} 
  \caption{The plot of $E^2$ obtained from the Bayesian four-state fit versus
    ${\bm{p}}^2$ for the $a09m130W$ (squares) and $a06m135$ (circles)
    ensembles. The slope gives the speed of light, $c^2$ listed in Table~\protect\ref{tab:disp}. It shows 
    significant deviation from unity for the $a06m135$ ensemble. Note that 
    the blue line lying above most square points is a consequence of including the
    full covariance matrix.}
  \label{fig:dispersion}
\end{figure}

To extract the nucleon spectrum, we make two kinds of fits to the 
spectral decomposition of $ C^\text{2pt}(t)$. 
The first is a frequentist (labeled
$F$) multiexponential fit, i.e., without any priors. It is a 
three-state fit for $a\approx 0.06, 0.09\,\fm$ ensembles, and 
two-state for $a\approx 0.12, 0.15\,\fm$ ensembles.  These
frequentist results ($n_s=2$ or $3$) are compared against empirical
Bayesian four-state fits ($n_s=4$) in
Table~\ref{tab:disp}, and their difference is shown in 
Fig.~\ref{fig:comp-M1} (bottom panel). We observe that
\begin{itemize}
\item
The ground state masses from the $F$- and $B$-fits given in 
Table~\ref{tab:disp} are consistent within one combined $\sigma$.
There is, however, a small but systematic shift with $M_0^{(4)} <
M_0^{(3)}$, indicating near convergence.  The
deviations are  $\approx 10\,\MeV$ on all except $a12m220S$ and $a06m310$
ensembles, where they are $20-30\,\MeV$.  Overall, the $B$-fit values
are smaller.\looseness-1
\item
For all except the $a15m310$ and $a06m135$ ensembles, the $E^2$
obtained from the four-state Bayesian fit satisfy the relativistic
dispersion relation, i.e., the speed of light, $c^2$, is consistent
with unity to within $1\sigma$. The fits for the $a09m130W$ and
$a06m135$ ensembles are shown in Fig.~\ref{fig:dispersion}.
\end{itemize}

\begin{table}[t!] %T07
  \centering
  \renewcommand{\arraystretch}{1.1}
  \begin{ruledtabular}
    \begin{tabular}{lcccccc}
      ID & $n_s$ & $aM_0^\text{2pt}$ & $aM_0^\text{Disp}$ & $c^2$ & $\widehat{\chi}^2/dof$ & $p$ \\ \hline
      $a15m310$ & 4 & 0.8302(21) & 0.8304(21) & 0.930(12) & 1.37 & 0.195 \\
      $a15m310$ & 2 & 0.8315(20) & 0.8319(19) & 0.936(11) & 0.96 & 0.474 \\ \hline
      $a12m310$ & 4 & 0.6660(27) & 0.6662(26) & 1.001(14) & 0.62 & 0.777 \\
      $a12m310$ & 2 & 0.6715(13) & 0.6716(13) & 1.001(09) & 0.73 & 0.685 \\ \hline
      $a12m220L$ & 4 & 0.6125(21) & 0.6135(17) & 0.995(15) & 0.39 & 0.940 \\
      $a12m220L$ & 2 & 0.6187(10) & 0.6187(10) & 1.013(07) & 0.67 & 0.741 \\ \hline
      $a12m220$ & 4 & 0.6080(31) & 0.6086(30) & 0.989(27) & 0.33 & 0.967 \\
      $a12m220$ & 2 & 0.6151(14) & 0.6152(14) & 1.001(10) & 0.91 & 0.515 \\ \hline
      $a12m220S$ & 4 & 0.6039(52) & 0.6110(41) & 0.970(29) & 1.19 & 0.297 \\
      $a12m220S$ & 2 & 0.6194(26) & 0.6204(24) & 0.997(21) & 0.69 & 0.718 \\ \hline
      $a09m310$ & 4 & 0.4951(14) & 0.4959(13) & 1.027(13) & 1.72 & 0.078 \\
      $a09m310$ & 3 & 0.4952(15) & 0.4961(13) & 1.024(14) & 0.96 & 0.473 \\ \hline
      $a09m220$ & 4 & 0.4495(20) & 0.4513(15) & 1.020(16) & 0.36 & 0.955 \\
      $a09m220$ & 3 & 0.4514(16) & 0.4528(13) & 1.021(14) & 0.53 & 0.857 \\ \hline
      $a09m130W$ & 4 & 0.4208(17) & 0.4221(16) & 0.978(31) & 0.77 & 0.647 \\
      $a09m130W$ & 3 & 0.4213(18) & 0.4225(17) & 0.981(31) & 1.12 & 0.342 \\ \hline
      $a06m310$ & 4 & 0.3248(30) & 0.3257(28) & 0.996(42) & 0.97 & 0.422 \\
      $a06m310$ & 3 & 0.3305(21) & 0.3319(19) & 1.059(25) & 0.80 & 0.524 \\ \hline
      $a06m310W$ & 4 & 0.3277(18) & 0.3296(16) & 1.025(22) & 2.11 & 0.077 \\
      $a06m310W$ & 3 & 0.3289(16) & 0.3303(14) & 1.030(19) & 2.14 & 0.073 \\ \hline
      $a06m220$ & 4 & 0.3036(19) & 0.3035(19) & 0.926(52) & 0.26 & 0.902 \\
      $a06m220$ & 3 & 0.3065(17) & 0.3060(16) & 0.987(42) & 1.22 & 0.299 \\ \hline
      $a06m220W$ & 4 & 0.3030(21) & 0.3045(17) & 1.033(40) & 0.51 & 0.730 \\
      $a06m220W$ & 3 & 0.3047(14) & 0.3053(13) & 1.027(25) & 0.33 & 0.858 \\ \hline
      $a06m135$ & 4 & 0.2714(24) & 0.2716(22) & 0.857(48) & 0.48 & 0.886 \\
      $a06m135$ & 3 & 0.2735(16) & 0.2737(16) & 1.008(35) & 0.33 & 0.967 
    \end{tabular}
  \end{ruledtabular}
  \caption{Comparison of the ground state nucleon mass obtained from fits to
    the dispersion relation, $E_{\bm{p}}^2 = (M_0^\text{Disp})^2 + c^2
    \bm{p}^2$ with $M_0^\text{2pt}$ from zero-momentum two-point
    correlator. Here $n_s$ is the number of states kept in the fits
    with $n_s=2$ or 3 implying a frequentist fit and $n_s=4$ implying
    an empirical Bayesian fit. The speed of light $c^2$, the
    ${\chi}^2/dof$ and $p$-value are for the fits to the dispersion
    relation, which are shown for the $a09m130W$ and $a06m135$ ensembles in
    Fig.~\protect\ref{fig:dispersion}.}
  \label{tab:disp}
\end{table}

%\clearpage

The analysis of the first excited state mass from fits to the
three-point correlations functions has been presented in
Sec.~\ref{ssec:ESS}. Here we study its extraction from the spectral
decomposition of $C^\text{2pt}(t)$: 
\begin{align}
  C^\text{2pt}(t) &= a_0 e^{-E_0 t} \left\{1 + \sum_{k=1}^\infty b_k e^{-(E_k-E_0)t}\right\}\,,
  \label{eq:def-corr-2pt-decomp}
\end{align}
where the coefficients $a_0$ and $b_k$ are positive definite since 
the same interpolating operator is used at the source and the sink. Starting from 
the definition of the effective mass 
\begin{align}
  m_\text{eff}(t) &= \log \frac{C(t)}{C(t+1)} 
  \label{eq:meff0-fwd}
\end{align}
one can derive, using the symmetric lattice derivative ${df(t)/dt}
\to (f(t+1) - f(t-1))/{2}$, 
a series of effective masses $m_\text{eff}^{(n)}$ 
\begin{equation}
 m_\text{eff}^{(0)} \equiv -\frac{d}{dt}\log C^\text{2pt}(t) 
\end{equation}
\begin{widetext}
\begin{align}
  m_\text{eff}^{(n)} &\equiv 
  m_\text{eff}^{(n-1)} -\frac{d}{dt} \log(m_\text{eff}^{(n-1)} - E_{n-1}) 
  \label{eq:meff-gen} \\
  &= E_n
  - \frac{d}{dt} \log \left\{ 1 + \sum_{k=n+1}^\infty \frac{b_k}{b_n}\frac{(E_k-E_0)\cdots(E_k-E_{n-1})}{(E_n-E_0)\cdots(E_n-E_{n-1})} e^{-(E_k-E_n)t} \right\} \,,\; (n=1,2,\ldots) \,.
\end{align}
\end{widetext}
that should approach a plateau from above at a sufficiently large time
$t$ and give the energy levels $E_n$. To determine
$m_\text{eff}^{(n)}(t)$, one could take the $E_n$ from a
multi-exponential fit, with $n$ limited by the statistical quality of
the data.  Note that no prior information of the overlap factors $a_0$
and $b_k$ is required to calculate $m_\text{eff}^{(n)}(t)$.

\begin{table}[!tb]   %T08
  \begin{ruledtabular}
    \begin{tabular}{lrcccc}
      Ensemble ID & $t_1/a$ & $4s$                          & $3s$                         & $M_1^{(4)}$ & $M_1^{(3)}$ \\
                  &         & $[t_\text{min},t_\text{max}]$ & $[t_\text{min},t_\text{max}]$&  GeV        &  GeV        \\
      \hline
      a15m310  &  4 & $[1,10]$ & $[1,10]$ & 2.04(06) & 2.22(03) \\
      a12m310  &  6 & $[2,15]$ & $[2,10]$ & 1.58(09) & 2.51(05) \\ 
      a12m220S &  6 & $[2,15]$ & $[2,10]$ & 1.50(08) & 2.40(08) \\ 
      a12m220  &  6 & $[2,15]$ & $[1,12]$ & 1.63(12) & 2.39(02) \\ 
      a12m220L &  7 & $[2,15]$ & $[2,14]$ & 1.69(18) & 2.40(03) \\ 
      a09m310  &  7 & $[2,18]$ & $[2,18]$ & 2.06(13) & 2.09(16) \\ 
      a09m220  &  9 & $[3,14]$ & $[2,20]$ & 1.72(09) & 2.00(10) \\ 
      a09m130W &  7 & $[4,20]$ & $[2,20]$ & 1.76(09) & 1.82(09) \\ 
      a06m310  & 12 & $[7,30]$ & $[3,30]$ & 1.65(11) & 2.09(14) \\ 
      a06m310W &  8 & $[4,25]$ & $[2,25]$ & 2.05(15) & 2.37(15) \\ 
      a06m220  & 13 & $[7,30]$ & $[3,30]$ & 1.87(08) & 2.10(06) \\ 
      a06m220W & 11 & $[4,20]$ & $[2,25]$ & 1.82(15) & 2.21(11) \\ 
      a06m135  & 12 & $[6,30]$ & $[2,25]$ & 1.69(11) & 1.85(05) 
    \end{tabular}
  \end{ruledtabular}
  \caption{Results for $M_1$ from the four-state empirical Bayesian
    fit (4s) and the three-state frequentist fit (3s). For 
    ensembles with $a \approx 0.12, 0.15\,\fm$, a two-state
    frequentist fit is performed, nevertheless, we keep the label
    ``3s'' for brevity. The second column gives the approximate time
    $t_1$ at which the $m_\text{eff}^{(1)}$ reaches the first excited
    state energy $E_1$ given by  the four-state fit.  The time interval used 
    in the four- (three-) state fits to $ C^\text{2pt}(t)$ is given in
    the third (fourth) column. These fit ranges were chosen individually
    for each case balancing between keeping the maximum number of points
    with signal in the effective mass plot and the $\chi^2/dof$. }
  \label{tab:meff-gen-fig-ref}
\vspace{0.3cm}
\end{table}

\begin{figure*}[!tbh]   %F23
  \includegraphics[trim=340 0 0 0, clip, width=0.49\textwidth]{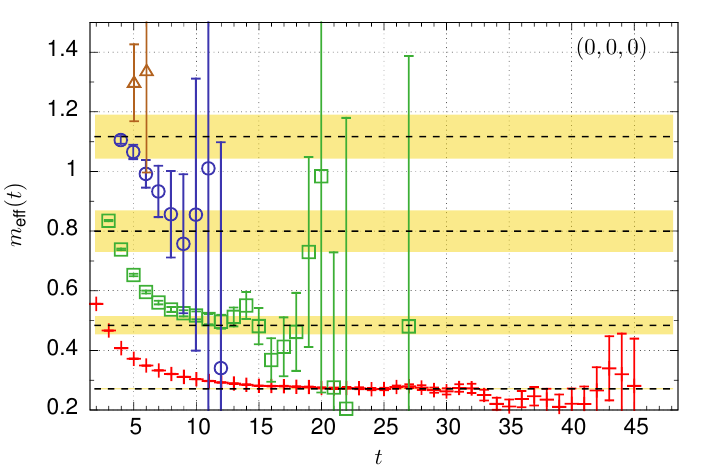} \hfill
  \includegraphics[trim=340 0 0 0, clip, width=0.49\textwidth]{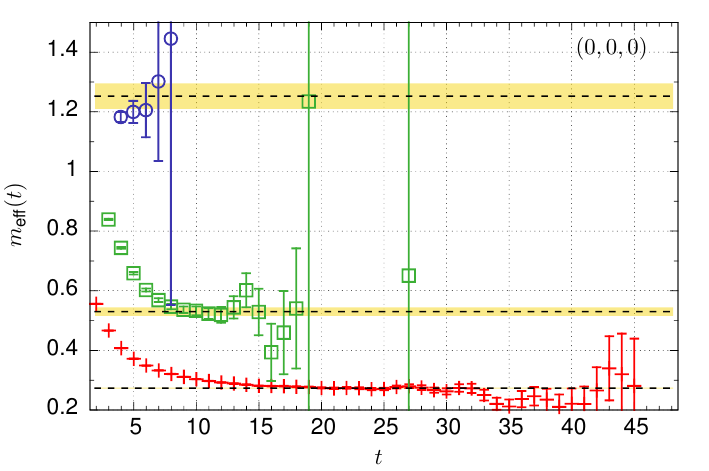}
  \includegraphics[trim=340 0 0 0, clip, width=0.49\textwidth]{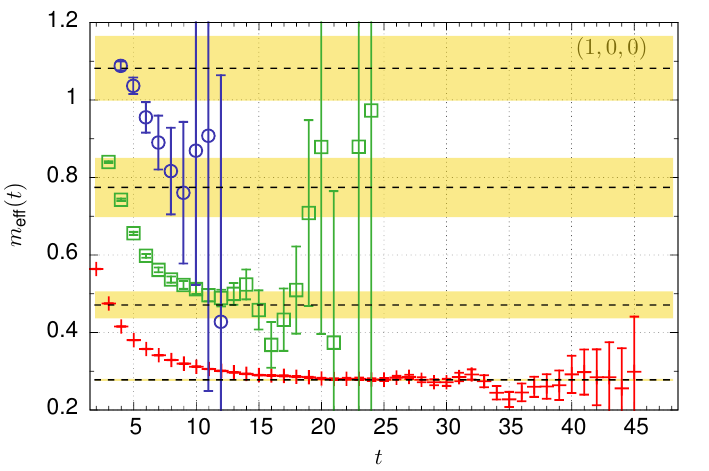}
  \includegraphics[width=0.49\textwidth]{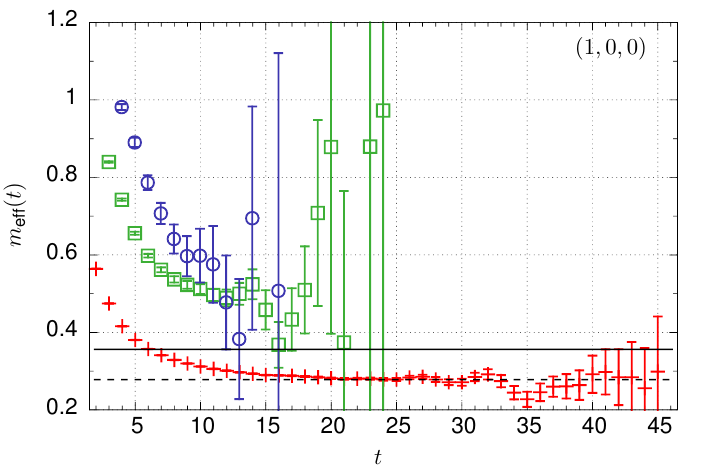}
  \caption{Data for the effective masses $m_\text{eff}^{(n)}$, defined
    in Eq.~\protect\eqref{eq:meff-gen}, from the a06m135 two-point
    correlators. Top panel shows results for $\bm{p}=\bm{0}$ with the
    $E_{n-1}$ in Eq.~\eqref{eq:meff-gen} taken from the four-state
    fit (left) and three-state fit (right). These input energy levels
    $E_n$ are shown by the dashed lines with yellow error bands.  The
    red plus, green square, and blue circle symbols correspond to
    $m_\text{eff}^{(n)}$ with $n=0, 1, 2$, respectively.  The bottom
    panel shows $m_\text{eff}^{(n)}$ for $\bm{p}=2\pi \bm{n}/L$,
    $\bm{n}=(1,0,0)$. In the bottom right panel, $E_0$ is taken from the
    four-state fit and $E_1=E_0+2M_\pi$ (solid black line) is
    assumed.}
  \label{fig:meff-gen-a06m135}
\end{figure*}

These effective masses for the a06m135 ensemble are shown in
Fig.~\ref{fig:meff-gen-a06m135} for the lowest two momenta and
compared with when the $E_i$ are taken from a four- (left
panels) versus three-state (right panels) fits with values 
given by the black dashed lines with yellow error bands.  The fit parameters 
and the first excited state masses, $M_1^{(4)}$ and
$M_1^{(3)}$, are given in Table~\ref{tab:meff-gen-fig-ref}. We note 
that 
\begin{itemize}
\item
The estimate of $E_1$ is slightly larger from the 3-state fits. Again, this
is expected since the fits give ``effective'' $E_i$ that partly incorporate
the contributions of all the higher states neglected in the fits.
\item
The time $t_1$ when $m_\text{eff}^{(1)}(t)$ reaches the estimate $E_1$ is roughly 
constant, $\approx 0.7$~fm. 
\item
The signal in $m_\text{eff}^{(1)}(t)$ becomes noisy for $t \gsim t_1$, i.e., before 
confirmation of it having plateaued. 
\item
Estimates of $M_1^{(4)}$ and $M_1^{(3)}$ for the two physical pion
mass ensembles (see Table~\ref{tab:meff-gen-fig-ref} and $M_1^{(4)} -
M_1^{(3)}$ plotted in Fig.~\ref{fig:comp-M1}) are consistent with the
$N(1710)$ excited state, or a combination of the $N(1440)$
and $N(1710)$ as they overlap within their widths, $\Gamma \approx 300$ and $100\,\MeV$. 
\item
Estimates of $E_2$ and $m_\text{eff}^{(2)}(t)$ from 3-state fits are not reliable.
\item
In the bottom right panel of Fig.~\ref{fig:meff-gen-a06m135}, we input
$E_1 = E_0 + 2M_\pi$ (solid black line) to study impact on
$m_\text{eff}^{(n)}(t)$. Estimates of $m_\text{eff}^{(1)}(t)$ are not
changed but $m_\text{eff}^{(2)}(t)$ shows a much more rapid fall. The
signal is, however, poor and dies before any conclusion can be
reached.
\end{itemize}

Overall, this analysis highlights the challenge of determining the
excited state energies $E_i$ from fits to $ C^\text{2pt}(t)$ and
making an association with physical states.

\clearpage
\onecolumngrid
\section{Extrapolation of the nucleon mass $M_N$ to the Physical Point}   %A0B
\label{sec:spectrum}
\vspace{0.2cm}
\twocolumngrid

\begin{figure}[!t]   %F24
  \centering
  \includegraphics[width=0.80\columnwidth]{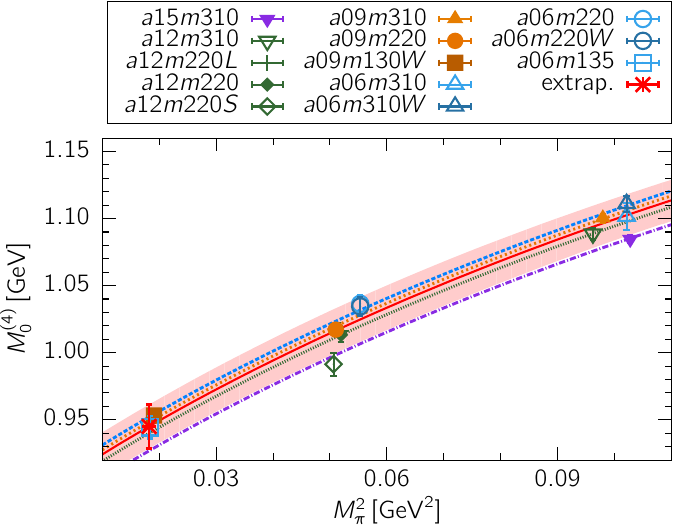} 
  \\ %\vspace{0.5em}
  \includegraphics[trim=0 0 0 63, clip, width=0.80\columnwidth]{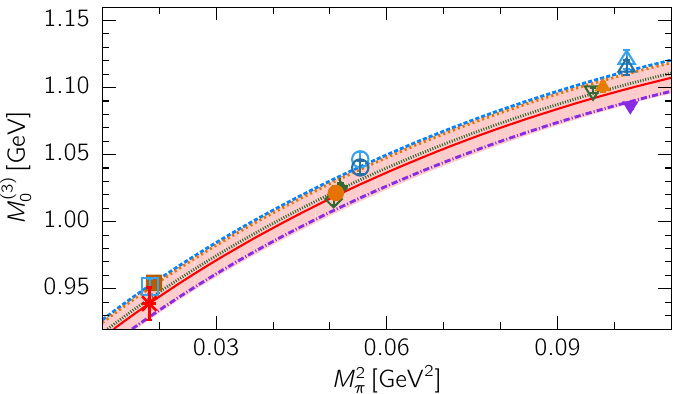} 
  \caption{The result of the chiral-continuum (CC) fit (no finite
    volume term) to the nucleon mass $M_0^{(4)}$ (B1 fit in
    Table~\ref{tab:ccfv-MN})(top panel) and $M_0^{(3)}$ (F1 fit in
    Table~\ref{tab:ccfv-MN}) (bottom panel) is shown by the red line
    with the error band.  The data for B1 and F1 fits given in
    Table~\ref{tab:meff-gen-fig-ref} are plotted versus $M_\pi^2$
    after shifting them in $a$ to $a = 0$ using the CC fits.  The CC
    fit is also shown versus $M_\pi^2$ with $a$ set to $a=0.06$~fm
    (blue dashed line), 0.09~fm (orange dotted line), 0.12~fm (green
    dotted line), and 0.15~fm (purple dash-dot line). In a perfect
    fit, these curves should pass through points with the same color,
    i.e., with the same lattice spacing $a$. }
  \label{fig:comp-M0}
\end{figure}

Here we revisit the extrapolation of the nucleon mass $M_N(a,M_\pi^2,
M_\pi L)$ given in Table~\ref{tab:disp} to the physical point and
extend the discussion in the Appendix~B in Ref.~\cite{Jang:2019jkn}.
We use the following CCFV ansatz:
\begin{align}
  M_N = c_0 + c_1 a + c_2 a^2 + c_3 M_\pi^2 + c_4 M_\pi^3 + c_5 M_\pi^2 e^{-M_\pi L} \,.
  \label{eq:ccfv-MN}
\end{align}
Results and the fit parameters $c_i$ for various truncations of this
ansatz are given in Table~\ref{tab:ccfv-MN}.  The AIC score is defined
as ${\rm AIC} = 2k - 2log(L) = 2k + \chi^2 + constant$ where k is the
number of parameters and $L$ is the likelihood function. When quoting
AIC scores, we drop the irrelevant constant.  The CCFV fits F1 and B1 are
shown in Fig.~\ref{fig:comp-M0}.  Our analysis indicates
\begin{itemize}
\item
The CCFV fits, F1-F4, to the $M_0^{(3)}$ data give slightly smaller
continuum $M_N$ than fits to $M_0^{(4)}$ even though $M_0^{(3)} >
M_0^{(4)}$ as shown in Fig.~\ref{fig:comp-M1} (bottom panel) for each
of the thirteen ensembles.
\item
Only F1 ($M_N=0.939(12)\,\GeV$) and B1 ($M_N=0.945(16)\,\GeV$) fits
give estimates consistent with the physical value of $M_N=939\,\GeV$.
The other fits give $\approx 25$~MeV higher values.
\item
The F3 and B3 fits, which include the higher order $M_\pi^3$ term give a $c_4$ that is 
roughly consistent with the $\chi$PT prediction $c_4=3g_A^2/(32\pi
F_\pi^2)=-5.716$.  On including $a^2$ and/or finite volume correction
terms in addition to the $M_\pi^3$ term, $c_4$ remains consistent with
the $\chi$PT prediction for F1, F2 and F4 fits but becomes smaller for B1,
B2 and B4.
\item
The finite volume coefficient, $c_5$, is not well determined in any of
the fits.  Without it, fits F1 and B1 have small $p$-value but give
results consistent with the experimental value. Including it, the $p$-value 
of F2 and B2 fits 
improves to an acceptable level, but the
coefficients of the lattice spacing dependence, $c_1$ and $c_2$,
become less well determined.  Neglecting the $c_2$ term (F4 and B4
fits), the $c_1$ becomes well determined, while the other $c_i$ are
essentially unchanged. In these cases, the $\sim 25\,\MeV$ shifts in
the $M_N$ from the F1 or B1 fits persist.
\end{itemize}
Overall, with the current data, we are not able to determine whether $M_0^{(3)}$ or $
M_0^{(4)}$ give better estimates of the ground state nucleon
mass. Also, we can at best make four-state fits to the
two-point function and three-state fits to the three-point functions.

\begin{table*}[!tbh]   %T09
  \centering
  \renewcommand{\arraystretch}{1.1}
  \begin{ruledtabular}
    \begin{tabular}{c c ccc cccccc}
      Fit & $M_N$ & $\chi^2/\text{DOF}$ & $p$ & \text{AIC} & $c_0[1]$ & $c_1[a]$ & $c_2[a^2]$ & $c_3[M_\pi^2]$ & $c_4[M_\pi^3]$ & $c_5[\text{FV}]$
      \\ 
      & GeV &  &  &  & GeV & GeV fm${}^{-1}$ & GeV fm${}^{-2}$ & GeV${}^{-1}$ & GeV${}^{-2}$  & GeV${}^{-1}$ 
      \\ \hline
      % 13pt M0_s2_or_s3
      % a_asq_mpisq_mpicub
       F1 & 0.939(12) &  2.187 &  0.025 & 27.5 &  0.878(013) &  0.41(25) & -3.2(1.2) &  4.24(38) & -6.5(1.0) & - \\
      % a_asq_mpisq_mpicub_fv
       F2 & 0.954(14) &  1.758 &  0.091 & 24.3 &  0.895(015) &  0.11(28) & -1.6(1.4) &  4.07(38) & -5.9(1.0) & -6.0(2.6) \\
      % a_asq_mpisq_mpicub_fv6
%       F3 & 0.958(14) &  1.611 &  0.127 & 23.3 &  0.899(015) &  0.00(30) & -0.9(1.5) &  4.03(39) & -5.7(1.0) & -4.5(1.8)$^\prime$ \\
      % a_mpisq_mpicub
       F3 & 0.968(04) &  2.686 &  0.004 & 32.2 &  0.904(008) & -0.23(04) & - &  4.53(36) & -7.3(9) & - \\
      % a_mpisq_mpicub_fv
       F4 & 0.969(04) &  1.686 &  0.096 & 23.5 &  0.908(008) & -0.19(04) & - &  4.13(38) & -6.0(1.0) & -7.4(2.3) \\
      % a_mpisq_mpicub_fv6
%       F6 &0.966(04) &  1.454 &  0.169 & 21.6 &  0.906(008) & -0.17(04) & - &  4.05(38) & -5.8(1.0) & -5.2(1.5)$^\prime$ \\
      %
       \hline
      % 13pt M0_s4
      % a_asq_mpisq_mpicub
       B1 & 0.945(16) &  1.109 &  0.353 & 18.9 &  0.896(017) &  0.27(32) & -2.6(1.5) &  3.18(46) & -3.6(1.2) & - \\
      % a_asq_mpisq_mpicub_fv
       B2 & 0.968(20) &  0.675 &  0.693 & 16.7 &  0.922(021) & -0.16(38) & -0.2(1.9) &  2.86(48) & -2.5(1.3) & -10.4(5.1) \\
      % a_asq_mpisq_mpicub_fv6
%       B3 & 0.969(20) &  0.706 &  0.667 & 16.9 &  0.923(022) & -0.23(41) &  0.3(2.1) &  2.83(49) & -2.4(1.4) & -6.8(3.4)$^\prime$ \\
      % a_mpisq_mpicub
       B3 & 0.972(05) &  1.318 &  0.221 & 19.9 &  0.921(009) & -0.28(04) & - &  3.35(44) & -4.2(1.2) & - \\
      % a_mpisq_mpicub_fv
       B4 & 0.970(05) &  0.592 &  0.785 & 14.7 &  0.924(009) & -0.20(05) & - &  2.86(48) & -2.5(1.3) & -10.7(4.0) \\
      % a_mpisq_mpicub_fv6
%       B6 & 0.966(06) &  0.620 &  0.762 & 15.0 &  0.920(009) & -0.18(06) & - &  2.83(49) & -2.4(1.3) & -6.5(2.5)$^\prime$ 
      %
    \end{tabular}
  \end{ruledtabular}
  \caption{Summary of CCFV fits to $M_N(a, M_\pi^2, M_\pi L)$ using
    Eq.~\eqref{eq:ccfv-MN}. Fits $F$1-$F$4 are to the frequentist
    (3-state or 2-state) data labeled $M_0^{(3)}$ in the text, and
    $B1$-$B4$ are to the 4-state empirical Bayesian fit data and
    labeled $M_0^{(4)}$. To make the interpretation of coefficients
    $c_i$ defined in Eq.~\eqref{eq:ccfv-MN} easier, we give both the
    functional dependance within square parentheses and the units. The
    results for $M_0^{(4)}$ are the same as in
    Ref.~\protect\cite{Jang:2019jkn}, except for a small change in $a06m135$
    value due to increased statistics. Fits corresponding to the B2
    and B4 were given in Table~XV in Ref.~\protect\cite{Jang:2019jkn} (labeled
    B1 and B2 there) and led to $M_N=0.976(20)\,\GeV$ and
    $0.972(6)\,\GeV$, respectively. The table also gives the AIC score and 
    the $p$ value of the fits.}
  \label{tab:ccfv-MN}
\end{table*}

\clearpage
\onecolumngrid
\section{Data for the Form Factors versus $Q^2$}   %A0C
\label{sec:ff-tabs}

The unrenormalized values of the form factors $G_A$, $\GP$, and $G_P$
at the various $Q^2$ values simulated on the thirteen ensembles and
extracted using the three excited-state analysis strategies $\Ssim$,
$\Sfour$, and $\Stwo$ defined in Sec.~\ref{sec:ff} are given in
Tables~\ref{tab:ff-a15m310}---~\ref{tab:ff-a06m135}.  
Since the behavior of the data for the four correlation
functions varies significantly with $Q^2$ and the 13 ensembles, we
could not develop a simple prescription for making the fits.  These $O(1000)$
fits were done individually over a three year period. In addition to the 
energies $E_i$ and amplitudes $A_i$, one has to select two additional parameters 
in each of the fits: (i) the number of points, 
$t_{\rm skip}$, skipped next to the source and sink at which the ESC
are the largest and (ii) the values of source-sink separation $\tau$ used in
the fit. For these we made the following common choices: 
(i)  $t_{\rm skip}$ was taken to be
the same for all values of $\tau$ used in a given fit; (ii) in all fits
we used data with the largest three values of $\tau$ for which the
errors were comparable to or smaller than the difference between the
central values. In most cases these were the largest three values of $\tau$ 
simulated. 

The full covariance matrix was used for all fits to the 3-point
functions with the $\Stwo$ and $\Sfour$ strategies and for the 2-point
functions. In the simultaneous fits ($\Ssim$ strategy), we restricted
the covariance matrix to be block diagonal in each correlation
function.

A rough estimate
of the size of contributions of the $N\pi$ state can be obtained from
the difference between results with $\Stwo$ and $\Ssim$ (or $\Sfour$)
strategies in the following tables.  In particular, the smallest $Q^2$
data for the two physical mass ensembles $a09m130W$ and $a06m135$ in
Tables~\ref{tab:ff-a09m130W} and~\ref{tab:ff-a06m135} show $\sim 5\%$,
$\sim 45\%$ and $\sim 45\%$ differences between the $\Stwo$ and $\Ssim$ 
values for $G_A$, $\GP$, and $G_P$, respectively. A $\sim 5\%$ level of effect in $G_A(Q^2)$ is also consistent
with what is observed in the axial charge $g_A$ extracted from the
forward matrix element between ``3-RD'' and ``$3^\ast$'' strategies as
shown in Table~\ref{tab:charge} in Appendix~\ref{sec:gASTdata}.
%% Tables~\ref{tab:ff-a15m310},
%% \ref{tab:ff-a12m310}, \ref{tab:ff-a12m220S}, \ref{tab:ff-a12m220},
%% \ref{tab:ff-a12m220L}, \ref{tab:ff-a09m310}, \ref{tab:ff-a09m220},
%% \ref{tab:ff-a09m130W}, \ref{tab:ff-a06m310W}, \ref{tab:ff-a06m310}, 
%% \ref{tab:ff-a06m220W}, \ref{tab:ff-a06m220}, \ref{tab:ff-a06m135}. 

The values of $Q^2 = |\vec p|^2 - (E-M)^2$ given in the first column
show that for a given gauge coupling $\beta$ (approximately constant $a$) and
keeping $M_\pi L$ constant, for example, in the three $a\approx
0.09$~fermi ensembles, the values of $Q^2$ decrease as $M_\pi \to
135$~MeV. This is because $\vec p = 2 \pi {\vec n}/La$ decreases as
$L$ is increased to keep the parameter controlling the finite-volume
effects, $M_\pi L$, constant. The same decrease happens if $L$ is
increased at fixed $a$ and $M_\pi$ to reduce finite-volume effects as
is evident from the data on the three ensembles,
$a12m220S$, $a12m220$, and $a12m220L$. This is a simple
kinematic effect, i.e., the $Q^2$ for a given ${\bf n}$
decreases as the lattice volume is increased. For fixed
$M_\pi$ and $M_\pi L$, the $Q^2$ remains roughly the same when $a$ is
decreased toward the continuum limit as we keep $L a$, the lattice
size in physical units (fermi) constant. This can be deduced from the
data on the $a12m310$, $a09m310$ and $a06m310$ ensembles. In short, as
lattice QCD calculations improve (larger $L$, $M_\pi \approx 135$~MeV,
and $a \to 0$), the $Q^2$ values (and $Q^2|_{\rm max}$ for fixed ${\vec
n}|_{\rm max} = (3,1,0) $ in our case) decrease. Thus, the standard method for
calculations of form factors used in this work will increasingly
give more precise form factors in the $Q^2 < 0.5$~GeV${}^2$ region. Further
algorithmic developments are needed to push calculations with  momentum transfer squared up to $Q^2
\sim 5$~GeV${}^2$ to meet the needs of the DUNE experiment. 

\vspace{0.5cm}

\iffalse

\begin{table}[!htb]
  \begin{ruledtabular}
    \begin{tabular}{lc}
      Ensemble ID & Form Factors Table \\
      \hline
      a15m310  & \ref{tab:ff-a15m310} \\
      a12m310  & \ref{tab:ff-a12m310} \\ 
      a12m220S & \ref{tab:ff-a12m220S} \\ 
      a12m220  & \ref{tab:ff-a12m220} \\ 
      a12m220L & \ref{tab:ff-a12m220L} \\ 
      a09m310  & \ref{tab:ff-a09m310} \\ 
      a09m220  & \ref{tab:ff-a09m220} \\ 
      a09m130W & \ref{tab:ff-a09m130W} \\ 
      a06m310W & \ref{tab:ff-a06m310W} \\ 
      a06m310  & \ref{tab:ff-a06m310} \\ 
      a06m220W & \ref{tab:ff-a06m220W} \\ 
      a06m220  & \ref{tab:ff-a06m220} \\ 
      a06m135  & \ref{tab:ff-a06m135} 
    \end{tabular}
  \end{ruledtabular}
  \caption{List of form factors table for the thirteen calculations.}
  \label{tab:list-ff-tabs}
\end{table}
\fi

\begin{table*}[!tbh]   %T10
  \centering
  \renewcommand{\arraystretch}{1.1}
  \begin{ruledtabular}
    \begin{tabular}{c ccc ccc ccc}
      & \multicolumn{3}{c}{$G_A$} & \multicolumn{3}{c}{$\GP$} & \multicolumn{3}{c}{$G_P$} \\ \cline{2-4}\cline{5-7}\cline{8-10}
      $Q^2 [\GeV]$ & $\Ssim$ & $\Sfour$ & $\Stwo$ & $\Ssim$ & $\Sfour$ & $\Stwo$ & $\Ssim$ & $\Sfour$ & $\Stwo$ \\ \hline
 0.252(00) & 1.007(008) & 1.010(009) & 0.993(006) & 12.53(22) & 12.89(23) & 11.01(14) & 15.12(24) & 15.56(23) & 13.14(15) \\
 0.483(01) & 0.842(007) & 0.842(007) & 0.822(007) &  6.388(126) &  6.433(102) &  5.827(093) &  8.115(143) &  8.162(105) &  7.386(097) \\
 0.703(02) & 0.720(007) & 0.720(010) & 0.690(008) &  3.967(071) &  4.046(104) &  3.647(072) &  5.255(095) &  5.357(105) &  4.939(081) \\
 0.911(06) & 0.648(017) & 0.657(020) & 0.614(017) &  2.890(132) &  2.915(148) &  2.553(094) &  4.080(172) &  4.071(160) &  3.559(097) \\
 1.102(07) & 0.593(011) & 0.587(013) & 0.561(011) &  2.144(082) &  2.065(090) &  1.978(061) &  3.184(133) &  3.046(105) &  2.856(088) \\
 1.297(09) & 0.520(006) & 0.524(011) & 0.491(013) &  1.560(045) &  1.613(076) &  1.590(078) &  2.341(063) &  2.418(090) &  2.369(117) \\
 1.637(22) & 0.450(017) & 0.476(025) & 0.469(023) &  1.018(062) &  1.119(087) &  1.119(103) &  1.662(082) &  1.793(118) &  1.640(121) \\
 1.803(22) & 0.439(029) & 0.449(034) & 0.452(023) &  0.999(106) &  1.028(124) &  1.050(104) &  1.444(133) &  1.489(148) &  1.290(207) \\
 1.790(29) & 0.543(114) & 0.458(039) & 0.395(056) &  1.215(356) &  1.002(135) &  0.959(204) &  1.557(428) &  1.121(319) &  0.995(972) \\
 1.917(31) & 0.428(108) & 0.437(024) & 0.379(046) &  0.911(353) &  0.938(085) &  1.225(153) &  1.455(505) &  1.354(166) &  1.293(322) \\
    \end{tabular}
  \end{ruledtabular}
  \caption{The bare form factors $G_A$, $\GP$, and $G_P$ versus $Q^2$
for the 3 strategies $\Ssim$, $\Sfour$, and $\Stwo$ on ensemble $a15m310$.}
  \label{tab:ff-a15m310}
\end{table*}

\begin{table*}[!tbh]   %T11
  \centering
  \renewcommand{\arraystretch}{1.1}
  \begin{ruledtabular}
    \begin{tabular}{c ccc ccc ccc}
      & \multicolumn{3}{c}{$G_A$} & \multicolumn{3}{c}{$\GP$} & \multicolumn{3}{c}{$G_P$} \\ \cline{2-4}\cline{5-7}\cline{8-10}
      $Q^2 [\GeV]$ & $\Ssim$ & $\Sfour$ & $\Stwo$ & $\Ssim$ & $\Sfour$ & $\Stwo$ & $\Ssim$ & $\Sfour$ & $\Stwo$ \\ \hline
 0.176(00) & 1.086(011) & 1.086(011) & 1.050(014) & 18.01(29) & 17.84(38) & 16.51(56)  & 22.32(36) & 22.08(41) & 21.15(86) \\
 0.342(01) & 0.948(010) & 0.937(009) & 0.905(013) &  9.817(153) &  9.501(169) &  9.493(281)  & 12.67(20) & 12.27(19) & 12.02(33) \\
 0.498(02) & 0.844(011) & 0.842(012) & 0.787(016) &  6.505(106) &  6.411(155) &  5.954(199)  &  8.635(154) &  8.499(206) &  8.309(252) \\
 0.646(03) & 0.764(013) & 0.763(014) & 0.686(022) &  4.664(087) &  4.600(138) &  4.124(189)  &  6.478(133) &  6.385(194) &  5.863(191) \\
 0.787(04) & 0.694(011) & 0.691(012) & 0.639(016) &  3.504(088) &  3.454(113) &  3.246(123)  &  4.957(127) &  4.891(157) &  4.301(162) \\
 0.920(05) & 0.649(011) & 0.666(013) & 0.576(019) &  2.882(118) &  3.006(076) &  2.399(168)  &  4.015(150) &  4.161(106) &  3.424(180) \\
 1.178(09) & 0.533(010) & 0.550(017) & 0.506(025) &  1.684(068) &  1.818(110) &  1.651(126)  &  2.606(074) &  2.731(163) &  2.433(222) \\
 1.293(10) & 0.472(031) & 0.463(040) & 0.465(025) &  1.296(193) &  1.295(214) &  1.469(150)  &  2.056(173) &  2.048(221) &  1.659(259) \\
 1.315(19) & 0.462(016) & 0.538(047) & 0.482(336) &  1.175(083) &  1.609(253) &  1.264(1.241)&  2.104(152) &  2.446(317) &  1.680(546) \\
 1.435(18) & 0.471(013) & 0.488(032) & 0.462(044) &  1.224(066) &  1.293(138) &  1.438(258)  &  1.908(087) &  1.984(223) &  0.840(515) \\
    \end{tabular}
  \end{ruledtabular}
  \caption{The bare form factors $G_A$, $\GP$, and $G_P$ versus $Q^2$
for the 3 strategies $\Ssim$, $\Sfour$, and $\Stwo$ on ensemble $a12m310$.}
  \label{tab:ff-a12m310}
\end{table*}

\begin{table*}[!tbh]  %T12
  \centering
  \renewcommand{\arraystretch}{1.1}
  \begin{ruledtabular}
    \begin{tabular}{c ccc ccc ccc}
      & \multicolumn{3}{c}{$G_A$} & \multicolumn{3}{c}{$\GP$} & \multicolumn{3}{c}{$G_P$} \\ \cline{2-4}\cline{5-7}\cline{8-10}
      $Q^2 [\GeV]$ & $\Ssim$ & $\Sfour$ & $\Stwo$ & $\Ssim$ & $\Sfour$ & $\Stwo$ & $\Ssim$ & $\Sfour$ & $\Stwo$ \\ \hline
 0.175(01) & 1.129(033) & 1.123(034) & 1.084(020) & 19.73(92)  & 19.75(1.08) & 16.14(69) & 26.03(1.19)& 26.27(1.49)& 21.67(96) \\
 0.339(03) & 0.918(037) & 0.920(033) & 0.912(020) &  9.517(656)  &  9.580(552)   &  8.373(527) & 14.44(1.11)& 14.83(98)  & 12.60(91) \\
 0.490(06) & 0.814(031) & 0.821(044) & 0.766(027) &  5.271(346)  &  5.634(304)   &  5.165(360) &  8.344(603)  &  9.064(487)  &  8.620(681) \\
 0.636(09) & 0.749(120) & 0.749(044) & 0.705(037) &  4.284(877)  &  4.166(328)   &  3.744(288) &  6.528(1.027)&  6.563(449)  &  6.237(399) \\
 0.773(10) & 0.629(039) & 0.646(041) & 0.640(024) &  3.016(409)  &  3.123(268)   &  2.743(171) &  7.386(2.229)&  6.484(705)  &  5.342(341) \\
 0.909(13) & 0.587(069) & 0.627(045) & 0.593(028) &  2.568(343)  &  2.763(309)   &  2.276(212) &  4.458(804)  &  4.516(556)  &  3.919(353) \\
 1.178(23) & 0.461(144) & 0.514(062) & 0.502(041) &  1.110(705)  &  1.689(269)   &  1.515(176) &  2.710(222)  &  3.422(587)  &  3.156(390) \\
 1.307(25) & 0.467(115) & 0.571(059) & 0.511(039) &  1.116(1.966)&  1.810(250)   &  1.480(193) &  2.183(5.142)&  2.986(644)  &  2.890(476) \\
 1.238(33) & 0.509(049) & 0.551(077) & 0.555(068) &  1.247(181)  &  1.605(316)   &  1.499(288) &  2.975(858)  &  3.966(821)  &  3.586(682) \\
 1.358(36) & 0.415(037) & 0.425(135) & 0.440(050) &  0.771(163)  &  0.712(764)   &  0.890(225) &  2.597(424)  &  3.625(1.536)&  2.818(614) \\
    \end{tabular}
  \end{ruledtabular}
  \caption{The bare form factors $G_A$, $\GP$, and $G_P$ versus $Q^2$
for the 3 strategies $\Ssim$, $\Sfour$, and $\Stwo$ on ensemble $a12m220S$.}
  \label{tab:ff-a12m220S}
\end{table*}

\begin{table*}[!tbh]  %T13
  \centering
  \renewcommand{\arraystretch}{1.1}
  \begin{ruledtabular}
    \begin{tabular}{c ccc ccc ccc}
      & \multicolumn{3}{c}{$G_A$} & \multicolumn{3}{c}{$\GP$} & \multicolumn{3}{c}{$G_P$} \\ \cline{2-4}\cline{5-7}\cline{8-10}
      $Q^2 [\GeV]$ & $\Ssim$ & $\Sfour$ & $\Stwo$ & $\Ssim$ & $\Sfour$ & $\Stwo$ & $\Ssim$ & $\Sfour$ & $\Stwo$ \\ \hline
 0.105(00) & 1.174(020) & 1.169(021) & 1.145(017) & 28.84(1.16)& 28.37(1.27)& 24.11(1.40)& 38.36(1.42)& 37.84(1.55)& 32.00(2.09)\\
 0.206(02) & 1.033(014) & 1.041(017) & 1.031(022) & 15.55(42)  & 15.93(52)  & 14.53(97) & 21.95(58)  & 22.23(66)  & 20.38(1.43)\\
 0.301(02) & 0.954(017) & 0.956(017) & 0.930(021) & 10.75(34)  & 10.93(34)  &  9.99(48) & 15.17(47)  & 15.30(49)  & 13.78(59)  \\
 0.391(03) & 0.908(017) & 0.911(017) & 0.900(021) &  7.975(219)  &  8.011(221)  &  7.498(345) & 11.51(34)  & 11.43(34)  & 10.44(41)  \\
 0.482(04) & 0.834(014) & 0.836(016) & 0.812(021) &  6.182(192)  &  6.168(240)  &  5.752(262) &  9.135(300)  &  9.236(356)  &  8.444(356)  \\
 0.568(05) & 0.784(018) & 0.787(021) & 0.749(024) &  4.988(200)  &  5.142(246)  &  4.601(249) &  7.746(282)  &  7.955(344)  &  6.803(348)  \\
 0.732(08) & 0.690(015) & 0.730(029) & 0.671(028) &  3.404(152)  &  3.746(303)  &  3.246(219) &  5.518(262)  &  6.051(497)  &  5.145(361)  \\
 0.808(10) & 0.686(028) & 0.705(033) & 0.644(031) &  3.213(252)  &  3.323(218)  &  2.850(237) &  5.606(475)  &  5.833(388)  &  4.690(377)  \\
 0.806(12) & 0.676(043) & 0.711(035) & 0.643(045) &  3.147(203)  &  3.077(357)  &  2.885(321) &  5.421(360)  &  5.058(541)  &  4.995(556)  \\
 0.884(12) & 0.655(023) & 0.671(025) & 0.622(036) &  2.579(184)  &  2.562(240)  &  2.693(266) &  4.085(261)  &  4.115(295)  &  3.753(470)  \\
    \end{tabular}
  \end{ruledtabular}
  \caption{The bare form factors $G_A$, $\GP$, and $G_P$ versus $Q^2$
for the 3 strategies $\Ssim$, $\Sfour$, and $\Stwo$ on ensemble $a12m220$.}
  \label{tab:ff-a12m220}
\end{table*}

\begin{table*}[!tbh]  %T14
  \centering
  \renewcommand{\arraystretch}{1.1}
  \begin{ruledtabular}
    \begin{tabular}{c ccc ccc ccc}
      & \multicolumn{3}{c}{$G_A$} & \multicolumn{3}{c}{$\GP$} & \multicolumn{3}{c}{$G_P$} \\ \cline{2-4}\cline{5-7}\cline{8-10}
      $Q^2 [\GeV]$ & $\Ssim$ & $\Sfour$ & $\Stwo$ & $\Ssim$ & $\Sfour$ & $\Stwo$ & $\Ssim$ & $\Sfour$ & $\Stwo$ \\ \hline
 0.067(0) & 1.235(20) & 1.259(16) & 1.199(09) & 39.90(1.21)& 40.42(1.12)& 30.88(2.09)& 52.61(1.52) & 52.68(1.37)& 40.29(3.15)\\
 0.132(0) & 1.129(10) & 1.150(09) & 1.118(08) & 24.34(55)  & 24.51(45)  & 20.55(1.17)& 32.11(63)  & 32.17(53)  & 27.08(1.63)\\
 0.195(0) & 1.061(09) & 1.078(08) & 1.052(09) & 17.24(36)  & 17.22(26)  & 15.09(68)  & 23.01(42)  & 23.01(32)  & 20.17(95)  \\
 0.257(1) & 0.982(11) & 1.008(09) & 0.983(13) & 12.68(41)  & 13.01(19)  & 11.58(44)  & 17.74(41)  & 17.98(27)  & 16.02(68)  \\
 0.316(1) & 0.944(08) & 0.961(09) & 0.936(12) & 10.34(21)  & 10.40(18)  &  9.30(27)  & 14.34(27)  & 14.43(22)  & 12.98(37)  \\
 0.374(1) & 0.907(10) & 0.921(13) & 0.890(14) &  8.719(155)  &  8.623(250)  &  7.765(203)  & 12.19(24)  & 12.11(31)  & 10.99(29)  \\
 0.487(2) & 0.822(29) & 0.839(09) & 0.808(14) &  6.264(153)  &  6.239(091)  &  5.576(124)  &  9.003(302)  &  9.046(132)  &  8.146(154)  \\
 0.541(3) & 0.782(15) & 0.802(12) & 0.771(16) &  5.503(129)  &  5.347(173)  &  4.894(126)  &  8.083(176)  &  7.852(241)  &  7.260(159)  \\
 0.541(3) & 0.776(15) & 0.797(10) & 0.770(17) &  5.152(168)  &  5.075(130)  &  4.883(130)  &  7.922(237)  &  7.606(199)  &  7.429(190)  \\
 0.595(3) & 0.739(13) & 0.766(09) & 0.741(16) &  4.549(105)  &  4.477(110)  &  4.264(120)  &  6.992(210)  &  6.920(168)  &  6.565(145)  \\
    \end{tabular}
  \end{ruledtabular}
  \caption{The bare form factors $G_A$, $\GP$, and $G_P$ versus $Q^2$
for the 3 strategies $\Ssim$, $\Sfour$, and $\Stwo$ on ensemble $a12m220L$.}
  \label{tab:ff-a12m220L}
\end{table*}

\begin{table*}[!tbh]  %T15
  \centering
  \renewcommand{\arraystretch}{1.1}
  \begin{ruledtabular}
    \begin{tabular}{c ccc ccc ccc}
      & \multicolumn{3}{c}{$G_A$} & \multicolumn{3}{c}{$\GP$} & \multicolumn{3}{c}{$G_P$} \\ \cline{2-4}\cline{5-7}\cline{8-10}
      $Q^2 [\GeV]$ & $\Ssim$ & $\Sfour$ & $\Stwo$ & $\Ssim$ & $\Sfour$ & $\Stwo$ & $\Ssim$ & $\Sfour$ & $\Stwo$ \\ \hline
 0.183(00) & 1.043(06) & 1.043(06) & 1.053(04) & 17.38(22) & 17.27(20) & 14.60(30) & 21.90(21) & 21.76(18) & 18.35(35) \\
 0.356(01) & 0.894(06) & 0.894(07) & 0.907(06) &  9.238(099) &  9.231(105) &  8.316(185) & 12.36(14) & 12.37(12) & 11.24(25) \\
 0.520(04) & 0.793(09) & 0.794(09) & 0.795(12) &  6.027(123) &  6.047(103) &  5.594(178) &  8.372(148) &  8.399(126) &  7.931(253) \\
 0.673(04) & 0.714(09) & 0.715(09) & 0.717(09) &  4.325(084) &  4.313(082) &  4.042(077) &  6.225(108) &  6.208(103) &  5.776(086) \\
 0.819(08) & 0.647(07) & 0.659(10) & 0.652(09) &  3.284(067) &  3.370(079) &  3.115(091) &  4.889(103) &  4.989(111) &  4.620(103) \\
 0.961(13) & 0.602(07) & 0.609(12) & 0.591(13) &  2.612(057) &  2.710(084) &  2.482(079) &  3.938(082) &  4.059(114) &  3.834(106) \\
 1.197(09) & 0.531(14) & 0.565(25) & 0.521(09) &  1.906(082) &  2.028(121) &  1.676(046) &  3.116(126) &  3.198(114) &  2.781(063) \\
 1.323(13) & 0.498(10) & 0.506(16) & 0.489(08) &  1.513(076) &  1.601(093) &  1.455(047) &  2.535(120) &  2.663(131) &  2.455(123) \\
 1.325(17) & 0.449(12) & 0.529(34) & 0.482(24) &  1.352(080) &  1.687(162) &  1.438(104) &  2.312(138) &  2.783(174) &  2.319(131) \\
 1.421(14) & 0.481(23) & 0.513(30) & 0.470(11) &  1.430(141) &  1.519(146) &  1.296(052) &  2.435(240) &  2.560(197) &  2.216(111) \\
    \end{tabular}
  \end{ruledtabular}
  \caption{The bare form factors $G_A$, $\GP$, and $G_P$ versus $Q^2$
for the 3 strategies $\Ssim$, $\Sfour$, and $\Stwo$ on ensemble $a09m310$.}
  \label{tab:ff-a09m310}
\end{table*}

\begin{table*}[!tbh]  %T16
  \centering
  \renewcommand{\arraystretch}{1.1}
  \begin{ruledtabular}
    \begin{tabular}{c ccc ccc ccc}
      & \multicolumn{3}{c}{$G_A$} & \multicolumn{3}{c}{$\GP$} & \multicolumn{3}{c}{$G_P$} \\ \cline{2-4}\cline{5-7}\cline{8-10}
      $Q^2 [\GeV]$ & $\Ssim$ & $\Sfour$ & $\Stwo$ & $\Ssim$ & $\Sfour$ & $\Stwo$ & $\Ssim$ & $\Sfour$ & $\Stwo$ \\ \hline
 0.086(0) & 1.185(17) & 1.181(17) & 1.169(10) & 34.73(1.00)& 34.12(92) & 27.06(88) & 47.06(1.16)& 46.25(1.03)& 36.91(1.17)\\
 0.169(0) & 1.063(11) & 1.060(11) & 1.074(09) & 19.29(39)  & 19.16(38) & 17.17(48) & 27.19(45)  & 27.01(45)  & 23.67(66)  \\
 0.248(1) & 0.978(10) & 0.974(10) & 0.986(09) & 13.11(25)  & 13.04(25) & 11.99(30) & 18.90(31)  & 18.85(30)  & 17.25(48)  \\
 0.324(1) & 0.925(11) & 0.922(11) & 0.920(11) &  9.940(202)  &  9.784(217) &  8.960(227) & 14.41(28)  & 14.19(27)  & 13.22(33)  \\
 0.398(2) & 0.862(08) & 0.859(09) & 0.858(10) &  7.665(141)  &  7.563(146) &  7.158(154) & 11.59(19)  & 11.46(19)  & 10.67(24)  \\
 0.470(2) & 0.807(09) & 0.806(09) & 0.802(11) &  6.165(119)  &  6.086(124) &  5.736(134) &  9.556(168)  &  9.451(172)  &  8.847(189)  \\
 0.608(4) & 0.726(10) & 0.725(10) & 0.719(12) &  4.401(081)  &  4.379(090) &  4.142(101) &  7.007(131)  &  7.020(136)  &  6.462(159)  \\
 0.674(4) & 0.687(11) & 0.694(11) & 0.683(13) &  3.807(084)  &  3.772(088) &  3.532(104) &  6.187(137)  &  6.169(135)  &  5.456(160)  \\
 0.671(5) & 0.706(09) & 0.711(14) & 0.690(16) &  3.882(070)  &  3.935(095) &  3.538(131) &  6.299(162)  &  6.333(165)  &  5.616(208)  \\
 0.736(5) & 0.664(11) & 0.677(12) & 0.654(13) &  3.351(105)  &  3.362(111) &  3.147(102) &  5.480(177)  &  5.473(176)  &  4.962(178)  \\
    \end{tabular}
  \end{ruledtabular}
  \caption{The bare form factors $G_A$, $\GP$, and $G_P$ versus $Q^2$
for the 3 strategies $\Ssim$, $\Sfour$, and $\Stwo$ on ensemble $a09m220$.}
  \label{tab:ff-a09m220}
\end{table*}

\begin{table*}[!tbh]  %T17
  \centering
  \renewcommand{\arraystretch}{1.1}
  \begin{ruledtabular}
    \begin{tabular}{c ccc ccc ccc}
      & \multicolumn{3}{c}{$G_A$} & \multicolumn{3}{c}{$\GP$} & \multicolumn{3}{c}{$G_P$} \\ \cline{2-4}\cline{5-7}\cline{8-10}
      $Q^2 [\GeV]$ & $\Ssim$ & $\Sfour$ & $\Stwo$ & $\Ssim$ & $\Sfour$ & $\Stwo$ & $\Ssim$ & $\Sfour$ & $\Stwo$ \\ \hline
  0.049(0) & 1.284(37) & 1.281(35) & 1.197(11) & 66.90(4.61)& 67.10(3.01)& 38.95(1.00)& 98.55(6.16)& 98.65(4.23)& 57.55(1.44)\\
  0.097(0) & 1.153(19) & 1.152(20) & 1.141(10) & 36.15(1.21)& 34.92(94)  & 25.08(55)  & 55.27(1.97)& 53.17(1.32)& 38.23(92)  \\
  0.143(0) & 1.091(17) & 1.092(17) & 1.094(10) & 24.32(73)  & 23.94(58)  & 18.69(42)  & 36.65(1.08)& 36.14(80)  & 28.41(65)  \\
  0.189(1) & 1.014(13) & 1.022(17) & 1.044(10) & 17.42(36)  & 17.56(41)  & 14.58(32)  & 26.35(65)  & 26.32(59)  & 22.00(50)  \\
  0.234(1) & 0.971(12) & 0.976(13) & 0.997(09) & 13.80(31)  & 13.76(28)  & 11.88(26)  & 21.34(49)  & 21.09(38)  & 18.08(34)  \\
  0.277(1) & 0.942(13) & 0.945(13) & 0.957(10) & 11.57(29)  & 11.55(24)  & 10.00(23)  & 17.75(37)  & 17.70(33)  & 15.51(32)  \\
  0.361(2) & 0.877(11) & 0.867(13) & 0.881(12) &  8.244(191)  &  8.044(197)  &  7.399(168)  & 12.84(24)  & 12.85(28)  & 11.71(26)  \\
  0.403(3) & 0.839(13) & 0.839(13) & 0.847(13) &  7.165(179)  &  7.100(180)  &  6.517(170)  & 11.25(28)  & 11.22(26)  & 10.30(25)  \\
  0.404(4) & 0.824(18) & 0.822(17) & 0.827(19) &  6.833(225)  &  6.806(241)  &  6.315(216)  & 11.02(41)  & 10.97(41)  & 10.21(35)  \\
  0.443(4) & 0.804(14) & 0.800(13) & 0.804(15) &  6.399(169)  &  6.150(203)  &  5.797(148)  & 10.24(25)  &  9.999(276)  &  9.110(269)  \\
    \end{tabular}
  \end{ruledtabular}
  \caption{The bare form factors $G_A$, $\GP$, and $G_P$ versus $Q^2$
for the 3 strategies $\Ssim$, $\Sfour$, and $\Stwo$ on ensemble $a09m130W$.}
  \label{tab:ff-a09m130W}
\end{table*}

\begin{table*}[!tbh]  %T18
  \centering
  \renewcommand{\arraystretch}{1.1}
  \begin{ruledtabular}
    \begin{tabular}{c ccc ccc ccc}
      & \multicolumn{3}{c}{$G_A$} & \multicolumn{3}{c}{$\GP$} & \multicolumn{3}{c}{$G_P$} \\ \cline{2-4}\cline{5-7}\cline{8-10}
      $Q^2 [\GeV]$ & $\Ssim$ & $\Sfour$ & $\Stwo$ & $\Ssim$ & $\Sfour$ & $\Stwo$ & $\Ssim$ & $\Sfour$ & $\Stwo$ \\ \hline
  0.190(1) & 1.022(21) & 1.010(23) & 1.033(17) & 16.99(64) & 17.19(68) & 14.40(49) & - & - & - \\
  0.365(2) & 0.868(15) & 0.857(18) & 0.870(15) &  8.793(230) &  8.997(256) &  7.911(221) & - & - & - \\
  0.528(3) & 0.780(22) & 0.774(22) & 0.770(20) &  6.151(188) &  6.149(210) &  5.048(228) & - & - & - \\
  0.690(5) & 0.669(36) & 0.696(34) & 0.670(25) &  3.913(403) &  4.303(242) &  3.543(257) & - & - & - \\
  0.840(6) & 0.599(29) & 0.614(26) & 0.614(19) &  3.080(168) &  3.243(205) &  2.932(192) & - & - & - \\
    \end{tabular}
  \end{ruledtabular}
  \caption{The bare form factors $G_A$, $\GP$, and $G_P$ versus $Q^2$
  for the 3 strategies $\Ssim$, $\Sfour$, and $\Stwo$ on ensemble $a06m310W$.
  Data for $G_P$ were, by accident, not saved.}
  \label{tab:ff-a06m310W}
\end{table*}
%
%\clearpage
\begin{table*}[!tbh]  %T19
  \centering
  \renewcommand{\arraystretch}{1.1}
  \begin{ruledtabular}
    \begin{tabular}{c ccc ccc ccc}
      & \multicolumn{3}{c}{$G_A$} & \multicolumn{3}{c}{$\GP$} & \multicolumn{3}{c}{$G_P$} \\ \cline{2-4}\cline{5-7}\cline{8-10}
      $Q^2 [\GeV]$ & $\Ssim$ & $\Sfour$ & $\Stwo$ & $\Ssim$ & $\Sfour$ & $\Stwo$ & $\Ssim$ & $\Sfour$ & $\Stwo$ \\ \hline
 0.189(01) & 1.001(15) & 1.007(23) & 1.020(17) & 15.67(72) & 16.28(84) & 15.11(64) & - & - & - \\
 0.365(03) & 0.853(10) & 0.856(14) & 0.880(17) &  8.451(313) &  8.635(228) &  8.236(291) & - & - & - \\
 0.532(07) & 0.743(12) & 0.745(17) & 0.723(29) &  5.441(168) &  5.546(188) &  4.964(297) & - & - & - \\
 0.683(10) & 0.677(12) & 0.718(28) & 0.663(32) &  3.926(125) &  4.409(216) &  3.967(368) & - & - & - \\
 0.846(12) & 0.599(14) & 0.618(21) & 0.554(42) &  2.929(088) &  3.120(105) &  2.762(240) & - & - & - \\
    \end{tabular}
  \end{ruledtabular}
  \caption{The bare form factors $G_A$, $\GP$, and $G_P$ versus $Q^2$
for the 3 strategies $\Ssim$, $\Sfour$, and $\Stwo$ on ensemble $a06m310$.
  Data for $G_P$ were, by accident, not saved.}
  \label{tab:ff-a06m310}
\end{table*}

\begin{table*}[!tbh]  %T20
  \centering
  \renewcommand{\arraystretch}{1.1}
  \begin{ruledtabular}
    \begin{tabular}{c ccc ccc ccc}
      & \multicolumn{3}{c}{$G_A$} & \multicolumn{3}{c}{$\GP$} & \multicolumn{3}{c}{$G_P$} \\ \cline{2-4}\cline{5-7}\cline{8-10}
      $Q^2 [\GeV]$ & $\Ssim$ & $\Sfour$ & $\Stwo$ & $\Ssim$ & $\Sfour$ & $\Stwo$ & $\Ssim$ & $\Sfour$ & $\Stwo$ \\ \hline
 0.109(0) & 1.161(48) & 1.152(37) & 1.124(21) & 29.89(1.99) & 30.67(1.62)& 22.81(89) & - & - & - \\
 0.213(1) & 1.022(29) & 0.999(23) & 1.008(20) & 15.51(71)  & 15.35(56)  & 13.64(40) & - & - & - \\
 0.313(2) & 0.909(24) & 0.898(25) & 0.901(27) &  9.762(387)  & 10.030(490)  &  8.987(396) & - & - & - \\
 0.412(6) & 0.860(34) & 0.860(31) & 0.801(52) &  7.669(431)  &  7.993(393)  &  6.817(437) & - & - & - \\
 0.504(6) & 0.777(27) & 0.774(27) & 0.762(35) &  5.701(264)  &  5.694(253)  &  5.539(305) & - & - & - \\
    \end{tabular}
  \end{ruledtabular}
  \caption{The bare form factors $G_A$, $\GP$, and $G_P$ versus $Q^2$
for the 3 strategies $\Ssim$, $\Sfour$, and $\Stwo$ on ensemble $a06m220W$.
  Data for $G_P$ were, by accident, not saved.}
  \label{tab:ff-a06m220W}
\end{table*}

\begin{table*}[!tbh]  %T21
  \centering
  \renewcommand{\arraystretch}{1.1}
  \begin{ruledtabular}
    \begin{tabular}{c ccc ccc ccc}
      & \multicolumn{3}{c}{$G_A$} & \multicolumn{3}{c}{$\GP$} & \multicolumn{3}{c}{$G_P$} \\ \cline{2-4}\cline{5-7}\cline{8-10}
      $Q^2 [\GeV]$ & $\Ssim$ & $\Sfour$ & $\Stwo$ & $\Ssim$ & $\Sfour$ & $\Stwo$ & $\Ssim$ & $\Sfour$ & $\Stwo$ \\ \hline
 0.110(0) & 1.186(36) & 1.149(34) & 1.124(16) & 30.75(1.54)& 30.47(1.37)& 21.36(66) & 43.30(2.01)& 42.85(1.66)& 29.59(67) \\
 0.216(1) & 1.005(15) & 0.973(20) & 1.007(15) & 16.04(47)  & 15.59(46)  & 13.08(36) & 22.74(66)  & 22.14(54)  & 18.43(36) \\
 0.318(2) & 0.910(24) & 0.862(20) & 0.918(18) & 10.53(42)  & 10.08(31)  &  9.228(312) & 14.83(52)  & 14.93(41)  & 13.39(32) \\
 0.414(5) & 0.823(23) & 0.807(22) & 0.850(22) &  7.449(289)  &  7.439(259)  &  6.917(272) & 11.14(33)  & 10.98(35)  & 10.37(32) \\
 0.509(6) & 0.757(18) & 0.754(19) & 0.777(20) &  5.807(188)  &  5.781(186)  &  5.295(199) &  8.463(317)  &  8.621(274)  &  8.081(259) \\
    \end{tabular}
  \end{ruledtabular}
  \caption{The bare form factors $G_A$, $\GP$, and $G_P$ versus $Q^2$
for the strategies $\Ssim$, $\Sfour$, and $\Stwo$ on ensemble $a06m220$.}
  \label{tab:ff-a06m220}
\end{table*}

%\clearpage
\begin{table*}[!tbh]  %T22
  \centering
  \renewcommand{\arraystretch}{1.1}
  \begin{ruledtabular}
    \begin{tabular}{c ccc ccc ccc}
      & \multicolumn{3}{c}{$G_A$} & \multicolumn{3}{c}{$\GP$} & \multicolumn{3}{c}{$G_P$} \\ \cline{2-4}\cline{5-7}\cline{8-10}
      $Q^2 [\GeV]$ & $\Ssim$ & $\Sfour$ & $\Stwo$ & $\Ssim$ & $\Sfour$ & $\Stwo$ & $\Ssim$ & $\Sfour$ & $\Stwo$ \\ \hline
 0.051(0) & 1.201(51) & 1.211(56) & 1.179(20) & 59.23(4.34)& 61.92(4.15)& 35.54(1.42)& 94.52(6.52)& 99.18(6.04)& 56.03(2.10)\\
 0.102(1) & 1.075(33) & 1.075(34) & 1.109(16) & 32.30(1.88)& 32.17(1.48)& 22.04(64)  & 54.37(2.91)& 54.56(2.25)& 37.84(1.41)\\
 0.151(2) & 0.966(31) & 0.966(34) & 1.041(17) & 20.25(1.02)& 20.52(97)  & 16.21(55)  & 36.14(1.68)& 36.60(1.55)& 28.47(1.20)\\
 0.198(2) & 0.940(24) & 0.948(25) & 1.008(18) & 15.71(74)  & 15.95(57)  & 13.47(48)  & 26.73(1.01)& 26.96(85)  & 22.41(73)  \\
 0.246(3) & 0.876(20) & 0.877(22) & 0.940(20) & 11.82(47)  & 11.75(41)  & 10.33(31)  & 21.32(60)  & 21.32(66)  & 18.53(59)  \\
 0.294(4) & 0.836(17) & 0.838(21) & 0.876(32) &  9.153(297)  &  9.397(327)  &  8.750(352)  & 16.88(49)  & 17.34(54)  & 15.85(55)  \\
 0.386(6) & 0.778(19) & 0.782(18) & 0.788(37) &  6.977(212)  &  6.977(182)  &  6.705(320)  & 12.84(37)  & 12.83(37)  & 11.73(44)  \\
 0.431(5) & 0.755(19) & 0.755(18) & 0.740(34) &  5.883(182)  &  5.862(188)  &  5.665(321)  & 11.17(34)  & 11.25(35)  & 10.39(46)  \\
 0.432(5) & 0.739(22) & 0.750(21) & 0.753(37) &  6.129(219)  &  6.163(202)  &  5.909(331)  & 11.38(41)  & 11.33(39)  & 10.79(56)  \\
 0.475(6) & 0.718(21) & 0.736(19) & 0.707(28) &  5.386(184)  &  5.369(168)  &  4.757(193)  &  9.941(344)  &  9.931(313)  &  9.117(407)  \\
    \end{tabular}
  \end{ruledtabular}
  \caption{The bare form factors $G_A$, $\GP$, and $G_P$ versus $Q^2$
for the 3 strategies $\Ssim$, $\Sfour$, and $\Stwo$ on ensemble $a06m135$.}
  \label{tab:ff-a06m135}
\end{table*}

\clearpage
\onecolumngrid
\section{Results for $g_A$,  $\expv{r_A^2}$, $g_P^\ast$ and $g_{\pi NN}$ }     %A0D
\label{sec:EnsembleResults}

The results for $g_A$, $\expv{r_A^2}$, $g_P^\ast$,  $g_{\pi NN}F_\pi$ and $\frac{g_{\pi NN}F_\pi}{M_N}$ 
from the thirteen ensembles are 
given in Tables~\ref{tab:Qsqfit-gA-1},~\ref{tab:Qsqfit-gPstar-PD} and~\ref{tab:Qsqfit-gPstar-2}. 

\vspace{1.5cm}

\twocolumngrid

%%%%%%%%%%%%%%%%%%%
%%%   Combination of tables 29 -- 32 for $z^2$ and $\Ssim$
%%%%%%%%%%%%%%%%%%

\begin{table}[!tbh]    %T23
  \centering
  \renewcommand{\arraystretch}{1.1}
  \begin{ruledtabular}
    \begin{tabular}{l ccc ccc}
      ID       & $g_A$     & $\expv{r_A^2}$ & $\chi^2/dof$ & $p$ \\ \hline
      a15m310  & 1.211(30) & 0.229(11)  & 1.07 & 0.38 \\
      a12m310  & 1.209(40) & 0.221(17)  & 0.29 & 0.94 \\
      a12m220L & 1.246(43) & 0.300(25)  & 2.39 & 0.01 \\
      a12m220  & 1.234(46) & 0.292(28)  & 0.87 & 0.56 \\
      a12m220S & 1.331(80) & 0.331(59)  & 0.25 & 0.96 \\
      a09m310  & 1.188(49) & 0.250(11)  & 0.81 & 0.56 \\
      a09m220  & 1.233(54) & 0.297(21)  & 1.33 & 0.21 \\
      a09m130W & 1.272(65) & 0.446(72)  & 1.30 & 0.22 \\
      a06m310  & 1.158(44) & 0.239(18)  & 0.56 & 0.74 \\
      a06m310W & 1.165(48) & 0.221(24)  & 0.59 & 0.71 \\
      a06m220  & 1.300(59) & 0.368(45)  & 0.69 & 0.63 \\
      a06m220W & 1.261(70) & 0.311(50)  & 0.42 & 0.83 \\
      a06m135  & 1.349(85) & 0.74(13)   & 0.63 & 0.71 
    \end{tabular}
  \end{ruledtabular}
  \caption{Results for $g_A$ and $\expv{r_A^2}$ given by $z^2$
    fits to the axial form factor, $G_A(Q^2)$, obtained with the
    $\Ssim$ strategy. The $\chi^2/dof$ and $p$-value of the fits are also given.}
  \label{tab:Qsqfit-gA-1}
\end{table}

%%%%%%%%%%%%%%%%%%%
%%%   Combination of tables 34 -- 47 for $z^2$ and $\Ssim$
%%%%%%%%%%%%%%%%%%

%\clearpage
\begin{table}[!tbh]   %T24
  \centering
  \renewcommand{\arraystretch}{1.1}
  \begin{ruledtabular}
    \begin{tabular}{l ccc cc}
      ID & $g_P^\ast$  & $g_{\pi NN}F_\pi$  & $\frac{g_{\pi NN}F_\pi}{M_N}$ & $\chi^2/dof$  & $p$  \\ \hline
      a15m310  & 2.16(07) & 1.24(05) & 1.15(04) & 0.92 & 0.43 \\
      a12m310  & 2.44(09) & 1.33(06) & 1.22(05) & 0.61 & 0.61 \\
      a12m220L & 4.02(16) & 1.23(05) & 1.21(05) & 1.34 & 0.23 \\
      a12m220  & 3.73(18) & 1.14(06) & 1.13(06) & 0.68 & 0.69 \\
      a12m220S & 4.71(36) & 1.47(13) & 1.48(13) & 0.46 & 0.71 \\
      a09m310  & 2.37(10) & 1.32(06) & 1.20(05) & 0.86 & 0.46 \\
      a09m220  & 3.98(18) & 1.21(06) & 1.19(06) & 1.15 & 0.33 \\
      a09m130W & 8.38(46) & 1.19(07) & 1.25(07) & 1.05 & 0.39 \\
      a06m310  & 2.20(13) & 1.28(09) & 1.16(07) & 0.01 & 0.99 \\
      a06m310W & 2.31(16) & 1.34(12) & 1.20(11) & 3.48 & 0.03 \\
      a06m220  & 4.38(29) & 1.48(12) & 1.43(12) & 0.27 & 0.77 \\
      a06m220W & 4.16(38) & 1.41(16) & 1.37(15) & 1.36 & 0.26 \\
      a06m135  & 8.35(70) & 1.16(11) & 1.23(11) & 1.44 & 0.23  
    \end{tabular}
  \end{ruledtabular}
  \caption{Results for $g_P^\ast$, $g_{\pi NN}F_\pi$ and $g_{\pi
      NN}F_\pi/M_N$ given by the ``PD'' fits (defined in
    Eq.~\eqref{eq:expandGP}) to $\widetilde{G}_P$ obtained using
    $\Ssim$ strategy.  The $\chi^2/dof$ and $p$-value of the fits are
    also given, and $F_\pi$ and $M_N$ are in units of GeV.}
  \label{tab:Qsqfit-gPstar-PD}
\end{table}

\begin{table}[!tbh]  %T25
  \centering
  \renewcommand{\arraystretch}{1.1}
  \begin{ruledtabular}
    \begin{tabular}{l ccc cc}
      ID       & $g_P^\ast$  &  $g_{\pi NN}F_\pi$ & $\frac{g_{\pi NN}F_\pi}{M_N}$ & $\chi^2/dof$  & $p$ \\ \hline
      a15m310  & 2.22(08) & 1.30(06) & 1.20(05) & 0.46   & 0.84   \\ 
      a12m310  & 2.46(09) & 1.36(06) & 1.25(06) & 0.45   & 0.85   \\
      a12m220L & 4.06(16) & 1.26(06) & 1.24(05) & 0.92   & 0.52   \\
      a12m220  & 3.81(20) & 1.18(08) & 1.16(07) & 0.47   & 0.91   \\
      a12m220S & 4.62(32) & 1.47(12) & 1.48(12) & 0.79   & 0.58   \\
      a09m310  & 2.42(10) & 1.38(06) & 1.25(06) & 0.43   & 0.86   \\
      a09m220  & 4.11(20) & 1.28(07) & 1.26(07) & 0.66   & 0.76   \\
      a09m130W & 8.78(58) & 1.28(10) & 1.34(10) & 0.73   & 0.70   \\
      a06m310  & 2.20(13) & 1.29(09) & 1.17(07) & 0.23   & 0.95   \\
      a06m310W & 2.29(11) & 1.34(08) & 1.20(07) & 1.49   & 0.19   \\
      a06m220  & 4.28(23) & 1.45(09) & 1.40(09) & 0.42   & 0.84   \\
      a06m220W & 3.95(26) & 1.33(10) & 1.29(09) & 0.91   & 0.48   \\
      a06m135  & 8.39(73) & 1.18(13) & 1.25(13) & 0.83   & 0.55   
    \end{tabular}
  \end{ruledtabular}
  \caption{The values of $g_P^\ast$, $g_{\pi NN}F_\pi$ and $g_{\pi
      NN}F_\pi/M_N$ given by $z^2$ fits to $\Ssim$ strategy data for
    ${\widetilde F}_P$ defined in Eq.~\eqref{eq:GP2FP}.  The
    $\chi^2/dof$ and $p$-value of the fits are also given, and $F_\pi$
    and $M_N$ are in units of GeV.}
  \label{tab:Qsqfit-gPstar-2}
\end{table}

%%%%%%%%%%%%%%%%%%%%%%

\clearpage
\onecolumngrid
\section{Summary of CCFV Fits}          %A0E
\label{sec:ccfv-summary}

This appendix presents results after chiral-continuum-finite-volume
extrapolation for $g_A$ and $\expv{r_A^2}$ in
Table~\ref{tab:ccfv-rA-13pt-Ssim}; $g_P^\ast$ in
Tables~\ref{tab:ccfv-gPstar-13pt-Ssim}
and~\ref{tab:ccfv-pole-gPstar-13pt-Ssim}; and $g_{\pi NN}$ in
Table~\ref{tab:ccfv-gpiNN-13pt-Ssim}. All the data for $G_A(Q^2)$ and
${\widetilde G}_P(Q^2)$ used were obtained with the $\Ssim$ strategy
to remove ESC, and the $Q^2$ behavior was fit using either the $z^2$
truncation ($G_A(Q^2)$) or the PD fit (${\widetilde G}_P(Q^2)$) given
in Eq.~\eqref{eq:expandGP}. The four parameters, $b_{0,1,2,3}$, define
the CCFV ansatz given in Eq.~\eqref{eq:ccfv}. The tables also give the
$\chi^2/\text{DOF}$, the $p$-value, and the Akaika Information
Criteria (AIC and AICc)~\cite{1100705} scores for the CCFV fit. The
definition of AIC is given in Appendix~\ref{sec:spectrum}, and including 
correction for small sample sizes, AICc is defined as ${rm AICc} = AIC +
(2k^2 + 2k) / (n-k-1)$ where $n$ is the number of data points and $k$
is the number of parameters.

\vspace{1.5cm}

%%%%%%%%%%%%%%%%%
% rA, sim_A
%%%%%%%%%%%%%%%%%

\begin{table*}[!tbh]  %T26
  \centering
  \renewcommand{\arraystretch}{1.1}
  \begin{ruledtabular}
    \begin{tabular}{c cccc cccc}
            & $\chi^2/\text{DOF}$ & $p$ & AIC & AICc & $b_0[1]$ & $b_1[a]$    & $b_2[M_\pi^2]$ & $b_3[\text{FV}]$ \\
            &                     &     &     &      &          & fm${}^{-1}$ & GeV${}^{-2}$   & GeV${}^{-2}$   \\
 \hline
      \hline \multicolumn{9}{c}{$g_A$ (obtained from $G_A$ with $\Ssim$ and  $z^2$ fit) extrapolated using the 13-point CCFV fit} \\ \hline
 1.296(050) &  0.254 &  0.986 & 10.3 & 15.3 &  1.332(058) &  0.002(477) & -1.967(719) & 41.370(37.926) \\
 1.277(047) &  0.348 &  0.968 &  9.5 & 12.1 &  1.303(052) &  0.284(402) & -1.402(498) &  - \\
 1.219(042) &  1.037 &  0.410 & 15.4 & 16.6 &  1.219(042) &  0.039(392) &  - &  - \\
 1.302(032) &  0.361 &  0.971 &  8.0 &  9.2 &  1.326(040) &  - & -1.325(486) &  - \\
 1.248(027) &  0.940 &  0.500 & 14.3 & 15.5 &  1.248(027) &  - &  - & -23.153(22.360) \\
 1.223(013) &  0.951 &  0.494 & 13.4 & 13.8 &  1.223(013) &  - &  - &  - \\
\hline
      \multicolumn{9}{c}{$\expv{r_A^2}$ (obtained from $G_A$ with $\Ssim$ and  $z^2$ fit) extrapolated using the 13-point CCFV fit} \\ 
\hline
 0.418(033) &  1.310 &  0.225 & 19.8 & 24.8 &  0.457(040) & -0.489(260) & -2.169(449) & 34.126(20.944) \\
 0.384(025) &  1.445 &  0.154 & 20.4 & 23.1 &  0.413(029) & -0.168(170) & -1.596(280) &  - \\
 0.287(019) &  4.267 &  0.000 & 50.9 & 52.1 &  0.287(019) & -0.332(167) &  - &  - \\
 0.369(021) &  1.403 &  0.164 & 19.4 & 20.6 &  0.399(025) &  - & -1.643(276) &  - \\
 0.298(013) &  3.241 &  0.000 & 39.6 & 40.8 &  0.298(013) &  - &  - & -38.490(9.863) \\
 0.251(006) &  4.240 &  0.000 & 52.9 & 53.2 &  0.251(006) &  - &  - &  - \\
    \end{tabular}
  \end{ruledtabular}
  \caption{Summary of the parameters in the 13-point CCFV fit (Eq.~\protect\eqref{eq:ccfv}) to $g_A$
    and $\expv{r_A^2}$. The data used are given in
    Table~\ref{tab:Qsqfit-gA-1}. These were obtained by fitting the
    $Q^2$ behavior of $G_A$, obtained with the $\Ssim$ strategy, using the $z^2$
    truncation. Details are given in Sec.~\ref{ssec:aff-extrap}.}
  \label{tab:ccfv-rA-13pt-Ssim}
\end{table*}

%\clearpage
%%%%%%%%%%%%%%%%%
% gPstar, sim_A
%%%%%%%%%%%%%%%%%

\begin{table*}[!tbh]   %T27
  \centering
  \renewcommand{\arraystretch}{1.1}
  \begin{ruledtabular}
    \begin{tabular}{c cccc cccc}
      $g_P^\ast$ & $\chi^2/\text{DOF}$ & $p$ & AIC & AICc & $b_0[1]$ & $b_1[a]$    & $b_2[M_\pi^2]$ & $b_3[\text{FV}]$ \\
                 &                     &     &     &      &          & fm${}^{-1}$ & GeV${}^{-2}$   & GeV${}^{-2}$   \\
 \hline
      \hline \multicolumn{9}{c}{$g_P^\ast$ (obtained from ${\widetilde F}_P$ with $\Ssim$ and fit using $z^2$) extrapolated using the 13-point CCFV fit} \\ \hline
 9.300(459) &  0.897 &  0.527 & 16.1 & 21.1 &  0.261(015) & -0.117(124) & -0.018(180) &  6.359(9.373) \\
 9.213(441) &  0.853 &  0.577 & 14.5 & 17.2 &  0.257(013) & -0.079(110) &  0.067(129) &  - \\
 9.301(408) &  0.800 &  0.640 & 12.8 & 14.0 &  0.261(011) & -0.066(107) &  - &  - \\
 8.969(281) &  0.822 &  0.618 & 13.0 & 14.2 &  0.251(010) &  - &  0.047(126) &  - \\
 8.968(240) &  0.815 &  0.625 & 13.0 & 14.2 &  0.251(007) &  - &  - &  2.739(5.929) \\
 9.062(124) &  0.765 &  0.687 & 11.2 & 11.5 &  0.254(003) &  - &  - &  - \\
      \hline \multicolumn{9}{c}{$g_P^\ast$ (obtained from ${\widetilde G}_P$ with $\Ssim$ and PD fit, Eq.~\protect\eqref{eq:expandGP}) extrapolated using the 13-point CCFV fit} \\ \hline
 9.248(484) &  1.182 &  0.301 & 18.6 & 23.6 &  0.258(015) & -0.178(135) &  0.075(181) &  4.550(9.448) \\
 9.167(454) &  1.087 &  0.368 & 16.9 & 19.5 &  0.255(013) & -0.148(119) &  0.138(127) &  - \\
 9.274(443) &  1.096 &  0.360 & 16.1 & 17.3 &  0.260(012) & -0.105(112) &  - &  - \\
 8.708(264) &  1.129 &  0.333 & 16.4 & 17.6 &  0.243(009) &  - &  0.086(119) &  - \\
 8.793(228) &  1.159 &  0.310 & 16.7 & 17.9 &  0.247(006) &  - &  - &  2.360(5.527) \\
 8.876(123) &  1.078 &  0.374 & 14.9 & 15.3 &  0.249(003) &  - &  - &  - \\
    \end{tabular}
  \end{ruledtabular}
  \caption{Summary of parameters values in the 13-point CCFV fit (see
    Eq.~\eqref{eq:ccfv}) for obtaining $g_P^\ast$. The data used are
    given in Tables~\ref{tab:Qsqfit-gPstar-PD}
    and~\ref{tab:Qsqfit-gPstar-2}.  In the top half,
    the quantity $({Q^\ast}^2 + M_\pi^2) g_P^\ast = 2 m_\mu M_N
    {\widetilde F}_P(Q^{\ast 2})$, with $\FP$ is defined in
    Eq.~\eqref{eq:GP2FP} and fit using $z^2$, is extrapolated, while
    in the bottom half $({Q^\ast}^2 + M_\pi^2) g_P^\ast = ({Q^\ast}^2
    + M_\pi^2) (m_\mu/2 M_N) {\widetilde G}_P(Q^{\ast 2})$ is used.
    The extrapolated results are then converted to $g_P^\ast$ by
    dividing by the physical value of $({Q^\ast}^2 + M_\pi^2)$.
    Details are given in Sec.~\ref{sec:aff-gpx}.}
  \label{tab:ccfv-gPstar-13pt-Ssim}
\end{table*}

\begin{table*}[!tbh]    %T28
  \centering
  \renewcommand{\arraystretch}{1.1}
  \begin{ruledtabular}
    \begin{tabular}{c cccc ccccc}
      $g_P^\ast$ & $\chi^2/\text{DOF}$ & $p$ & AIC & AICc & $b_0[1]$ & $b_1[a]$ & $b_2[M_\pi^2]$ & $b_3[\text{FV}]$ & $b_4[\text{pole}]$ \\
                 &                     &     &     &      &          & fm${}^{-1}$ & GeV${}^{-2}$& GeV${}^{-2}$ & GeV${}^{2}$   \\
\hline
      \hline \multicolumn{10}{c}{$g_P^\ast$ (obtained from ${\widetilde F}_P(Q^{\ast 2})$ using $\Ssim$ data and fit using $z^2$) extrapolated using the 13-point CCFV plus pole fit} \\ \hline
 8.763(479) &  0.978 &  0.451 & 17.8 & 26.4 &  0.917(944) & -1.327(1.523) & -6.027(6.740) & 83.890(129.644) &  0.223(035) \\
 8.770(479) &  0.916 &  0.510 & 16.2 & 21.2 &  0.692(878) & -0.655(1.114) & -3.825(5.817) &  - &  0.228(034) \\
 9.019(292) &  0.867 &  0.563 & 14.7 & 17.3 &  0.127(182) & -0.679(1.113) &  - &  - &  0.249(011) \\
 8.714(469) &  0.859 &  0.572 & 14.6 & 17.3 &  0.631(872) &  - & -3.935(5.814) &  - &  0.229(034) \\
 8.950(294) &  0.900 &  0.532 & 15.0 & 17.7 &  0.086(226) &  - &  - & -18.315(86.479) &  0.249(014) \\
 8.969(281) &  0.822 &  0.618 & 13.0 & 14.2 &  0.047(126) &  - &  - &  - &  0.250(011) \\
      \hline \multicolumn{10}{c}{$g_P^\ast$ (obtained from ${\widetilde G}_P$ with $\Ssim$ and ``PD'' fit, Eq.~\protect\eqref{eq:expandGP}) extrapolated using the 13-point CCFV plus pole fit} \\ \hline
 8.590(418) &  1.273 &  0.252 & 20.2 & 28.8 &  1.075(905) & -2.081(1.715) & -6.246(6.684) & 83.651(136.326) &  0.214(031) \\
 8.585(418) &  1.174 &  0.307 & 18.6 & 23.6 &  0.813(798) & -1.322(1.186) & -3.821(5.390) &  - &  0.220(030) \\
 8.806(277) &  1.107 &  0.352 & 17.1 & 19.7 &  0.265(195) & -1.376(1.184) &  - &  - &  0.240(011) \\
 8.468(404) &  1.180 &  0.298 & 17.8 & 20.5 &  0.698(791) &  - & -4.209(5.378) &  - &  0.220(030) \\
 8.645(279) &  1.191 &  0.291 & 17.9 & 20.6 &  0.219(222) &  - &  - & -58.710(82.451) &  0.236(013) \\
 8.708(264) &  1.129 &  0.333 & 16.4 & 17.6 &  0.086(119) &  - &  - &  - &  0.242(010) \\
    \end{tabular}
  \end{ruledtabular}
  \caption{Summary of parameters values in the 13-point CCFV fit
    (Eq.~\eqref{eq:ccfv} plus an additional ``pole'' term $b_4/({Q^\ast}^2 +
    M_\pi^2)$) for obtaining $g_P^\ast$. The data used are given in
    Tables~\ref{tab:Qsqfit-gPstar-PD} and~\ref{tab:Qsqfit-gPstar-2}.
    In the top half, the quantity $g_P^\ast = ( m_\mu / 2M_N)
    {\widetilde F}_P(Q^{\ast 2})/({Q^\ast}^2 + M_\pi^2)$, is
    extrapolated, while in the bottom half $g_P^\ast = (m_\mu/2 M_N)
    {\widetilde G}_P(Q^{\ast 2})$ is used.  Details are given in
    Sec.~\ref{sec:aff-gpx}.}
  \label{tab:ccfv-pole-gPstar-13pt-Ssim}
\end{table*}

%%%%%%%%%%%%%%%%%
% gpiNN, sim_A
%%%%%%%%%%%%%%%%%

\begin{table*}[!tbh]     %T29
  \centering
  \renewcommand{\arraystretch}{1.1}
  \begin{ruledtabular}
    \begin{tabular}{c cccc cccc}
      $g_{\pi NN}$ & $\chi^2/\text{DOF}$ & $p$ & AIC & AICc & $b_0[1]$ & $b_1[a]$    & $b_2[M_\pi^2]$ & $b_3[\text{FV}]$ \\
                   &                     &     &     &      &          & fm${}^{-1}$ & GeV${}^{-2}$   & GeV${}^{-2}$   \\
 \hline
      \hline \multicolumn{9}{c}{$g_{\pi NN}$ (obtained with $\Ssim$ and fit using $z^2$) extrapolated using the 13-point CCFV fit} \\ \hline
14.491(857) &  0.878 &  0.544 & 15.9 & 20.9 &  1.330(090) & -0.942(752) &  0.339(1.093) & 55.002(56.824) \\
14.273(827) &  0.884 &  0.547 & 14.8 & 17.5 &  1.296(083) & -0.637(683) &  1.080(779) &  - \\
14.713(764) &  0.979 &  0.463 & 14.8 & 16.0 &  1.357(070) & -0.448(669) &  - &  - \\
13.666(511) &  0.883 &  0.556 & 13.7 & 14.9 &  1.243(060) &  - &  0.935(763) &  - \\
13.777(435) &  0.886 &  0.553 & 13.7 & 14.9 &  1.270(040) &  - &  - & 44.246(36.550) \\
14.225(228) &  0.934 &  0.511 & 13.2 & 13.6 &  1.312(021) &  - &  - &  - \\
      \hline \multicolumn{9}{c}{$g_{\pi NN}$ (obtained from ${\widetilde G}_P$ with $\Ssim$ and PD fit, Eq.~\protect\eqref{eq:expandGP}) extrapolated using the 13-point CCFV fit} \\ \hline
14.135(852) &  1.202 &  0.288 & 18.8 & 23.8 &  1.283(087) & -1.230(786) &  1.098(1.026) & 28.874(53.442) \\
13.975(799) &  1.111 &  0.349 & 17.1 & 19.8 &  1.261(077) & -1.036(699) &  1.497(711) &  - \\
14.240(788) &  1.412 &  0.159 & 19.5 & 20.7 &  1.313(073) & -0.519(655) &  - &  - \\
12.986(438) &  1.209 &  0.274 & 17.3 & 18.5 &  1.177(051) &  - &  1.127(666) &  - \\
13.271(381) &  1.349 &  0.190 & 18.8 & 20.0 &  1.224(035) &  - &  - & 35.810(31.073) \\
13.637(210) &  1.347 &  0.184 & 18.2 & 18.5 &  1.257(019) &  - &  - &  - \\
    \end{tabular}
  \end{ruledtabular}
  \caption{Summary of the 13-point CCFV fit parameters for the
    extraction of $g_{\pi NN}$ as described in
    Sec.~\ref{sec:gpiNN}. The data used are given in
    Tables~\ref{tab:Qsqfit-gPstar-PD} and~\ref{tab:Qsqfit-gPstar-2}.
    In the top table, the product $g_{\pi NN}F_\pi = M_N
    \FP(-M_\pi^2)$ is extrapolated, and the result, in the continuum,
    is divided by $F_\pi = 92.9\,\MeV$. In the bottom table,
    $\FP(-M_\pi^2)$ is extrapolated and the result in the continuum
    multiplied by $M_N/F_\pi$.  }
  \label{tab:ccfv-gpiNN-13pt-Ssim}
\end{table*}

\clearpage
\onecolumngrid
\section{Data and fits for the extraction of isovector charges $g_{A,S,T}$ from forward matrix elements}     %A0F
\label{sec:gASTdata}

This appendix gives the mass gaps of excited state for the 3-RD-$N\pi$
fit in units of the lattice pion mass for each ensemble. The $M_1$ is
fixed to the noninteracting energy of the $N(\bm{n})+\pi(-\bm{n})$
state with $\bm{n}=(1,0,0)$. The $M_2$ is constrained to be near the
first excited state mass given by the two-point correlator by using
the narrow prior shown in the last column in
Table~\ref{tab:charge-Npi-hard-params}. The results for the bare isovector charges
$g_{A,S,T,V}$ from the forward matrix elements and the $p$-value of the fit are 
given in Table~\ref{tab:charge} for the three strategies 3-RD, 3$^\ast$, and 3-RD-$N\pi$. 

The parameters of the fits defined in Eq.~\eqref{eq:charge-s3} and the
mass gaps for the 3-RD strategy are given in
Table~\ref{tab:charge-esc-params}. Summary of the various CCFV fits to
the 3-RD data for the renormalized isovector charges $g_{A,S,T}$ are given
in Tables~\ref{tab:ccfv-gA-3RD},~\ref{tab:ccfv-gS-3RD}, and~\ref{tab:ccfv-gT-3RD}.
The extraction of the final values at the physical point from these data are 
discussed in the main text.

\vspace{1.5cm}

% summary table for the 3-state-no-diag-trans-Npi fit
\begin{table*}[!tbh]        %T30
  \centering
  \renewcommand{\arraystretch}{1.1}
  \begin{ruledtabular}
    \begin{tabular}{l c ccccc}
      \multirow{2}{*}{ID} & \multirow{2}{*}{$M_1-M_0$} & \multicolumn{5}{c}{$M_2-M_0$} \\ \cline{3-7}
                          &                            & Axial & Scalar & Tensor & Vector & prior  \\
      \hline
      a15m310    & 2.2 &  3.50(01) &  3.52(01) &  3.52(03) &  3.50(02) &  3.52(12) \\
      a12m310    & 2.0 &  4.32(11) &  4.51(07) &  3.78(14) &  4.46(08) &  4.55(26) \\
      a12m220L   & 1.7 &  5.19(30) &  6.26(11) &  6.01(17) &  6.01(08) &  6.05(44) \\
      a12m220    & 2.0 &  5.93(12) &  6.02(08) &  5.55(23) &  5.88(12) &  6.00(44) \\
      a12m220S   & 2.5 &  6.09(04) &  6.12(03) &  5.97(12) &  6.10(05) &  6.13(36) \\
      a09m310    & 2.0 &  2.91(16) &  3.24(11) &  3.29(08) &  3.43(10) &  3.15(21) \\
      a09m220    & 1.8 &  3.90(66) &  4.43(20) &  4.49(14) &  2.07(07) &  4.35(40) \\
      a09m130W   & 2.1 &  5.83(39) &  7.27(29) &  6.48(23) &  2.37(05) &  5.84(66) \\
      a06m310    & 2.0 &  3.01(02) &  3.03(02) &  3.11(03) &  3.05(01) &  3.04(11) \\
      a06m310W   & 2.0 &  3.95(04) &  3.94(03) &  3.89(06) &  3.95(02) &  3.94(21) \\
      a06m220    & 2.0 &  4.20(13) &  4.40(06) &  4.50(11) &  4.47(06) &  4.49(29) \\
      a06m220W   & 2.0 &  4.91(06) &  4.96(04) &  5.04(10) &  4.95(03) &  4.98(29) \\
      a06m135    & 2.2 &  6.43(07) &  6.84(16) &  6.93(27) &  6.71(08) &  6.60(51) \\
    \end{tabular}
  \end{ruledtabular}
  \caption{Mass gaps of excited state for the 3-RD-$N\pi$ fit 
    in units of the lattice pion mass for each ensemble. The $M_1$ is
    fixed to the noninteracting energy of the
    $N(\bm{n})+\pi(-\bm{n})$ state with $\bm{n}=(1,0,0)$. The
    $M_2$ is constrained to be near the first excited state mass
    given by the two-point correlator by using the narrow prior shown in the last
    column. These mass gaps can be compared with the 3-RD fit results
    given in the Table~\ref{tab:charge-esc-params}. }
  \label{tab:charge-Npi-hard-params}
\end{table*}

\begin{table*}[!tbh]    %T31
  \centering
  \renewcommand{\arraystretch}{1.1}
  \begin{ruledtabular}
    \begin{tabular}{c cc cc cc cc}
      ID & $g_A$ & $p$ & $g_S$ & $p$ & $g_T$ & $p$ & $g_V$ & $p$ \\
      \hline
% charge fit summary table for the 3-state-no-diag-trans fit
a15m310    &  1.266(017) &  0.780 &  0.834(018) &  0.040 &  1.133(006) &  0.819 &  1.073(004) &  0.528 \\
           &  1.250(007) &  0.591 &  0.868(028) &  0.002 &  1.121(006) &  0.641 &  1.069(004) &  0.000 \\
           &  1.243(005) &  0.607 &  0.838(019) &  0.031 &  1.132(004) &  0.811 &  1.070(003)  &  0.394 \\
           \hline
a12m310    &  1.256(006) &  0.247 &  0.929(031) &  0.180 &  1.068(009) &  0.089 &  1.055(005) &  0.097 \\
           &  1.283(018) &  0.436 &  1.091(083) &  0.007 &  1.034(020) &  0.060 &  1.061(008) &  0.106 \\
           &  1.241(005) &  0.047 &  0.910(015) &  0.215 &  1.083(005) &  0.000 &  1.053(002)  &  0.069 \\
           \hline
a12m220L   &  1.275(005) &  0.175 &  0.829(025) &  0.038 &  1.090(007) &  0.679 &  1.068(003) &  0.053 \\
           &  1.289(013) &  0.410 &  0.873(042) &  0.000 &  1.069(011) &  0.194 &  1.067(004) &  0.165 \\
           &  1.266(007) &  0.005 &  0.865(016) &  0.089 &  1.092(003) &  0.690 &  1.064(002)  &  0.035 \\
           \hline
a12m220    &  1.253(010) &  0.252 &  0.987(056) &  0.561 &  1.080(011) &  0.363 &  1.063(004) &  0.892 \\
           &  1.265(021) &  0.173 &  1.113(095) &  0.401 &  1.048(018) &  0.243 &  1.071(009) &  0.622 \\
           &  1.239(007) &  0.157 &  0.929(029) &  0.445 &  1.084(006) &  0.266 &  1.061(003)  &  0.846 \\
           \hline
a12m220S   &  1.257(017) &  0.715 &  0.908(213) &  0.113 &  1.103(027) &  0.982 &  1.065(006) &  0.775 \\
           &  1.266(044) &  0.631 &  1.003(260) &  0.015 &  1.065(039) &  0.754 &  1.081(018) &  1.000 \\
           &  1.245(012) &  0.627 &  0.967(097) &  0.100 &  1.110(011) &  0.861 &  1.061(004)  &  0.713 \\
           \hline
a09m310    &  1.275(017) &  0.593 &  1.000(019) &  0.305 &  1.029(004) &  0.620 &  1.047(002) &  0.034 \\
           &  1.238(008) &  0.426 &  1.016(027) &  0.170 &  1.027(007) &  0.375 &  1.036(004) &  0.080 \\
           &  1.212(004) &  0.000 &  1.006(011) &  0.291 &  1.025(004) &  0.463 &  1.067(008) &  0.000 \\
           \hline
a09m220    &  1.282(016) &  0.173 &  0.987(025) &  0.570 &  1.018(004) &  0.809 &  1.051(002) &  0.449 \\
           &  1.279(013) &  0.440 &  1.056(046) &  0.222 &  1.001(011) &  0.634 &  1.049(004) &  0.325 \\
           &  1.216(006) &  0.000 &  0.989(015) &  0.531 &  1.007(005) &  0.379 &  1.040(007) &  0.000 \\
           \hline
a09m130W   &  1.320(034) &  0.132 &  1.049(023) &  0.542 &  1.010(006) &  0.869 &  1.054(002) &  0.045 \\
           &  1.271(015) &  0.021 &  1.049(061) &  0.069 &  1.000(011) &  0.648 &  1.052(006) &  0.090 \\
           &  1.231(006) &  0.006 &  1.135(024) &  0.068 &  0.990(007) &  0.250 &  1.011(008) &  0.000 \\
           \hline
a06m310    &  1.271(057) &  0.439 &  1.172(082) &  0.873 &  0.992(007) &  0.217 &  1.041(005) &  0.823 \\
           &  1.243(027) &  0.840 &  1.239(108) &  0.352 &  0.982(020) &  0.738 &  1.033(010) &  0.773 \\
           &  1.181(008) &  0.098 &  1.121(003) &  0.829 &  0.980(006) &  0.058 &  1.054(011)  &  0.586 \\
           \hline
a06m310W   &  1.264(089) &  0.397 &  1.115(065) &  0.288 &  0.979(016) &  0.438 &  1.036(005) &  0.886 \\
           &  1.216(021) &  0.669 &  1.122(073) &  0.501 &  0.975(016) &  0.094 &  1.035(011) &  0.413 \\
           &  1.208(012) &  0.358 &  1.144(049) &  0.280 &  0.985(009) &  0.407 &  1.036(005)  &  0.883 \\
           \hline
a06m220    &  1.336(065) &  0.009 &  1.183(157) &  0.625 &  0.975(011) &  0.668 &  1.048(005) &  0.373 \\
           &  1.235(018) &  0.012 &  1.109(066) &  0.275 &  0.975(012) &  0.372 &  1.050(007) &  0.328 \\
           &  1.190(011) &  0.000 &  1.026(028) &  0.484 &  0.975(007) &  0.664 &  1.059(007)  &  0.359 \\
           \hline
a06m220W   &  1.383(079) &  0.751 &  0.818(065) &  0.539 &  0.977(012) &  0.078 &  1.039(006) &  0.908 \\
           &  1.257(024) &  0.643 &  0.769(089) &  0.770 &  0.962(022) &  0.084 &  1.039(009) &  0.724 \\
           &  1.212(012) &  0.303 &  0.866(055) &  0.454 &  0.971(008) &  0.104 &  1.037(004)  &  0.882 \\
           \hline
a06m135    &  1.281(061) &  0.518 &  1.025(050) &  0.460 &  0.966(010) &  0.277 &  1.039(005) &  0.354 \\
           &  1.242(021) &  0.641 &  1.108(110) &  0.382 &  0.950(014) &  0.208 &  1.039(006) &  0.303 \\
           &  1.198(010) &  0.272 &  1.154(073) &  0.312 &  0.942(009) &  0.072 &  1.075(007)  &  0.003 
% charge fit summary table for the 3-state fit
    \end{tabular}
  \end{ruledtabular}
  \caption{Summary of bare charges $g_A$, $g_S$, $g_T$, and $g_V$
    obtained from forward matrix elements along with the $p$-value of the 
    three fits used to remove ESC: 3-RD (first row), 3$^\ast$ (or
    2-state for $g_S$) (second row), and 3-RD-$N\pi$ (third row)
    described in the text.  The mass gap $M_2-M_0$ output by the
    3-RD-$N\pi$ fits is summarized in
    Table~\ref{tab:charge-Npi-hard-params}.}
  \label{tab:charge}
\end{table*}

% summary table for the 3-state-no-diag-trans fit parameters
\begin{table*}[!tbh]       %T32
  \centering
  \renewcommand{\arraystretch}{1.1}
  \begin{ruledtabular}
    \begin{tabular}{lc ccc ccc}
Ensemble ID & Charge & $r_2 b_{02}$ & $r_1 r_2 b_{12}$ & $a\Delta M_2$ & 
                       $\frac{M_1-M_0}{M_\pi}$ & $\frac{M_2-M_0}{M_\pi}$ & $\frac{M(N\pi)-M_0}{M_\pi}$ \\
      \hline
      a15m310    & A  & -0.063(008) & -0.00(00) & -0.37(11) &  3.5(1) &  2.0(4) & 2.2\\
      a12m310    & A  & -0.047(008) & -4.0(2.9) & -0.15(10) &  4.6(2) &  3.7(5) & 2.0\\
      a12m220L   & A  & -0.051(011) & -3.4(1.7) & -0.16(08) &  6.1(1) &  4.9(6) & 1.7\\
      a12m220    & A  & -0.055(011) & -2.9(3.4) & -0.07(19) &  6.0(1) &  5.5(1.4) & 2.0\\
      a12m220S   & A  & -0.041(021) & -2.7(4.8) & -0.18(23) &  6.1(3) &  4.8(1.7) & 2.5\\
      a09m310    & A  & -0.076(013) &  0.12(03) & -0.26(06) &  3.2(5) &  1.3(1) & 2.0\\
      a09m220    & A  & -0.091(012) &  0.18(03) & -0.23(04) &  4.3(4) &  2.0(1) & 1.8\\
      a09m130W   & A  & -0.111(024) &  0.20(04) & -0.19(03) &  5.8(6) &  2.8(3) & 2.1\\
      a06m310    & A  & -0.127(038) &  0.12(14) & -0.17(06) &  3.0(4) &  1.2(3) & 2.0\\
      a06m310W   & A  & -0.070(073) &  0.19(29) & -0.27(06) &  3.9(5) &  1.1(5) & 2.0\\
      a06m220    & A  & -0.191(041) &  0.17(15) & -0.19(03) &  4.5(2) &  1.7(2) & 2.0\\
      a06m220W   & A  & -0.183(058) &  0.42(30) & -0.24(03) &  5.0(4) &  1.5(2) & 2.0\\
      a06m135    & A  & -0.149(052) &  0.32(09) & -0.13(03) &  6.6(3) &  3.3(5) & 2.2\\
           \hline
      a15m310    & S  & -0.190(026) &  1.13(66) &  0.04(08) &  3.5(1) &  3.7(3) & 2.2\\
      a12m310    & S  & -0.273(045) & -5.6(5.7) & -0.14(16) &  4.6(2) &  3.8(8) & 2.0\\
      a12m220L   & S  & -0.224(050) &  10(13) &  0.07(03) &  6.1(1) &  6.5(2) & 1.7\\
      a12m220    & S  & -0.242(034) & -27(18) &  0.04(07) &  6.0(1) &  6.3(5) & 2.0\\
      a12m220S   & S  & -0.256(086) & 18(56) & -0.12(34) &  6.1(3) &  5.3(2.5) & 2.5\\
      a09m310    & S  & -0.325(007) & -0.04(11) &  0.03(09) &  3.2(5) &  3.4(3) & 2.0\\
      a09m220    & S  & -0.324(009) &  0.43(18) & -0.00(00) &  4.3(4) &  4.3(4) & 1.8\\
      a09m130W   & S  & -0.316(037) & -0.35(55) &  0.36(10) &  5.8(6) & 11.8(1.5) & 2.1\\
      a06m310    & S  & -0.402(026) & -0.9(1.3) & -0.06(08) &  3.0(4) &  2.4(7) & 2.0\\
      a06m310W   & S  & -0.401(086) &  7(17) & -0.02(06) &  3.9(5) &  3.8(7) & 2.0\\
      a06m220    & S  & -0.482(079) &  0.2(1.3) & -0.12(06) &  4.5(2) &  2.7(9) & 2.0\\
      a06m220W   & S  & -0.424(351) & 57(64) &  0.09(13) &  5.0(4) &  6.3(1.8) & 2.0\\
      a06m135    & S  & -0.130(267) & -14.0(8.7) &  0.33(10) &  6.6(3) & 15.0(2.5) & 2.2\\
           \hline
      a15m310    & T  &  0.136(007) &  0.10(13) &  0.03(07) &  3.5(1) &  3.6(3) & 2.2\\
      a12m310    & T  &  0.172(007) & -1.17(69) & -0.30(05) &  4.6(2) &  3.0(2) & 2.0\\
      a12m220L   & T  &  0.190(020) & -2.06(92) & -0.12(10) &  6.1(1) &  5.2(7) & 1.7\\
      a12m220    & T  &  0.187(008) & -2.3(1.2) & -0.18(08) &  6.0(1) &  4.7(6) & 2.0\\
      a12m220S   & T  &  0.189(033) & -4.4(2.7) & -0.25(12) &  6.1(3) &  4.3(9) & 2.5\\
      a09m310    & T  &  0.200(002) &  0.33(07) &  0.04(07) &  3.2(5) &  3.4(1) & 2.0\\
      a09m220    & T  &  0.206(003) &  0.62(08) &  0.07(04) &  4.3(4) &  5.0(2) & 1.8\\
      a09m130W   & T  &  0.214(006) &  0.61(07) &  0.11(03) &  5.8(6) &  7.6(3) & 2.1\\
      a06m310    & T  &  0.215(029) &  1.25(71) &  0.11(05) &  3.0(4) &  4.2(6) & 2.0\\
      a06m310W   & T  &  0.227(026) &  0.15(43) & -0.05(07) &  3.9(5) &  3.4(7) & 2.0\\
      a06m220    & T  &  0.191(014) &  0.17(54) &  0.01(04) &  4.5(2) &  4.7(6) & 2.0\\
      a06m220W   & T  &  0.230(028) & -0.15(78) & -0.01(06) &  5.0(4) &  4.8(9) & 2.0\\
      a06m135    & T  &  0.226(016) &  1.91(67) &  0.12(04) &  6.6(3) &  9.7(1.1) & 2.2\\
           \hline
      a15m310    & V  & -0.012(001) &  0.16(08) & -0.23(07) &  3.5(1) &  2.6(3) & 2.2\\
      a12m310    & V  & -0.008(001) & -0.2(2.1) & -0.17(22) &  4.6(2) &  3.7(1.1) & 2.0\\
      a12m220L   & V  & -0.009(001) &  0.04(07) & -0.45(13) &  6.1(1) &  2.7(9) & 1.7\\
      a12m220    & V  & -0.009(001) &  0.06(18) & -0.26(12) &  6.0(1) &  4.1(9) & 2.0\\
      a12m220S   & V  & -0.009(002) &  0.04(18) & -0.36(25) &  6.1(3) &  3.5(1.9) & 2.5\\
      a09m310    & V  & -0.006(000) &  0.36(08) &  0.02(01) &  3.2(5) &  3.3(5) & 2.0\\
      a09m220    & V  & -0.006(000) &  0.53(08) &  0.01(00) &  4.3(4) &  4.5(4) & 1.8\\
      a09m130W   & V  & -0.006(000) &  0.48(03) &  0.01(00) &  5.8(6) &  6.0(7) & 2.1\\
      a06m310    & V  & -0.005(001) &  0.66(28) &  0.01(04) &  3.0(4) &  3.2(7) & 2.0\\
      a06m310W   & V  & -0.006(004) & -0.4(1.3) &  0.13(14) &  3.9(5) &  5.3(1.5) & 2.0\\
      a06m220    & V  & -0.009(002) &  1.29(42) &  0.07(02) &  4.5(2) &  5.5(5) & 2.0\\
      a06m220W   & V  & -0.004(001) &  0.03(17) & -0.17(16) &  5.0(4) &  2.6(2.3) & 2.0\\
      a06m135    & V  & -0.003(001) &  0.78(16) &  0.01(02) &  6.6(3) &  6.8(6) & 2.2
    \end{tabular}
  \end{ruledtabular}
  \caption{Outputs $r_2 b_{02}$, $r_1 r_2 b_{12}$, and the excited
    state mass gap $a\Delta M_2$ of the 3-RD fit for the axial (A),
    scalar (S), tensor (T) and vector (V) charges.  The mass gaps in
    columns 6--8 are in units of $M_\pi$ for that ensemble. The mass
    gap $M(N\pi)-M_0 \equiv M_N(\bm{n})+M_\pi(-\bm{n}) -M_0$ with
    $\bm{n}=(1,0,0)$ is close to $M_2-M_0$ for the axial channel and
    much smaller for the other charges.  Note that $M_1-M_0$ and
    $M(N\pi)-M_0$ are the same for the four charges. }
  \label{tab:charge-esc-params}
\end{table*}

\begin{table*}[!tbh]   %T33
  \centering
  \renewcommand{\arraystretch}{1.1}
  \begin{ruledtabular}
    \begin{tabular}{c cccc cccc}
      $g_A$ & $\chi^2/\text{DOF}$ & $p$-value & AIC & AICc & $c_0[1]$ & $c_1[a]$    & $c_2[M_\pi^2]$ & $c[\text{FV}]$ \\
            &                     &           &     &      &          & fm${}^{-1}$ & GeV${}^{-2}$   & GeV${}^{-2}$   \\
\hline
      \hline \multicolumn{9}{c}{$Z_A g_A^\text{(bare)}$, 13-pt} \\ \hline
1.281(052) &  0.296 &  0.987 &  7.3 &  8.5 &  1.281(052) & -0.59(45) & - & - \\
 1.228(029) &  0.430 &  0.943 &  8.7 &  9.9 &  1.232(036) & - & -0.22(48) & - \\
 1.281(052) &  0.325 &  0.975 &  9.3 & 11.9 &  1.280(054) & -0.60(49) &  0.03(52) & - \\
 1.285(054) &  0.348 &  0.959 & 11.1 & 16.1 &  1.287(057) & -0.69(55) & -0.08(60) & 10.0(28.7) \\
      \hline \multicolumn{9}{c}{$Z_A g_A^\text{(bare)}$, 11-pt} \\ \hline
 1.264(057) &  0.214 &  0.993 &  5.9 &  7.4 &  1.264(057) & -0.45(49) & - & - \\
 1.222(029) &  0.295 &  0.976 &  6.7 &  8.2 &  1.225(037) & - & -0.18(49) & - \\
 1.264(057) &  0.241 &  0.983 &  7.9 & 11.4 &  1.263(058) & -0.47(55) &  0.04(55) & - \\
 1.268(059) &  0.265 &  0.967 &  9.9 & 16.5 &  1.269(062) & -0.54(62) & -0.04(62) &  7.7(28.8) \\
      \hline \multicolumn{9}{c}{$Z_A g_A^\text{(bare)}$, 10-pt} \\ \hline
 1.308(072) &  0.112 &  0.999 &  4.9 &  6.6 &  1.308(072) & -0.93(68) & - & - \\
 1.226(031) &  0.313 &  0.961 &  6.5 &  8.2 &  1.232(040) & - & -0.32(60) & - \\
 1.316(074) &  0.105 &  0.998 &  6.7 & 10.7 &  1.320(078) & -0.90(68) & -0.25(61) & - \\
 1.317(075) &  0.115 &  0.995 &  8.7 & 16.7 &  1.321(078) & -0.88(69) & -0.20(64) & -6.7(31.8) \\
      \hline \multicolumn{9}{c}{$Z_A/Z_V \times g_A^\text{(bare)}/g_V^\text{(bare)}$, 13-pt} \\ \hline
 1.317(037) &  0.334 &  0.979 &  7.7 &  8.9 &  1.317(037) & -0.62(33) & - & - \\
 1.252(014) &  0.651 &  0.786 & 11.2 & 12.4 &  1.253(018) & - & -0.09(26) & - \\
 1.317(038) &  0.367 &  0.961 &  9.7 & 12.3 &  1.317(039) & -0.62(33) &  0.00(26) & - \\
 1.316(038) &  0.403 &  0.934 & 11.6 & 16.6 &  1.315(040) & -0.61(34) &  0.03(31) & -3.0(13.8) \\
      \hline \multicolumn{9}{c}{$Z_A/Z_V \times g_A^\text{(bare)}/g_V^\text{(bare)}$, 11-pt} \\ \hline
 1.310(039) &  0.268 &  0.983 &  6.4 &  7.9 &  1.310(039) & -0.56(34) & - & - \\
 1.250(014) &  0.555 &  0.835 &  9.0 & 10.5 &  1.252(018) & - & -0.07(26) & - \\
 1.309(039) &  0.302 &  0.966 &  8.4 & 11.8 &  1.309(040) & -0.56(35) &  0.01(27) & - \\
 1.308(040) &  0.335 &  0.939 & 10.3 & 17.0 &  1.307(041) & -0.54(36) &  0.05(31) & -3.7(13.8) \\
      \hline \multicolumn{9}{c}{$Z_A/Z_V \times g_A^\text{(bare)}/g_V^\text{(bare)}$, 10-pt} \\ \hline
 1.320(044) &  0.266 &  0.977 &  6.1 &  7.8 &  1.320(044) & -0.66(39) & - & - \\
 1.250(014) &  0.617 &  0.764 &  8.9 & 10.7 &  1.251(019) & - & -0.05(28) & - \\
 1.321(045) &  0.302 &  0.953 &  8.1 & 12.1 &  1.322(047) & -0.66(39) & -0.03(28) & - \\
 1.323(045) &  0.312 &  0.931 &  9.9 & 17.9 &  1.322(047) & -0.65(39) &  0.03(31) & -7.3(14.8) \\
      \hline \multicolumn{9}{c}{$(Z_A g_A^\text{(bare)} + Z_A/Z_V \times g_A^\text{(bare)}/g_V^\text{(bare)})/2$, 13-pt} \\ \hline
 1.292(041) &  0.354 &  0.973 &  7.9 &  9.1 &  1.292(041) & -0.56(36) & - & - \\
 1.238(019) &  0.551 &  0.869 & 10.1 & 11.3 &  1.241(024) & - & -0.18(33) & - \\
 1.292(041) &  0.389 &  0.952 &  9.9 & 12.6 &  1.292(042) & -0.56(38) &  0.00(35) & - \\
 1.294(042) &  0.429 &  0.920 & 11.9 & 16.9 &  1.294(044) & -0.59(41) & -0.03(39) &  3.0(17.6) \\
      \hline \multicolumn{9}{c}{$(Z_A g_A^\text{(bare)} + Z_A/Z_V \times g_A^\text{(bare)}/g_V^\text{(bare)})/2$, 11-pt} \\ \hline
 1.280(044) &  0.268 &  0.983 &  6.4 &  7.9 &  1.280(044) & -0.46(38) & - & - \\
 1.235(019) &  0.411 &  0.930 &  7.7 &  9.2 &  1.238(024) & - & -0.15(33) & - \\
 1.280(044) &  0.301 &  0.966 &  8.4 & 11.8 &  1.280(044) & -0.46(41) & -0.00(36) & - \\
 1.281(045) &  0.343 &  0.934 & 10.4 & 17.1 &  1.282(047) & -0.48(44) & -0.02(40) &  1.5(17.7) \\
      \hline \multicolumn{9}{c}{$(Z_A g_A^\text{(bare)} + Z_A/Z_V \times g_A^\text{(bare)}/g_V^\text{(bare)})/2$, 10-pt} \\ \hline
 1.311(054) &  0.183 &  0.993 &  5.5 &  7.2 &  1.311(054) & -0.77(49) & - & - \\
 1.237(020) &  0.456 &  0.888 &  7.6 &  9.4 &  1.240(026) & - & -0.20(38) & - \\
 1.316(056) &  0.187 &  0.988 &  7.3 & 11.3 &  1.319(058) & -0.75(49) & -0.15(39) & - \\
 1.317(056) &  0.195 &  0.978 &  9.2 & 17.2 &  1.319(058) & -0.73(50) & -0.10(41) & -7.1(19.3) \\
    \end{tabular}
  \end{ruledtabular}
  \caption{Summary of CCFV fits to $g_A$ using Eq.~\protect\eqref{eq:ccfv}. We
    show results for (i) four different trunctions of the
    four-parameter CCFV ansatz; (ii) fits with three different cuts on the 13
    points labeled ``13-pt'', ``11-pt'' and ``10-pt''; and (iii) fits to data obtained with three different renormalization procedures
    defined in the text.}
  \label{tab:ccfv-gA-3RD}
\end{table*}

\begin{table*}[!tbh]  %T34
  \centering
  \renewcommand{\arraystretch}{1.1}
  \begin{ruledtabular}
    \begin{tabular}{c cccc cccc}
      $g_T$ & $\chi^2/\text{DOF}$ & $p$-value & AIC & AICc & $c_0[1]$ & $c_1[a]$    & $c_2[M_\pi^2]$ & $c[\text{FV}]$ \\
            &                     &           &     &      &          & fm${}^{-1}$ & GeV${}^{-2}$   & GeV${}^{-2}$   \\
\hline
      \hline \multicolumn{9}{c}{$Z_T g_T^\text{(bare)}$, 13-pt} \\ \hline
0.990(029) &  0.281 &  0.989 &  7.1 &  8.3 &  0.990(029) &  0.35(32) & - & - \\
 1.001(019) &  0.270 &  0.991 &  7.0 &  8.2 &  0.994(024) & - &  0.38(33) & - \\
 0.980(031) &  0.222 &  0.994 &  8.2 & 10.9 &  0.974(033) &  0.29(33) &  0.32(34) & - \\
 0.982(031) &  0.145 &  0.998 &  9.3 & 14.3 &  0.982(034) &  0.11(38) &  0.02(46) & 30.4(31.8) \\
      \hline \multicolumn{9}{c}{$Z_T g_T^\text{(bare)}$, 11-pt} \\ \hline
 0.983(034) &  0.329 &  0.966 &  7.0 &  8.5 &  0.983(034) &  0.40(35) & - & - \\
 0.999(020) &  0.303 &  0.974 &  6.7 &  8.2 &  0.991(026) & - &  0.45(36) & - \\
 0.978(034) &  0.266 &  0.977 &  8.1 & 11.6 &  0.971(036) &  0.29(37) &  0.35(38) & - \\
 0.981(034) &  0.176 &  0.990 &  9.2 & 15.9 &  0.980(037) &  0.10(42) &  0.05(49) & 30.1(31.9) \\
      \hline \multicolumn{9}{c}{$Z_T g_T^\text{(bare)}$, 10-pt} \\ \hline
 1.012(040) &  0.156 &  0.996 &  5.2 &  7.0 &  1.012(040) &  0.00(46) & - & - \\
 1.004(020) &  0.117 &  0.999 &  4.9 &  6.6 &  1.000(026) & - &  0.22(39) & - \\
 1.005(042) &  0.133 &  0.996 &  6.9 & 10.9 &  1.001(045) & -0.02(47) &  0.22(40) & - \\
 1.000(044) &  0.124 &  0.993 &  8.7 & 16.7 &  0.998(046) & -0.04(47) &  0.09(50) & 16.3(37.5) \\
      \hline \multicolumn{9}{c}{$Z_T/Z_V \times g_T^\text{(bare)}/g_V^\text{(bare)}$, 13-pt} \\ \hline
 1.008(025) &  0.172 &  0.999 &  5.9 &  7.1 &  1.008(025) &  0.38(28) & - & - \\
 1.024(015) &  0.208 &  0.997 &  6.3 &  7.5 &  1.018(019) & - &  0.33(27) & - \\
 0.999(026) &  0.089 &  1.000 &  6.9 &  9.6 &  0.994(028) &  0.33(28) &  0.27(27) & - \\
 1.000(026) &  0.072 &  1.000 &  8.7 & 13.7 &  0.997(029) &  0.27(31) &  0.14(38) & 13.1(26.7) \\
      \hline \multicolumn{9}{c}{$Z_T/Z_V \times g_T^\text{(bare)}/g_V^\text{(bare)}$, 11-pt} \\ \hline
 1.002(029) &  0.192 &  0.995 &  5.7 &  7.2 &  1.002(029) &  0.43(30) & - & - \\
 1.024(016) &  0.227 &  0.991 &  6.0 &  7.5 &  1.017(020) & - &  0.37(29) & - \\
 0.997(029) &  0.105 &  0.999 &  6.8 & 10.3 &  0.992(031) &  0.35(32) &  0.28(30) & - \\
 0.998(029) &  0.088 &  0.999 &  8.6 & 15.3 &  0.995(031) &  0.28(35) &  0.15(40) & 12.8(26.8) \\
      \hline \multicolumn{9}{c}{$Z_T/Z_V \times g_T^\text{(bare)}/g_V^\text{(bare)}$, 10-pt} \\ \hline
 1.009(033) &  0.198 &  0.991 &  5.6 &  7.3 &  1.009(033) &  0.34(38) & - & - \\
 1.026(016) &  0.188 &  0.993 &  5.5 &  7.2 &  1.020(021) & - &  0.29(31) & - \\
 1.000(035) &  0.118 &  0.997 &  6.8 & 10.8 &  0.995(037) &  0.32(39) &  0.27(31) & - \\
 0.995(037) &  0.099 &  0.997 &  8.6 & 16.6 &  0.992(038) &  0.30(39) &  0.14(41) & 15.3(31.7) \\
      \hline \multicolumn{9}{c}{$(Z_T g_T^\text{(bare)} + Z_T/Z_V \times g_T^\text{(bare)}/g_V^\text{(bare)})/2$, 13-pt} \\ \hline
 0.998(020) &  0.383 &  0.963 &  8.2 &  9.4 &  0.998(020) &  0.37(22) & - & - \\
 1.012(013) &  0.387 &  0.962 &  8.3 &  9.5 &  1.005(016) & - &  0.36(22) & - \\
 0.989(021) &  0.249 &  0.991 &  8.5 & 11.2 &  0.984(023) &  0.30(23) &  0.30(23) & - \\
 0.991(021) &  0.167 &  0.997 &  9.5 & 14.5 &  0.990(024) &  0.19(25) &  0.08(32) & 21.8(22.0) \\
      \hline \multicolumn{9}{c}{$(Z_T g_T^\text{(bare)} + Z_T/Z_V \times g_T^\text{(bare)}/g_V^\text{(bare)})/2$, 11-pt} \\ \hline
 0.993(023) &  0.440 &  0.914 &  8.0 &  9.5 &  0.993(023) &  0.41(24) & - & - \\
 1.011(013) &  0.432 &  0.919 &  7.9 &  9.4 &  1.004(017) & - &  0.41(24) & - \\
 0.987(023) &  0.296 &  0.967 &  8.4 & 11.8 &  0.982(025) &  0.31(25) &  0.32(25) & - \\
 0.990(024) &  0.202 &  0.985 &  9.4 & 16.1 &  0.988(026) &  0.19(28) &  0.10(34) & 21.5(22.1) \\
      \hline \multicolumn{9}{c}{$(Z_T g_T^\text{(bare)} + Z_T/Z_V \times g_T^\text{(bare)}/g_V^\text{(bare)})/2$, 10-pt} \\ \hline
 1.012(028) &  0.303 &  0.965 &  6.4 &  8.1 &  1.012(028) &  0.16(32) & - & - \\
 1.014(013) &  0.214 &  0.989 &  5.7 &  7.4 &  1.009(018) & - &  0.26(26) & - \\
 1.003(029) &  0.217 &  0.982 &  7.5 & 11.5 &  0.999(031) &  0.14(32) &  0.25(26) & - \\
 0.998(030) &  0.205 &  0.975 &  9.2 & 17.2 &  0.996(031) &  0.13(32) &  0.13(34) & 14.7(27.1) \\
    \end{tabular}
  \end{ruledtabular}
  \caption{Summary of CCFV fits to $g_T$ using Eq.~\protect\eqref{eq:ccfv}. The rest is same as in Table~\protect\ref{tab:ccfv-gA-3RD}. }
  \label{tab:ccfv-gT-3RD}
\end{table*}

\begin{table*}[!tbh]  %T35
  \centering
  \renewcommand{\arraystretch}{1.1}
  \begin{ruledtabular}
    \begin{tabular}{c cccc cccc}
      $g_S$ & $\chi^2/\text{DOF}$ & $p$-value & AIC & AICc & $c_0[1]$ & $c_1[a]$    & $c_2[M_\pi^2]$ & $c[\text{FV}]$ \\
            &                     &           &     &      &          & fm${}^{-1}$ & GeV${}^{-2}$   & GeV${}^{-2}$   \\
\hline
      \hline \multicolumn{9}{c}{$Z_S g_S^\text{(bare)}$, 13-pt} \\ \hline
0.991(046) &  2.073 &  0.019 & 26.8 & 28.0 &  0.991(046) & -1.37(46) & - & - \\
 0.876(021) &  2.809 &  0.001 & 34.9 & 36.1 &  0.882(027) & - & -0.32(37) & - \\
 0.991(046) &  2.278 &  0.012 & 28.8 & 31.4 &  0.990(047) & -1.40(49) &  0.07(40) & - \\
 1.001(047) &  2.435 &  0.009 & 29.9 & 34.9 &  1.008(050) & -1.61(54) & -0.37(62) & 38.7(41.6) \\
      \hline \multicolumn{9}{c}{$Z_S g_S^\text{(bare)}$, 11-pt} \\ \hline
 1.044(051) &  1.055 &  0.393 & 13.5 & 15.0 &  1.044(051) & -1.83(50) & - & - \\
 0.888(022) &  2.317 &  0.013 & 24.9 & 26.4 &  0.897(028) & - & -0.52(38) & - \\
 1.045(051) &  1.180 &  0.307 & 15.4 & 18.9 &  1.043(051) & -1.88(56) &  0.10(42) & - \\
 1.066(053) &  1.056 &  0.389 & 15.4 & 22.1 &  1.076(056) & -2.27(62) & -0.56(63) & 60.4(42.2) \\
      \hline \multicolumn{9}{c}{$Z_S g_S^\text{(bare)}$, 10-pt} \\ \hline
 1.077(070) &  1.130 &  0.339 & 13.0 & 14.8 &  1.077(070) & -2.20(75) & - & - \\
 0.882(022) &  2.179 &  0.026 & 21.4 & 23.1 &  0.886(028) & - & -0.21(42) & - \\
 1.076(070) &  1.288 &  0.251 & 15.0 & 19.0 &  1.075(071) & -2.23(77) &  0.06(43) & - \\
 0.993(086) &  1.043 &  0.395 & 14.3 & 22.3 &  1.016(080) & -1.67(84) & -1.23(89) & 132.3(79.7) \\
      \hline \multicolumn{9}{c}{$Z_S/Z_V \times g_S^\text{(bare)}/g_V^\text{(bare)}$, 13-pt} \\ \hline
 0.999(053) &  2.456 &  0.005 & 31.0 & 32.2 &  0.999(053) & -1.43(49) & - & - \\
 0.832(027) &  3.202 &  0.000 & 39.2 & 40.4 &  0.826(034) & - &  0.31(47) & - \\
 0.988(054) &  2.396 &  0.008 & 30.0 & 32.6 &  0.973(055) & -1.73(52) &  0.87(50) & - \\
 0.994(054) &  2.513 &  0.007 & 30.6 & 35.6 &  0.990(057) & -1.78(52) &  0.21(76) & 44.6(38.6) \\
      \hline \multicolumn{9}{c}{$Z_S/Z_V \times g_S^\text{(bare)}/g_V^\text{(bare)}$, 11-pt} \\ \hline
 1.072(064) &  1.696 &  0.084 & 19.3 & 20.8 &  1.072(064) & -2.03(56) & - & - \\
 0.846(027) &  3.132 &  0.001 & 32.2 & 33.7 &  0.846(035) & - &  0.04(48) & - \\
 1.083(064) &  1.443 &  0.173 & 17.5 & 21.0 &  1.064(064) & -2.59(63) &  1.05(54) & - \\
 1.093(064) &  1.312 &  0.240 & 17.2 & 23.8 &  1.090(066) & -2.68(64) &  0.16(79) & 59.8(38.9) \\
      \hline \multicolumn{9}{c}{$Z_S/Z_V \times g_S^\text{(bare)}/g_V^\text{(bare)}$, 10-pt} \\ \hline
 1.103(074) &  1.825 &  0.067 & 18.6 & 20.3 &  1.103(074) & -2.35(69) & - & - \\
 0.837(028) &  3.211 &  0.001 & 29.7 & 31.4 &  0.829(037) & - &  0.40(54) & - \\
 1.097(074) &  1.630 &  0.122 & 17.4 & 21.4 &  1.079(075) & -2.71(72) &  1.00(56) & - \\
 0.981(089) &  0.987 &  0.432 & 13.9 & 21.9 &  1.001(082) & -1.69(84) & -1.10(1.05) & 174.6(74.5) \\
      \hline \multicolumn{9}{c}{$(Z_S g_S^\text{(bare)} + Z_S/Z_V \times g_S^\text{(bare)}/g_V^\text{(bare)})/2$, 13-pt} \\ \hline
 1.004(043) &  2.563 &  0.003 & 32.2 & 33.4 &  1.004(043) & -1.46(40) & - & - \\
 0.867(022) &  3.727 &  0.000 & 45.0 & 46.2 &  0.871(028) & - & -0.24(37) & - \\
 1.003(043) &  2.737 &  0.002 & 33.4 & 36.0 &  0.996(044) & -1.62(44) &  0.37(41) & - \\
 1.014(044) &  2.883 &  0.002 & 33.9 & 38.9 &  1.017(047) & -1.81(47) & -0.17(60) & 40.7(34.1) \\
      \hline \multicolumn{9}{c}{$(Z_S g_S^\text{(bare)} + Z_S/Z_V \times g_S^\text{(bare)}/g_V^\text{(bare)})/2$, 11-pt} \\ \hline
 1.057(048) &  1.547 &  0.125 & 17.9 & 19.4 &  1.057(048) & -1.90(44) & - & - \\
 0.880(022) &  3.408 &  0.000 & 34.7 & 36.2 &  0.888(028) & - & -0.47(38) & - \\
 1.064(049) &  1.600 &  0.119 & 18.8 & 22.2 &  1.055(049) & -2.18(51) &  0.47(44) & - \\
 1.085(050) &  1.420 &  0.192 & 17.9 & 24.6 &  1.090(053) & -2.49(55) & -0.28(62) & 58.2(34.4) \\
      \hline \multicolumn{9}{c}{$(Z_S g_S^\text{(bare)} + Z_S/Z_V \times g_S^\text{(bare)}/g_V^\text{(bare)})/2$, 10-pt} \\ \hline
 1.095(063) &  1.636 &  0.109 & 17.1 & 18.8 &  1.095(063) & -2.32(64) & - & - \\
 0.870(023) &  3.294 &  0.001 & 30.4 & 32.1 &  0.871(029) & - & -0.04(43) & - \\
 1.093(063) &  1.755 &  0.092 & 18.3 & 22.3 &  1.085(064) & -2.47(66) &  0.40(45) & - \\
 0.983(080) &  1.212 &  0.296 & 15.3 & 23.3 &  1.007(073) & -1.63(76) & -1.32(89) & 163.2(72.9) \\
    \end{tabular}
  \end{ruledtabular}
  \caption{Summary of CCFV fits to $g_S$ using Eq.~\protect\eqref{eq:ccfv}. The rest is same as in Table~\protect\ref{tab:ccfv-gA-3RD}. }
  \label{tab:ccfv-gS-3RD}
\end{table*}

%-----------
% reference
%-----------
%\bibliographystyle{abbrv} %%% physical review
%% \makeatletter
%% \ifx\@bibitemShut\undefined\let\@bibitemShut\relax\fi
%% \makeatother
%\bibliography{refs} %%% ref.bib file

\end{document}